\documentclass{tlp}
\usepackage{aopmath}
\usepackage[usenames,dvipsnames,svgnames]{xcolor}
\usepackage{graphics}
\usepackage{graphicx}
\usepackage{enumerate}
\usepackage{hyperref}
\usepackage{comment}

\newtheorem{definition}{Definition} 
\newtheorem{example}{Example} 
\newtheorem{problem}{Problem} 

\def\causes{\; {\bf causes} \;}
\def\lif{\; {\bf if} \;}

\def\iff{\; {\bf iff} \;}
\def\impossible{{\bf impossible} \;}

\newcommand{\no}{\mbox{not }}
\newcommand{\st}{\medskip\noindent}
\newcommand{\is}{\ \ \mbox{:-} \ \ }
\newcommand{\br}{\ | \ }

\begin{document}
\bibliographystyle{acmtrans}

\submitted{1 December 2013}
\revised{15 October 2014, 6 April 2015}
\accepted{12 May 2015}

\title{Modular Action Language ${\cal ALM}$}

\author[Daniela Inclezan and Michael Gelfond]
{DANIELA INCLEZAN\\
Department of Computer Science and Software Engineering, Miami University\\
Oxford, OH 45056, USA\\
\email{inclezd@MiamiOH.edu}
\and MICHAEL GELFOND\\
Department of Computer Science, Texas Tech University\\
Lubbock, TX 79409, USA\\
\email{michael.gelfond@ttu.edu}
}

\pagerange{\pageref{firstpage}--\pageref{lastpage}}
\setcounter{page}{1}

\maketitle

\label{firstpage}

\begin{abstract}
The paper introduces a new modular action language, ${\cal ALM}$,
and illustrates the methodology of its use.
It is based on the approach of Gelfond and Lifschitz \citeyear{gl93,gl98}
in which a high-level action language is used as a front end for a logic programming
system description. The resulting logic programming representation is used to
perform various computational tasks.
The methodology based on existing action languages works well for
small and even medium size systems,
but is not meant to deal with larger systems that require
\emph{structuring of knowledge}. $\mathcal{ALM}$ is meant to
remedy this problem. Structuring of knowledge in ${\cal ALM}$ is
  supported by the concepts of \emph{module} (a formal description
of a specific piece of knowledge packaged as a unit),
\emph{module hierarchy}, and \emph{library},
and by the division of
a system description of ${\cal ALM}$ into two parts: \emph{theory} and
\emph{structure}. A \emph{theory} consists of one or more modules
with a common theme, possibly organized into a module hierarchy based on a \emph{dependency
  relation}. It contains
declarations of sorts, attributes, and properties of the domain together
with axioms describing them.  \emph{Structures} are used to describe the
domain's objects.
These features, together with the means for defining classes
of a domain as special cases of previously defined ones,
facilitate the stepwise development,
testing, and readability of a knowledge base, as well as the creation
of knowledge representation libraries.
\end{abstract}



\section{Introduction}
\label{sec:intro}
In this paper we introduce a new modular action language, ${\cal ALM}$,
and illustrate the principles of its use.
Our work builds upon the methodology for representing knowledge about
discrete dynamic systems introduced by Gelfond and Lifschitz \citeyear{gl93,gl98}.
In this approach, a system is viewed as a \emph{transition diagram} whose nodes
correspond to possible states of the system and whose arcs are labeled
by actions. The diagram is defined by a \emph{system description} --
a collection of statements in a high-level \emph{action language}
 expressing the
direct and indirect effects of actions as well as their executability
conditions
(see, for instance, action languages
${\cal A}$ \cite{gl93},
${\cal B}$ \cite{gl98};
${\cal AL}$ \cite{tur97,bg00};
the non-modular extension of ${\cal AL}$ with multi-valued fluents
\cite{dfp07};
${\cal C}$ \cite{giul98};
${\cal C}+$ \cite{gllmt04};
${\cal K}$ \cite{eflpp04};
${\cal D}$ \cite{st12};
${\cal E}$ \cite{km97};
${\cal H}$ \cite{cgw05,sc12}).
Such languages allow concise representations of very large diagrams.
In order to reason about the system, its action language description is
often translated into a logic program under the answer set semantics (\citeNP{gl88}; \citeyearNP{gl91}).
This allows for the use of Answer Set Programming (ASP) \cite{gl91,nie98,mt99} to perform
complex reasoning tasks such as planning, diagnosis, etc.
This methodology was successfully used in a number of interesting
medium size applications, but does not seem to be fully adequate
for applications requiring a larger body of knowledge about actions
and their effects, step-wise design, and multiple use of, possibly
previously designed, pieces of knowledge.
(The phenomenon is of course well known in Computer Science.
Similar considerations led to the early development of notions of
subroutine and module in procedural programming.
In logic programming, early solutions were based on the concepts of macro and template \cite{bdt06,ci06}.)
Just a few examples of domains that we consider large enough to 
benefit from the above-mentioned practices are: 
the Zoo World and Traffic World examples proposed by Erik Sandewall \cite{slmw99}
and modeled in \cite{ht99,aellt04}; 
the Monkey and Banana Problem by John McCarthy \cite{mc63,mc68}
and formalized in \cite{el06,thesiser08};
the Missionaries and Cannibals Problem by John McCarthy \cite{mccarthy98} 
represented in \cite{gk04,thesiser08}.

This inadequacy is due to the fact that most
  action languages, with some notable exceptions
like $MAD$ \cite{lr06,el06,ds07}
and TAL-C \cite{gk04}, have {\em no built-in features for supporting the
description of a domain's ontology and its objects, and for structuring
knowledge and creating knowledge-based libraries.
${\cal ALM}$ is designed to address these problems.} It is based on an
earlier action language, ${\cal AL}$, introduced in \cite{gi09} where it
is called ${\cal AL}_d$, which so far has been the
authors' language of choice (see, for instance, \cite{gk14}).
However, the basic ideas presented in the paper can be
used for defining versions of ${\cal ALM}$ based on other action languages.

${\cal ALM}$ has constructs for representing  \emph{sorts} (i.e., classes, kinds, types,
  categories) of objects relevant to a given domain, their
  \emph{attributes,}\footnote{Attributes are intrinsic properties of a
    sort of objects. In  ${\cal ALM}$ they are represented by {\em possibly} partial
functions defined on elements of that sort.} and a subsort
  relation that can be viewed
  as a directed acyclic graph (DAG).  We refer to this relation as a \emph{sort hierarchy}.
These constructs support a methodology of knowledge representation that
starts with determining the sorts of objects in a domain and
formulating the
domain's causal laws and other axioms in terms of these sorts.
The \emph{specialization} construct of the language,
which corresponds to the links of the sort hierarchy,
allows to define new sorts (including various sorts of actions) in
terms of other, previously defined sorts.

The definition of particular objects populating the sorts is usually
given only when the domain knowledge is used to solve a particular
task, e.g., predicting the effects of some particular sequences of
actions, planning, diagnosis, etc.

It is worth noting that allowing definitions of actions as special
cases of other, previously defined actions was one of the main goals of actions languages
like ${\cal ALM}$ and $MAD$. Such definitions are not allowed
in traditional action languages.
${\cal ALM}$'s solution consists in allowing action sorts, which do
not exist in $MAD$. We believe that the ${\cal ALM}$  solution
is simpler than the one in $MAD$, where special cases of
actions are described using import statements (similar to bridge rules
in ${\cal C}+$).

${\cal ALM}$ also facilitates the introduction of
particular domain objects (including particular actions)
that are defined as \emph{instances} of the corresponding sorts.
For example, an action $go(bob,london,paris)$ can be defined as an
instance of action sort $move$ with attributes $actor$, $origin$, and
$destination$ set to $bob$, $london$, and $paris$ respectively;
action $go(bob,paris)$ is another instance of the same sort
in which the origin of the $move$ is absent.
Note that, since
axioms of the domain are formulated in terms of sorts and their
attributes, they are applicable to both of these actions.
This is very different from the traditional action language
representation of objects as \emph{terms}, which requires separate axioms
for $go(bob,london,paris)$ and $go(bob,paris)$.

Structuring of knowledge in ${\cal ALM}$ is
  supported by the concepts of \emph{module}, \emph{module hierarchy}, and \emph{library},
and by the division of
a system description of ${\cal ALM}$ into two parts: \emph{theory} and
\emph{structure}. \emph{Theories} contain
declarations of sorts, attributes, and properties of the domain together
with axioms describing them, while \emph{structures} are used to describe the
domain's objects.
Rather traditionally, ${\cal ALM}$
views a module as a formal description of a specific piece of knowledge
packaged as a unit.
A theory consists of one or more modules with a common theme, possibly
organized into a module hierarchy based on a \emph{dependency
  relation}.
Modules of a theory can be developed and tested independently,
which facilitates the reuse of knowledge and stepwise development 
and
refinement \cite{Wirth71} of knowledge bases, and increases their \emph{elaboration
  tolerance} \cite{mccarthy98}.

Theories describing recurrent knowledge
may be stored in libraries and used in different applications.
The \emph{structure} part of an ${\cal ALM}$ system description
contains definitions of objects of the domain together with their sorts, values of their
attributes, and \emph{statics} - relations between objects that cannot be
changed by actions. If a system description of ${\cal ALM}$ satisfies
some natural consistency requirements and
provides complete information about the membership of its objects in the
system's sorts then it describes the unique transition diagram
containing all possible trajectories of the system.
In this sense ${\cal ALM}$ is semantically similar to ${\cal AL}$.
There are also some substantial differences.
First, if no complete information about membership of objects in sorts
is given, then the system description specifies the \emph{collection} of
transition diagrams corresponding to various
possible placements of objects in the system's sorts.
This has no analog in ${\cal AL}$. Second, in addition to the
semantics of its system descriptions, ${\cal ALM}$ provides semantics
for its theories. Informally, a theory of ${\cal ALM}$ can be viewed as
a function taking as an input objects of the domain, their sort
membership, and the values of static relations, and returning
the corresponding transition diagram -- a possible \emph{model} of the theory.
(This definition has some
similarity with the notions of module developed for logic programs
under the answer set semantics, e.g. \cite{oj06} and \cite{lt13}.
Accurate mathematical analysis of these
   similarities and their use for automatic reasoning in ${\cal ALM}$
   is a matter for future research.)
The availability of a formal semantics clarifies the notion of
an ${\cal ALM}$ theory
 and allows us to define an entailment relation ($T$ entails $q$ if $q$ is true
in every model of $T$).

To accurately define the semantics of ${\cal ALM}$ theories, we introduce the
notion of a \emph{basic action theory} (${\cal BAT}$) --- a pair
consisting of a specific type of sorted signature (which we call an
\emph{action signature}), and a set of axioms over this
signature. An interpretation $I$ of the signature of a ${\cal BAT}$ theory
$T$ defines: objects, their sort membership, and statics; while $T$ can be
viewed as a function that takes $I$ as input and returns the
transition diagram $T(I)$ defined by $I$.
In a sense, $T(I)$ is very similar to system descriptions of ${\cal AL}$ and
other traditional action languages. The difference is in the forms of their
signatures and axioms. As in ${\cal AL}$,
the precise definition of states and transitions of $T(I)$
is given in terms of its translation into logic programs
under the answer set semantics.

{\em A system description $D$ of ${\cal ALM}$ can be viewed as a formal
definition of a particular ${\cal BAT}$ theory $T$, and a class of its
interpretations.} The latter is given by the structure of $D$, the
former by its theory.
If the structure of $D$ is complete, i.e., defines
exactly one interpretation $I$, then $D$ represents $T(I)$.


An earlier version of ${\cal ALM}$
has been tested in the context of a real-life application,
as part of our collaboration on Project Halo.
Project Halo is a research effort by Vulcan Inc. aimed towards the development of
a Digital Aristotle -- \emph{``an application containing large volumes of
scientific knowledge and capable of applying sophisticated
problem-solving methods to answer novel questions''}
\cite{projecthalo10}.
The Digital Aristotle uses the knowledge representation
language called SILK (Semantic Inferencing on Large Knowledge) \cite{gdk09},
which is based on the well-founded semantics \cite{vgrs91} and
transaction logic with defaults and argumentation theories \cite{fk11}.
Our first contribution to Project Halo consisted in
creating an ${\cal ALM}$ formalization of an important biological process, \emph{cell division} \cite{ig11}.
The use of ${\cal ALM}$  \emph{allowed us to create libraries of knowledge and reuse information}
when representing the cell division domain.
As a second step, we created a question answering system capable of answering
complex temporal projection questions about this biological process \cite{di10}.
Our model of cell division represented in the higher level language ${\cal ALM}$ served as a front end
for the question answering system, which was implemented both in ASP and
in the language of the Digital Aristotle.

Our language has evolved since our collaboration on Project Halo.
The version of ${\cal ALM}$ presented here differs from that
described in previous papers \cite{gi09,ig11} in various ways.
We simplified and generalized the basic concepts of our language,
as well as its syntax and semantics.
(We say more about the new features of $\mathcal{ALM}$ in the
conclusion section of the paper.)
The reasoning in $\mathcal{ALM}$ is based on the reduction of
temporal projection, planning, diagnosis, etc. to the problem of
computing the answer sets of logic programs (for a general description
see, for instance, \cite{bar03}) by ASP solvers (see
\cite{nl97}, \cite{asppractice}, or \cite{EiterGLPPSW2006}).

The rest of this paper is organized as follows: we first introduce the
concept of \emph{basic action theory}, which is a fundamental concept in this work.
We then describe language ${\cal ALM}$ and the methodology of ${\cal ALM}$'s use.
We end with conclusions and future work. There are three appendices
containing the grammar of ${\cal ALM}$ (\ref{grammar}), the description of the use of
${\cal ALM}$ in Digital Aristotle (\ref{aristotle}), 
and a comparison between ${\cal ALM}$ and $MAD$ (\ref{ALMandMAD}).

\section{Basic Action Theories}
In this section we give the definition of a fundamental concept of
${\cal ALM}$ called \emph{basic action theory} (${\cal
  BAT}$). A ${\cal BAT}$ consists of a collection of axioms over a
so called \emph{action signature} --- a special type of sorted
signature providing suitable vocabulary for representing knowledge
about dynamic domains. Sorted signatures needed for our
purpose are somewhat atypical. They allow partial functions and contain means for
describing a hierarchy of sorts and attributes of their elements.
We start with the precise
definition of sorted signatures and their interpretations.

\subsection{Sorted Signatures and Their Interpretations}
\label{sorted_signature}

By \emph{sorted signature} we mean a tuple
$$\Sigma=\langle \mathcal{C}, \mathcal{O}, \mathcal{H}, \mathcal{F}\rangle$$
where $\mathcal{C}$, $\mathcal{O}$, and $\mathcal{F}$ are sets of strings over some fixed
alphabet. The strings are used to name
\emph{sorts}, \emph{objects}, and (possibly partial) \emph{functions}
respectively. Each function symbol $f \in {\cal F}$ is assigned a positive integer 
$n$ (called $f$'s arity),
sorts $c_0,\dots,c_n$ for its parameters, and sort $c$ for its values. 
We refer to $c$ as the \emph{range} of $f$ and 
use the standard mathematical notation 
$f : c_0\times\dots\times c_{n} \rightarrow c$
for this assignment. 

Finally, $\mathcal{H}$ is a
\emph{sort hierarchy} --- a directed acyclic graph with two types of nodes:
\emph{sort nodes}  labeled
by sort names from $\mathcal{C}$, and \emph{object nodes} labeled by
object names from $\mathcal{O}$. 
Whenever convenient we
identify nodes of the hierarchy with their labels.
A link from sort $c_1$ to sort $c_2$, denoted by $\langle c_1, c_2
\rangle$, indicates that elements of sort $c_1$ are also
elements of sort $c_2$. We refer to $c_2$ as a \emph{parent} of
$c_1$. 
A link from object $o$ to a sort $c$, denoted by $\langle o,c\rangle$,
 indicates that object $o$ is of sort $c$.
For simplicity, we assume that 
the graph has exactly one sink node, which corresponds to the sort 
containing all the elements of the hierarchy.
A triple $\langle \mathcal{C}, \mathcal{O}, \mathcal{H} \rangle$ will be sometimes
referred to as an \emph{ontology}.


Sorts, object constants, and functions of a sorted signature are normally partitioned
into \emph{user-defined}, \emph{pre-defined}, and \emph{special}.

The collection of \emph{pre-defined} symbols may include
names for some commonly used sorts and functions, such as: sorts
$booleans$ and $integers$; a sort $[m \mbox{..} n]$ for every pair of
natural numbers $m$ and $n$ such that $m < n$,
denoting the set of natural numbers in the closed interval $[m,n]$; standard
object constants {\em true}, {\em false},  $0$, $1$, $2$, etc.,
denoting elements of these sorts; standard arithmetic
functions and relations $+$, $-$, $*$, $/$, $mod$, $<$, $\leq$, etc.
(The list is not exhaustive. When needed we may introduce other
similar symbols.)  All these symbols are  pre-interpreted, i.e.,
come with their usual mathematical interpretations.

The collection of \emph{special} symbols consists of:
\begin{itemize}
\item sorts and function symbols
pertinent to sort hierarchies of sorted signatures:
\begin{itemize}
\item Sort $nodes$ denoting the collection of sorts 
  labeling the sort nodes of ${\cal H}$. This sort is never used 
as a label of a node in ${\cal H}$.
\item Sort $object\_constants$ denoting the collection of 
constants  labeling the object nodes of ${\cal H}$. This sort is never used 
as a label of a node in ${\cal H}$.
\item Sort $universe$ denoting the collection of elements of
sorts from ${\cal H}$.
\item Function symbol $link : nodes \times nodes \rightarrow 
  booleans$
where $link(c_1,c_2)$  returns \emph{true}
iff $\mathcal{H}$ contains a link from sort $c_1$ to sort $c_2$.
\item Function symbol $is\_a : universe \times nodes \rightarrow
  booleans$ where $is\_a(x,c)$ returns \emph{true}
if $c$ is a source node of $\mathcal{H}$ (i.e., $c$ has no subsorts in $\mathcal{H}$) 
and object $x$ from the universe 
is of the sort denoted by $c$.
\item Function symbol $instance : universe \times nodes
\rightarrow booleans$ denoting the membership relation between objects
of the universe and the sorts of the domain. This function will be
later defined in terms of function $is\_a$.
\item Function symbols $subsort : nodes \times nodes  \rightarrow
booleans$,\\
$has\_child, has\_parent, sink, source : nodes  \rightarrow booleans$ \\ describing
properties of sorts of $\mathcal{H}$ and their members.
All these functions (with their self-explanatory meaning) will be later defined
in terms of function $link$.
\end{itemize}
\item Function symbol $dom_f : c_0\times\dots\times
c_{n} \rightarrow booleans$ (read as \emph{domain of $f$}) for every
user-defined function
symbol $f : c_0\times\dots\times c_{n} \rightarrow c$ with $n > 0$.
\end{itemize}

\medskip
\emph{Terms} of a sorted signature are defined as usual:
\begin{itemize}
\item A variable and an object constant is a term.
\item If $f : c_0\times\dots\times c_{n} \rightarrow c$ is a function
  symbol and $t_0,\dots,t_n$ are terms then $f(t_0,\dots,t_n)$ is a term.
\end{itemize}

Expressions of the form
\begin{equation}\label{atom_1}
t_1 = t_2 \ \ \ \ \ \ \mbox{and}\ \ \ \ \ \  t_1 \not= t_2
\end{equation}
are called \emph{literals}. Positive literals are also referred to as $\emph{atoms}$.
(For simplicity of presentation we use standard shorthands and write $t$ and
$\neg t$ instead of $t=true$ and $t=false$, respectively; $3 \leq 5$ instead of
$\leq(3,5)$; etc.)
Terms and literals not containing variables are called {\em ground}.
Our notion of an interpretation of a sorted signature is slightly
different from the traditional one.

\begin{definition}[Interpretation]\label{interp}
An \emph{interpretation} ${\cal I}$ of $\Sigma$
consists of 
\begin{itemize}
\item A non-empty set $|{\cal I}|$ of strings called the \emph{universe}
of ${\cal I}$. 
\item An assignment that maps 
\begin{itemize}
\item every user-defined 
sort $c$ of $\mathcal{H}$ into a subset ${\cal I} (c)$ of $|{\cal I}|$
and user defined  object constant $o$ into an element from $|{\cal I}|$;

\item every user-defined function symbol 
$f : c_0 \times \dots \times c_n \rightarrow c$ of $\Sigma$ into 
a ({\em possibly partial}) function 
${\cal I}(f) : {\cal I}(c_0) \times \dots \times {\cal I}(c_n) 
\rightarrow {\cal I}(c)$;

\item the special function $is\_a$ into function $\mathcal{I}(is\_a)$ such that:
\begin{itemize}
\item for every 
$x \in |\mathcal{I}|$ and every sort $c$ of $\mathcal{H}$, $\mathcal{I}(is\_a)(x,c)$
is \emph{true} iff $c$ is a source node of $\mathcal{H}$ and $x \in {\cal I}(c)$ and

\item for every object $o$ and sort $c$ of $\mathcal{H}$,
$\mathcal{I}(is\_a)(\mathcal{I}(o),c)$ is \emph{true} iff
$\langle o, c \rangle \in \mathcal{H}$;
\end{itemize}

\item the special function $link$ into function $\mathcal{I}(link)$ such that for every 
two sort nodes $c_1,c_2, $ $\mathcal{I}(link)(c_1,c_2)$
is \emph{true} iff $\langle c_1,c_2 \rangle \in {\cal H}$;

\item the special function $dom_f$ for user-defined function 
$f : c_0 \times \dots \times c_n \rightarrow c$
into function ${\cal I}(dom_f)$ such that for every 
$\bar{x} \in {\cal I}(c_0) \times \dots \times {\cal I}(c_n)$,
${\cal I}(dom_f)(\bar{x})$ is \emph{true}
iff $\bar{x}$ belongs to the domain of ${\cal I}(f)$.

\end{itemize}
\item On pre-defined symbols, ${\cal I}$ is identified 
with the symbols' standard 
interpretations.
\end{itemize}

\end{definition}

\medskip\noindent
An interpretation ${\cal I}$ of $\Sigma$ 
can be naturally extended to ground
terms: if ${\cal I}$ is defined on terms
$t_1,\dots,t_n$  and ${\cal I}(f)$ is defined on
the tuple $({\cal I}(t_1), \dots, {\cal I}(t_n))$
then
$${\cal I}(f(t_1,\dots,t_n)) =_{def} {\cal I}(f)({\cal
  I}(t_1),\dots,{\cal I}(t_n)).$$
Otherwise ${\cal I}(f(t_1,\dots,t_n))$ is undefined.

\medskip
Finally, we say that an atom $t_1 = t_2$ is
\begin{itemize}
\item {\em true} in ${\cal I}$ if  both ${\cal I}(t_1)$ and ${\cal I}(t_2)$ are defined
and have the same value;
\item {\em false} in
${\cal I}$ if  both ${\cal I}(t_1)$ and ${\cal I}(t_2)$ are defined
and have different values; and
\item {\em undefined}  in ${\cal I}$ otherwise.
\end{itemize}
Similarly, a literal $t_1 \not= t_2$ is {\em true}  in ${\cal I}$  if $t_1 =
t_2$ is {\em false}  in ${\cal I}$; it is  {\em false}  in ${\cal I}$  if $t_1 =
t_2$ is {\em true}  in ${\cal I}$; and {\em undefined} otherwise.

\st
Note that every interpretation ${\cal I}$ 
 can be uniquely represented by the collection of atoms 
that are true in this interpretation. 
For instance, for every sort $c$ of $\mathcal{H}$,
  ${\cal I}(c)$ can be represented as the set $\{instance(o,c) : {\cal I}(o) \in 
  {\cal I}(c)\}$; for a unary function $f$, ${\cal I}(f)$ can be viewed as the set 
$\{f(x)=y : {\cal I}(x) \in {\cal I}(dom_f) \mbox{ and } {\cal I}(f)({\cal I}(x)) = {\cal I}(y)\}$, etc.


\subsection{Action Signature and Axioms of a ${\cal BAT}$}
\bigskip
Since ${\cal ALM}$ is a language for specifying properties of
actions, in what follows we limit ourselves to
\emph{action signatures} --- sorted signatures that
\begin{itemize}
\item contain a special
sort \emph{actions} and
\item have their user-defined and special
function symbols  divided
into three disjoint categories: \emph{attributes}, \emph{statics}, and \emph{fluents}.
Attributes describe intrinsic properties of objects of a given
sort; statics and fluents describe relations between
objects. Values of attributes and statics are constants -- they
cannot be changed by actions. The values of fluents can.
Both statics and fluents are further divided into \emph{basic} and
\emph{defined}. The latter are \emph{total boolean functions} that can be defined
in terms of the former. They are used primarily for the brevity of representation.
\end{itemize}
A literal (atom) in which $f$ is an attribute is called an
\emph{attribute literal (atom)}.
Similarly for \emph{static} and \emph{fluent} literals that are, in
turn, divided into \emph{basic} and \emph{defined}.
We assume that all special functions of an action signature, except
$dom_f$, are defined statics;  
$dom_f$ is a basic fluent when $f$ is a basic fluent and
a defined static otherwise.
Since the semantics of ${\cal ALM}$ will be defined in terms of
a version of ASP with function symbols, ASP\{f\} \cite{bal13}, 
which does not allow terms with nested user-defined functions, we
\emph{limit atoms of an action signature to those constructed from terms
with at most one user-defined function symbol}. (This is not a serious
limitation and can easily be avoided by viewing nested terms as shorthands.)

We can now define the syntax and informal semantics of \emph{statements} of
a $\mathcal{BAT}$ over a fixed action signature $\Sigma$.
Variables in these statements are universally quantified.

\begin{definition}[Statements of a $\mathcal{BAT}$]
\begin{itemize}
\item A {\em dynamic causal law} is an expression of the form
\begin{equation}\label{dcl}
\begin{array}{l}
occurs(a) \ \causes \ f(\bar{x})= o\ \lif \ instance(a,c), cond
\end{array}
\end{equation}
where $a$ and $o$ are variables or object constants, $f$ is a basic fluent,
$c$ is the sort
$actions$ or a subsort of it, and $cond$ is a collection of literals.
The law says that
{\em an occurrence of an action $a$ of the sort $c$ in a
  state satisfying property $cond$ causes the value of $f(\bar{x})$ to
become $o$ in any resulting state}.
\item A {\em state constraint} is an expression of the form
\begin{equation}\label{sc}
\begin{array}{l}
 f(\bar{x})= o\ \lif \ cond
\end{array}
\end{equation}
where $o$ is a variable or an object constant,
$f$ is any function except a defined
function, and $cond$ is a collection of literals.
The law says that
{\em the value of  $f(\bar{x})$ in any state satisfying condition
  $cond$ must be $o$}.
Additionally, $f(\bar{x}) = o$ can also be replaced by the object constant $false$,
in which case the law says that
{\em there is no state satisfying condition $cond$}.
\item The {\em definition} of a defined function $p$ is an expression of the form
\begin{equation}\label{fd}
\begin{array}{l}
p(\overline{t}_1) \ \lif \ cond_1\\
\dots \\
p(\overline{t}_k) \ \lif \ cond_k
\end{array}
\end{equation}
where $\overline{t}$s are sequences of terms, and
$cond_1,\dots,cond_k$ are collections of
literals. Moreover, if $p$ is a static then $cond_1,\dots,cond_k$ can not contain fluent
literals. Statements of the definition will be often
  referred to as its \emph{clauses}.
The statement says that, for every $\overline{Y}$, $p(\overline{Y})$
is true in a state $\sigma$ iff
there is $1 \leq m \leq k$ such that statements $cond_m$ and $\overline{t}_m = \overline{Y}$
are true in $\sigma$.

\item  An {\em executability condition for actions} is an expression of the form
\begin{equation}\label{ec}
\begin{array}{l}
\impossible \ occurs(a) \ \lif \ instance(a,c),cond
\end{array}
\end{equation}
where $a$ is a variable
or an object constant,
$c$ is the sort $actions$ or a subsort of
it, and $cond$ is a collection of literals and expressions
of the form $occurs(t)$ or $\neg occurs(t)$ where $t$ is a variable or
an object constant of the sort $actions$.
The law says that {\em an occurrence of an action $a$ of the sort $c$
is impossible when condition $cond$ holds}.
\end{itemize}

\smallskip\noindent
Dynamic causal laws and constraints will be sometimes referred to as
\emph{causal laws}. We use the term \emph{head} to refer to $l$ in (\ref{dcl}) and (\ref{sc}),
and to any of the $p(\overline{t_i})$, $1 \leq i \leq k$, in (\ref{fd}).
We call \emph{body} the expression to the right of the keyword ${\bf if}$ in
statements (\ref{dcl}), (\ref{sc}), (\ref{ec}), or in any of the statements of (\ref{fd}).
Statements not containing variables will be referred to as \emph{ground}.
\end{definition}

\begin{definition}[Basic Action Theory -- ${\cal BAT}$]\label{def-of-bat}
 A {\em Basic Action Theory} (${\cal BAT}$) is a pair consisting of
 an action signature $\Sigma$ and a collection $T$
of statements over $\Sigma$ (called \emph{axioms} of the theory) such
that:
\begin{itemize}
\item If $f$ is a basic fluent then
\begin{itemize}
\item $T$ contains a state constraint:
\begin{equation}\label{dom1}
\begin{array}{l}
dom_f(X_0,\dots,X_n) \lif f(X_0,\dots,X_n) = Y\\
\end{array}
\end{equation}
\item No dynamic causal law of $T$ contains an atom formed by $dom_f$ in the head.
\end{itemize}
\item If $f$ is a defined fluent, a static, or an attribute then $T$
  contains the definition:
\begin{equation}\label{dom1a}
\begin{array}{l}
dom_f(X_0,\dots,X_n) \lif f(X_0,\dots,X_n) = Y\\
\end{array}
\end{equation}




\item $T$ contains definitions of special statics of the hierarchy
given in terms of  functions $is\_a$ and $link$: 
\begin{equation}\label{instance}
\begin{array}{rll}
instance(O,C) & \lif & is\_a(O,C)\\
instance(O,C_2) & \lif & instance(O,C_1),link(C_1,C_2)\\
has\_child(C_2) & \lif & link(C_1, C_2)\\
has\_parent(C_1)   & \lif & link(C_1, C_2)\\
source(C) & \lif & \neg has\_child(C)\\
sink(C)   & \lif & \neg has\_parent(C)\\
subsort(C_1, C_2) & \lif & link(C_1, C_2)\\
subsort(C_1, C_2) & \lif & link(C_1, C), subsort(C, C_2)\\

\end{array}
\end{equation}

\end{itemize}
\end{definition}
To simplify the notation, in what follows we will often identify a theory with the collection of its axioms.
Axioms (\ref{dom1})--(\ref{instance}) above are self-explanatory, with the possible exception
of the restriction prohibiting the appearance of
$dom_f$ in the head of dynamic causal laws. To
understand the latter requirement it is sufficient to notice that it is not enough
to include object $O$ in the domain of basic fluent $f$ --- it is also necessary to
specify the value of $f(O)$. Otherwise the causal law making $dom_f(O)$ true
would become non-deterministic,\footnote{To see why, consider, for instance, a
basic fluent $f$ declared as 
$f : \{0,1\} \rightarrow \{0,1\}$
and a dynamic causal law
``$occurs(a) \ {\bf causes} \ dom_f(1)\mbox{.}$''
Intuitively, the axiom says that after $a$ is executed $f(1)$ must be defined, i.e.,
$f(1) = 0$ or $f(1) = 1$, 
which is non-deterministic.} which is not allowed in the current version of 
${\cal ALM}$. The presence of a law assigning a value to $f(O)$ makes
dynamic causal laws with $dom_f$ in the head unnecessary.
It is however useful to allow dynamic causal laws with $\neg dom_f(O)$ in the
head as a simple way of removing $O$ from the domain of $f$.
%

The following is an example of a basic action theory.
\begin{example}[A Basic Action Theory $T^0$]
\label{bat_ex}
Let us consider an action signature $\Sigma^0$ with three sorts, $c_1$,
$c_2$ and $c_3$, the special sorts $universe$ and $actions$, and the pre-defined sort $booleans$,
organized in a hierarchy ${\cal H}^0$ 
in which $universe$ is the parent of $c_1$,
$c_1$ is the parent of $c_2$, $c_3$,
$actions$, and $booleans$, and 
object constant $o$ is of sort $c_3$;  
\begin{figure}[!htbp]
\centering 
\includegraphics[scale=0.5]{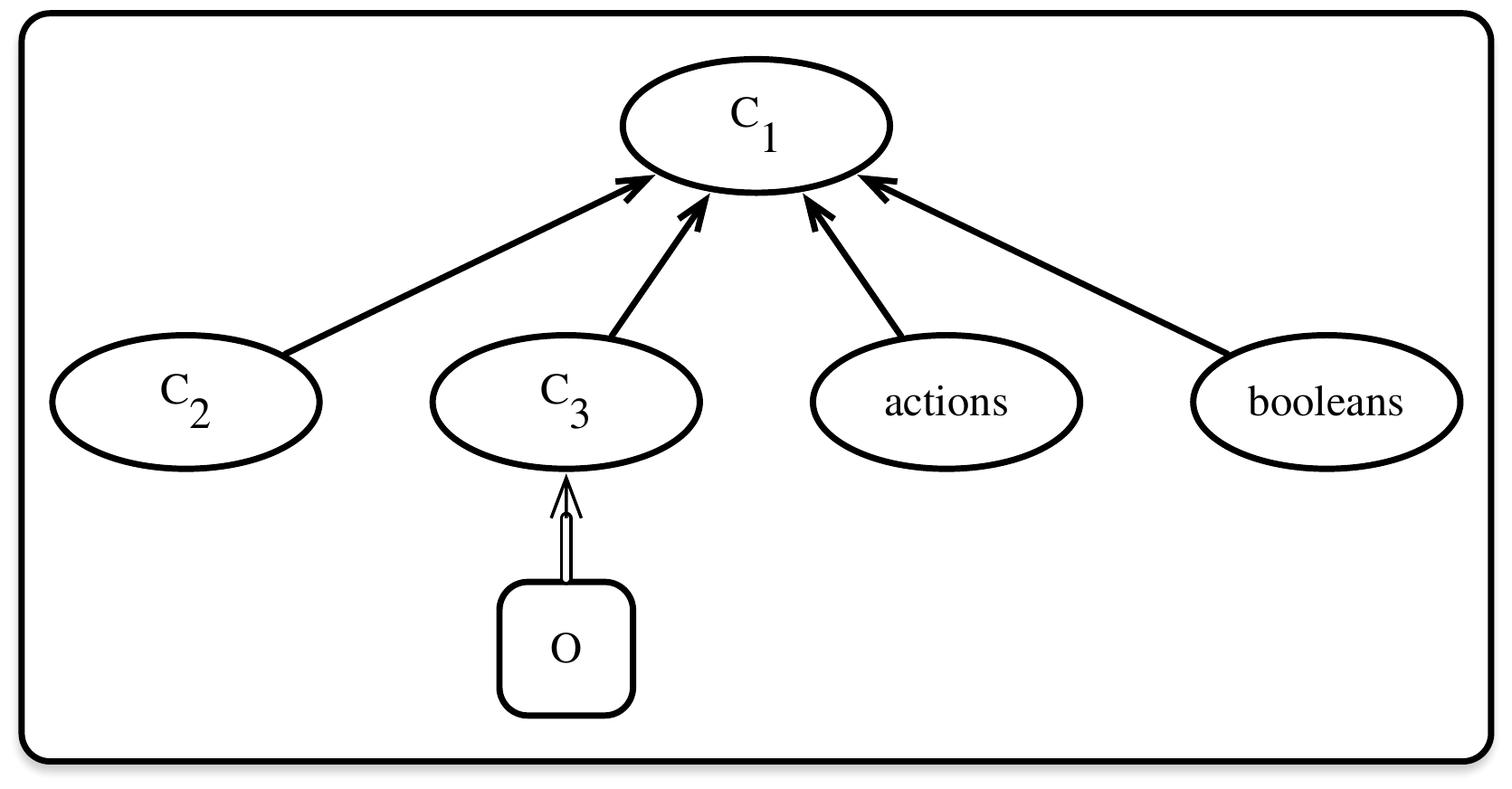}
\caption{Hierarchy ${\cal H}^0$ of $T^0$}
\label{fig3:hierarchy}
\end{figure}
attributes $attr_1, attr_2 : actions \rightarrow c_3$;
basic fluents $f,g : c_2 \rightarrow c_3$;
and special functions like $link$, $is\_a$, 
$dom_f$, $dom_g$. The hierarchy ${\cal H}^0$ can be seen in Figure \ref{fig3:hierarchy},
but we omitted from the picture the sort $universe$ whose only child is $c_1$.\\
The basic action theory $T^0$ over $\Sigma^0$ consists of the causal laws

\bigskip
$
\begin{array}{ll}
occurs(A) \causes \ f(X) = Y \ \lif & instance(A, actions),\\
                                                 & attr_1(A) = Y,\\
                                                 & g(X) = o
\end{array}
$

$
\begin{array}{ll}
occurs(A) \causes \ \neg dom_f(X) \ \lif & instance(A, actions),\\
                                                              & attr_2(A) = o
\end{array}
$

$
\begin{array}{lll}
false & \lif & \neg dom_g(X),\\
      &      & instance(X,c_2)\mbox{.}
\end{array}
$

\st
The third axiom requires function $g$ to be total.

\st
In addition, $T^0$ contains standard $\mathcal{BAT}$ axioms:

\st State constraints for the basic fluents:

\st
$
\begin{array}{l}
dom_f(X) \lif f(X) = Y \\
dom_g(X) \lif g(X) = Y
\end{array}
$

\st Definitions for the domains of attributes:

\st
$
\begin{array}{l}
dom_{attr_1}(X)\lif attr_1(X) = Y \\
dom_{attr_2}(X) \lif attr_2(X) = Y
\end{array}
$

\st and the collection of axioms from (8).
\end{example}

\subsection{Semantics of $\mathcal{BAT}$s}
\label{sec_semantics}
Intuitively, a basic action theory $T$ defines the collection of
discrete dynamic systems satisfying its axioms. The semantics of $T$ will
describe such systems by specifying their transition diagrams, often
referred to as \emph{models} of $T$.
Nodes of a transition diagram represent possible states of the dynamic system;
arcs of the diagram are labeled by actions.
A transition $\langle \sigma_0, a, \sigma_1\rangle$ says
that the execution of action $a$ in state $\sigma_0$
may take the system to state $\sigma_1$.

A state of the diagram will be
defined by the universe --- a collection of objects of the sorts of
$T$, and by a physically possible assignment of values to $T$'s functions.
Moreover, we assume that the sorted universe and the values of statics and
attributes are the same in all states, i.e., states only differ by the
values of fluents.

\medskip
To make this precise it is convenient to partition an interpretation
${\cal I}$ of an action signature $\Sigma$ into two parts: \emph{fluent part} consisting
of the universe of ${\cal I}$ and the restriction of ${\cal I}$ on the sets of
fluents,  and \emph{static part} consisting of the same universe and the
restriction of ${\cal I}$ on the remaining elements of the signature.
Sometimes we will refer to the latter as a \emph{static interpretation}
of $\Sigma$.

\st We also need the following notation:
Given an action signature $\Sigma$ and a collection $U$ of
strings in some fixed alphabet, we denote by $\Sigma_U$ the signature
obtained from $\Sigma$ by expanding its set of object constants by elements of $U$,
which we assume to be of sort $universe$.

\begin{definition}[Pre-model]\label{pre-model}
Let $T$ be a basic action theory with
signature $\Sigma$ and  $U$ be a collection of strings in some fixed alphabet. 
A static interpretation ${\cal M}$ of $\Sigma_U$ is called a \emph{pre-model} of
$T$ (with the universe $U$) if ${\cal M}(universe) = U$ and 
for every object constant $o$ of $\Sigma_U$ that is not an
  object constant of $\Sigma$, ${\cal M}(o) = o$.
\end{definition}
Given a pre-model ${\cal M}$ with the universe $U$ we will often
denote signature $\Sigma_U$ by $\Sigma_{\cal M}$.

To illustrate this notion let us consider a pre-model of theory $T$
from Example \ref{bat_ex}:

\begin{example}[A pre-model of Basic Action Theory $T^0$]
\label{pre-model_ex}

To define a pre-model of basic action theory $T^0$
from Example \ref{bat_ex} let us consider a static interpretation
${\cal M}$ with the universe $U_{\cal M}= \{x,y,z,a,b,true,false\}$
such that:

\st
${\cal M}(universe) = {\cal M}(c_1) = \{x,y,z,a,b,true,false\}$; \\
${\cal M}(c_2) = \{x\}$;\\
${\cal M}(c_3) = \{y,z\}$,\\
${\cal M}(actions) = \{a,b\}$;\\
${\cal M}(o) = \{y\}$; and \\
${\cal M} (attr_1)(a) = {\cal M}(attr_2)(b) =y$. 

\st
In addition: every symbol from $U_{\cal M}$ is added to $\Sigma^0_U$
and mapped into itself; $dom_{attr_1} = \{a\}$, $dom_{attr_2} = \{b\}$;
the interpretation of special function $link$ is
determined by the hierarchy from Figure \ref{fig3:hierarchy}; the interpretation of
$is\_a$ is extracted from the interpretation of the hierarchy's sorts.

\noindent
Clearly, ${\cal M}$ satisfies the conditions in Definition
\ref{pre-model} and hence is a pre-model of $T^0$.
\end{example}



\medskip
A pre-model  $\mathcal{M}$ of $T$ uniquely defines a model
 $T_\mathcal{M}$ of $T$ if such a model exists.
The definition of  $T_\mathcal{M}$ will be
  given in two steps: first we define $T_\mathcal{M}$'s states and
  then its transitions.


\medskip
Intuitively, if theory $T$ does not contain 
definitions,
then a {\em state} of $T_\mathcal{M}$ is an interpretation
$\mathcal{I}$ with static part $\mathcal{M}$ that satisfies the state constraints of $T$.
The situation is less simple for theories containing 
definitions (especially recursive ones).
Similar to the case of
$\mathcal{AL}$, the definition of a state will be given using logic programs under
the answer set semantics; specifically, we will use logic programs
with non-Herbrand partial functions in the language ASP\{f\} \cite{bal13}.\footnote{
Other approaches for introducing non-Herbrand functions in ASP can be seen, for
instance, in \cite{pc11,vl12,bl13}. 
}

\medskip\noindent
Let ${\cal M}$ be a pre-model of action theory $T$.

\medskip\noindent
{\bf Program $S_{\cal M}$}:

\medskip
By $S_{\cal M}$ we denote a logic program that consists of:
\begin{enumerate}[a)]
\item rules obtained from the state constraints and
definitions of $T$ by replacing variables with properly
typed object constants of $\Sigma_{\cal M}$,
replacing object constants with their corresponding interpretations in ${\cal M}$,
removing the constant $false$ from the head of state constraints,
and replacing the keyword $\lif$ with $\leftarrow$,

\item the Closed World Assumption:
$$\neg d(t_0, \dots, t_n) \leftarrow \no d(t_0, \dots, t_n)$$
for every defined function $d: c_0 \times \dots \times c_n \rightarrow booleans$
and $t_i \in {\cal M}(c_i)$, $0 \leq i \leq n$.
\end{enumerate}
{\bf end of $S_{\cal M}$}:

\medskip
Finally, we define a program $S_{\cal I}$ used in the
definition of states of the transition diagram defined by ${\cal M}$.

\medskip\noindent
{\bf Program $S_{\cal I}$}:

\smallskip
For every interpretation ${\cal I}$ of $\Sigma$ with static part
${\cal M}$, by $S_{\cal I}$ we denote
the logic program obtained by adding to $S_{\cal M}$
the set of atoms obtained from ${\cal I}$ by removing
 the defined atoms.

\smallskip\noindent
{\bf end of $S_{\cal I}$}

\begin{definition}[State]
Let $\mathcal{M}$ be a pre-model of a $\mathcal{BAT}$ theory
$T$. An interpretation $\sigma$ with static part $\mathcal{M}$
 is a
{\em state} of the transition diagram $T_{\cal M}$ defined by  ${\cal M}$ if
$\sigma$ is \emph{the only} answer set of $S_\sigma$.
\end{definition}

Notice that $\sigma$ is \emph{not} a state if $S_\sigma$ has multiple answer sets,
a situation that would only occur when the value of some defined function is not completely determined
by the values of basic functions.
We will return to this issue later, in Section \ref{solving_comp_tasks}.

\begin{example}[States of the diagram]
\label{state_ex}
 Let ${\cal M}$ be the pre-model of theory $T^0$ from Example \ref{pre-model_ex}.
The program $S_{\cal M}$ for this ${\cal M}$ looks as follows:

\bigskip

$
\begin{array}{l}
\leftarrow \neg dom_g(x), instance(x,c_2)
\end{array}
$

\st
$
\begin{array}{l}
dom_f(x) \leftarrow f(x) = y \\
dom_f(x) \leftarrow f(x) = z \\
dom_g(x) \leftarrow g(x) = y\\
dom_g(x) \leftarrow g(x) = z
\end{array}
$

\st
$
\begin{array}{l}
dom_{attr_1}(a) \leftarrow attr_1(a) = y \\
dom_{attr_1}(a) \leftarrow attr_1(a) = z \\
dom_{attr_2}(a) \leftarrow attr_2(a) = y\\
dom_{attr_2}(a) \leftarrow attr_2(a) = z\\
dom_{attr_1}(b) \leftarrow attr_1(b) = y \\
dom_{attr_1}(b) \leftarrow attr_1(b) = z \\
dom_{attr_2}(b) \leftarrow attr_2(b) = y\\
dom_{attr_2}(b) \leftarrow attr_2(b) = z
\end{array}
$

\st
and the Closed World Assumptions for the special functions.
Recall that, according to the definition of an interpretation of a
sorted signature, for every $x \in |{\cal I}|$, 
${\cal I}(is\_a)(x,c)$ is true iff $c$ is a source node of the sort
hierarchy and ${\cal I}(x) \in {\cal I}(c)$, and 
for every object $o$ and sort $c$,
$\mathcal{I}(is\_a)(\mathcal{I}(o),c)$ is \emph{true} iff
$\langle o, c \rangle$ is a link in our hierarchy.
This, together with the
condition on the interpretation of $link$ guarantees that every
state of $T_{\cal M}$ contains
atoms $is\_a(x,c_2)$, $is\_a(y,c_3)$, and other atoms formed by $is\_a$
and $link$ that define our hierarchy. The collection of these atoms
together with the closed world assumptions for $is\_a$, $link$ and the other
defined statics uniquely determine their values.
It is easy to check that every state of
${\cal M}$ contains literals formed by these special fluents.
Every state of $T_{\cal M}$ also contains 
$attr_1(a) = y$, $attr_2(b) = y$, and $dom_g(x)$.
Overall, $T_{\cal M}$ has the following six states (for each state, we
only show non-special fluents):

\st
$
\begin{array}{ll}
\sigma_1 = \{f(x) = y,g(x) = y\} \ &\ \sigma_2 = \{f(x) = z,g(x) =
y\}\\
\sigma_3 = \{f(x) = y,g(x) =z\}\  & \ \sigma_4 = \{f(x) = z,g(x) =
z\}\\
\sigma_5 = \{g(x) = y\}\ &\ \sigma_6 = \{g(x) = z\}.
\end{array}
$

\st
In addition, states $\sigma_1, \sigma_2, \sigma_3$, and $\sigma_4$
contain $dom_f(x)$ while states $\sigma_5$ and $\sigma_6$, in which
$f$ is undefined on $x$, contain
$\neg dom_f(x)$.

\end{example}

\medskip
To define transitions of the diagram that corresponds to a pre-model
$\mathcal{M}$ with the universe $U$, we construct a logic program $P_{\cal M}$
whose signature is obtained from the signature of program $S_{\cal M}$
defined above by
\begin{itemize}
\item adding a new sort, $step$, ranging over 0 and 1;
\item replacing every fluent $f : c_0\times\dots\times c_{n}
  \rightarrow c$ by function\\ $f : c_0\times\dots\times c_{n}\times step
  \rightarrow c$;
\item adding a function symbol $occurs : actions \times step
  \rightarrow booleans$.

\end{itemize}

\medskip\noindent
{\bf Program $P_{\cal M}$}:

\medskip
Program $P_{\cal M}$ is obtained from a theory $T$ and pre-model
${\cal M}$ by
\begin{enumerate}[a)]
\item replacing variables by properly typed
object constants of $\Sigma_{\cal M}$;
\item replacing object constants by their corresponding
  interpretations in ${\cal M}$;
\item removing the object constant $false$ from the head of state constraints;
\item replacing every occurrence of a fluent term $f(\overline{t})$ in the head of
a dynamic causal law 
by $f(\overline{t},I+1)$;
\item replacing every other occurrence of a fluent term $f(\overline{t})$
by $f(\overline{t},I)$;
\item removing ``$occurs(a) \ {\bf causes}$''
from every dynamic causal law 
and adding $occurs(a)$ to the body;
\item replacing ``$\impossible occurs(a)$'' in every executability condition 
by $\neg occurs(a)$;
\item replacing $occurs(a)$ by $occurs(a, I)$ and $\neg occurs(a)$ by $\neg occurs(a, I)$;
\item replacing the keyword $\lif$  by $\leftarrow$;
\item adding the Closed World Assumption:
$$\neg d(t_0, \dots, t_n, I) \leftarrow \no d(t_0, \dots, t_n, I)$$
for every defined fluent $d: c_0 \times \dots \times c_n \rightarrow booleans$
and $t_i \in {\cal M}(c_i)$, $0 \leq i \leq n$;

\item adding the rule:
$$\neg f(t_0, \dots, t_n) \leftarrow \no f(t_0, \dots, t_n)$$
for every defined static of the form $f: c_0 \times \dots \times c_n \rightarrow booleans$
and $t_i \in {\cal M}(c_i)$, $0 \leq i \leq n$;

\item adding the Inertia Axiom:
$$
\begin{array}{lll}
dom_f(t_0, \dots, t_n, I+1) & \leftarrow &
    dom_f(t_0, \dots, t_n,I),\\
& & \no \neg dom_f(t_0, \dots, t_n,I+1)\\
\neg dom_f(t_0, \dots, t_n, I+1) & \leftarrow &
    \neg dom_f(t_0, \dots, t_n,I),\\
& & \no dom_f(t_0, \dots, t_n,I+1)\\
\end{array}
$$
for every basic fluent $dom_f: c_0 \times \dots \times c_n \rightarrow booleans$,
and $t_i \in {\cal M}(c_i)$, $0 \leq i \leq n$;

\item adding the Inertia Axiom:
$$
\begin{array}{lll}
f(t_0, \dots, t_n, I+1) = t & \leftarrow & dom_f(t_0, \dots, t_n,
I+1),\\
                                     &                 & f(t_0, \dots, t_n, I) = t, \\
                                     &                 &  \no f(t_0, \dots, t_n, I+1) \neq t
\end{array}
$$
for every basic fluent $f: c_0 \times \dots \times c_n \rightarrow c$
not formed by $dom$,
and $t_i \in {\cal M}(c_i)$, $0 \leq i \leq n$, and $t \in {\cal M}(c)$.
\end{enumerate}
{\bf end of $P_{\cal M}$}

\st Note that the last axiom is a modification of the standard logic
programming version of the Inertia Axiom (see, for instance,
\cite{gk14}), which is stated for total (boolean) functions. The main
difference is the addition of the domain statements in the body.
The inertia axiom for the function $dom_f$ is of the standard form.

\st
{\bf Program $P({\cal M}, \sigma_0, a)$}:
Let $\sigma_0$ be a state of the transition diagram
defined by a pre-model ${\cal M}$,
and let $a \subseteq {\mathcal M}(actions)$.
By $P({\cal M}, \sigma_0, a) $ we denote the logic program formed by
adding to $P_{\cal M}$
the set of atoms obtained from $\sigma_0$ by replacing every fluent atom
$f(t_0, \dots, t_n) = t$ by $f(t_0, \dots, t_n, 0) = t$ and adding the set of atoms
$\{ occurs(x, 0) : x \in a\}$.

\noindent
{\bf end of $P({\cal M}, \sigma_0, a)$}

\begin{definition}[Transition]\label{trans}
Let $\sigma_0$ and $\sigma_1$ be states of the transition diagram
defined by a pre-model ${\cal M}$ and let $a \subseteq {\mathcal M}(actions)$.
The triple $\langle \sigma_0, a, \sigma_1 \rangle$ is a
{\em transition} of the transition diagram defined by a pre-model
$\mathcal{M}$ of a $\mathcal{BAT}$ theory $T$ if program $P({\cal M}, \sigma_0, a)$
has an answer set $A$ such that $f(t_0, \dots, t_n) = t\ \in\
\sigma_1$ iff
\begin{itemize}
\item $f$ is an attribute or a static and $f(t_0, \dots, t_n) = t\ \in\ A$, or
\item $f$ is a fluent and $f(t_0, \dots, t_n,1) = t\ \in\ A$.
\end{itemize}
\end{definition}

\begin{definition}[Model]\label{model}
A transition diagram $T_\mathcal{M}$ defined by a pre-model
$\mathcal{M}$ of a basic action theory $T$ is called a \emph{model}
of $T$ if it has a non-empty collection of states.
\end{definition}

The following example illustrates the definition.

\begin{example}[A Model of Basic Action Theory $T^0$]
\label{model_ex}
To define a model of theory $T^0$ from Example \ref{bat_ex}
let us consider the pre-model ${\cal M}$ from Example \ref{pre-model_ex}.
States of the diagram defined by this pre-model were given in Example
\ref{state_ex}. To define the transitions of the model defined by ${\cal M}$
we use Definition \ref{trans}. Let us illustrate this by showing that  a triple
$\langle \sigma_1,b,\sigma_5 \rangle$ is a transition. To do that we
need first to construct a program $P({\cal M}, \sigma_1, b)$ (we are
only showing rules relevant to our argument):


\st
$
\begin{array}{llll}
[1] &f(x,1) = y & \leftarrow & instance(b,actions),\\
       &        &                & occurs(b,0),\\
        &      &                & attr_1(b) = y,\\
        &    &                & g(x,0) = y \mbox{.}
\end{array}
$

\st
$
\begin{array}{llll}
[2] & \neg dom_f(x,1) & \leftarrow & instance(b,actions),\\
      &         &                & occurs(b,0),\\
        &      &                & attr_2(b) = y \mbox{.}
\end{array}
$

\st
$
\begin{array}{ll}
[3] & dom_f(x,0) \leftarrow f(x,0) = y \mbox{.}\\
& dom_f(x,1) \leftarrow f(x,1) = y \mbox{.}\\
 & dom_g(x,0) \leftarrow g(x,0) = y \mbox{.} \\
 & dom_g(x,1) \leftarrow g(x,1) = y \mbox{.}
\end{array}
$


\st
$
\begin{array}{llll}
[4] & f(x,1) = y & \leftarrow & dom_f(x,1),\\
    &           &                & f(x,0) = y,\\
      &         &                & \no f(x,1) \not= y \mbox{.} \\
       & & & \\
&g(x,1) = y & \leftarrow & dom_g(x,1),\\
  &             &                & g(x,0) = y,\\
    &           &                & \no g(x,1) \not= y \mbox{.}
\end{array}
$

\st
$
\begin{array}{llll}
[5] & dom_f(x,1)& \leftarrow & dom_f(x,0),\\
        &       &                & \no \neg dom_f(x,1) \mbox{.} \\
        & & & \\
& dom_g(x,1)& \leftarrow & dom_g(x,0),\\
   &            &                & \no \neg dom_g(x,1) \mbox{.}
\end{array}
$

\st
$
\begin{array}{ll}
[6] & f(x,0) = y \mbox{.} \\
     & g(x,0) = y \mbox{.} \\
     & occurs(b,0)  \mbox{.}
\end{array}
$

\st
It is easy to see that the program has a unique answer set, say, $S$.
Since $\sigma_5 = \{g(x) = y\}$ we need to show that the only fluent
atom with the step parameter $1$ belonging to $S$ is $g(x,1) = y$.
By the second rule from group $[5]$, $dom_g(x,1) \in S$. By the second
rule of $[4]$ we have that $g(x,1) = y \in S$. As expected, function
$g$ maintains its value by inertia. The situation is different for
$f$. By rule $[2]$ we have that $\neg dom_f(x,1) \in S$ and hence
neither rule $[5]$ nor $[4]$ for $f$ are applicable. Rule $[1]$ is
also not applicable since $attr_1$ is not defined for $b$. Therefore
the state defined by $S$ is exactly $\sigma_5 = \{g(x) = y\}$.
(Note that the argument would not be possible if we were to use the
traditional version of the Inertia Axiom. The modification related
to the treatment of $dom$ presented in axioms $[4]$ and $[5]$ is essential.)

 \st
Using the same method one can easily verify that
triples $\langle \sigma_2,a,\sigma_1\rangle$, $\langle \sigma_5,a,\sigma_1 \rangle$,
$\langle \sigma_5,b,\sigma_5 \rangle$, etc. are
transitions of the transition diagram defined by ${\cal M}$.

\end{example}

\subsection{Entailment Relation}\label{er}
Let us consider a fixed action theory $T$ with action signature $\Sigma$, and define
an entailment relation between $T$ and statements of $\Sigma$.

Let $\mathcal{I}$ be an interpretation of $\Sigma$.
A \emph{ground instance} of a statement $\alpha$  of $\Sigma$ with respect to
$\mathcal{I}$ is a statement obtained by
replacing variables of $\alpha$ by properly typed object constants in $\Sigma_{\cal I}$ and
replacing object constants  of $\alpha$
by their interpretations in $\mathcal{I}$.

\medskip
Now let us consider a model $T_{\mathcal{M}}$ of a basic action theory $T$
defined by a pre-model
  $\mathcal{M}$ with the universe $U$ and let  $\sigma$ be a state of $T_{\mathcal{M}}$.


\begin{definition}[Satisfiability Relation for Ground Statements of a $\mathcal{BAT}$]
\begin{itemize}
\item A state $\sigma$ of $T_{\mathcal{M}}$ satisfies a ground state
constraint $\alpha$
if $\sigma$ contains the head of $\alpha$ whenever it contains its body.
\item A state $\sigma$ of $T_{\mathcal{M}}$ satisfies a ground
  definition $\alpha$ if $\sigma$ contains the head of a clause in $\alpha$ iff
$\alpha$ contains a clause with the same head and the body belonging
to $\sigma$.

\item A transition $\langle \sigma_0, a, \sigma_1 \rangle$ of $T_{\mathcal{M}}$ satisfies
a ground dynamic causal law $\alpha$ that starts with the expression
``$occurs(e)\ {\bf causes}$''
if $a$ contains action $e$ and
$\sigma_1$ contains the head of $\alpha$ whenever $\sigma_0$ contains its body.

\item A transition $\langle \sigma_0, a, \sigma_1 \rangle$ of $T_{\mathcal{M}}$ satisfies
a ground executability condition $\alpha$ 
that starts with the expression ``${\bf impossible}\ e$''
if either (1) $a$ does not contain $e$ or (2)
the body of $\alpha$ contains:
\begin{itemize}
\item a ground literal $l$ such that $l \notin \sigma_0$, or
\item an expression ``$occurs(e_1)$'' such that $e_1 \notin a$, or
\item an expression ``$\neg occurs(e_2)$'' such that $e_2 \in a$.
\end{itemize}
\begin{comment}
if $a$ does not contain $e$ 
whenever $\sigma_0$ contains the ground literals in the body of $\alpha$.
Additionally, if the body of $\alpha$ contains an expression of the form ``$occurs(e_1)$''
then the requirement above applies only if $a$ contains action $e_1$;
if the body contains ``$\neg occurs(e_2)$''
then the requirement applies only if $a$ does \emph{not} contain action $e_2$.
\end{comment}
\end{itemize}
\end{definition}

\begin{definition}[Satisfiability Relation for Arbitrary Statements of a
  $\mathcal{BAT}$]
Let $T_{\mathcal{M}}$ be a model of a basic action theory $T$ defined by a pre-model
  $\mathcal{M}$ with the universe $U$.
\begin{itemize}
\item $T_{\mathcal{M}}$ satisfies a constraint $\alpha$ over signature
  $\Sigma$ of $T$ if every state of $T_{\mathcal{M}}$ satisfies all ground instances
  of $\alpha$ with respect to $U$.
Similarly for definitions.

\item $T_{\mathcal{M}}$ satisfies a dynamic causal law $\alpha$ over signature
  $\Sigma$ of $T$ if every transition of $T_{\mathcal{M}}$ satisfies all ground instances
  of $\alpha$ with respect to $U$.
Similarly for executability conditions.
\end{itemize}
\end{definition}
\begin{definition}[Entailment]
{
A statement $\alpha$ is entailed by a theory $T$ ($T \models \alpha$) if $\alpha$ is true in every model of $T$.
}
\end{definition}
Having the notion of entailment allows us to investigate the relationship between
causal laws. For instance we can show that
$$\{occurs(A) \causes f \lif p, q;\ occurs(A) \causes f \lif \neg p\}
\models occurs(A) \causes f \lif q$$
$$\{occurs(A) \causes f \lif p,q;\ q \lif p\}
\models occurs(A) \causes f \lif p$$
etc. Our notion of entailment is somewhat similar to the notion of
\emph{subsumption} from \cite{Eiter10} -- a relation between an action
description and a query (including queries having the form of causal
laws and executability conditions). Our entailment relation can be
viewed as a generalization of subsumption from system descriptions to
theories. It allows variables and, unlike that of subsumption, is
defined in terms of multiple transition diagrams specified by the
theory. There are also related formalisms that allow entailment of
causal laws and executability conditions (see, for instance
\cite{Turner99} and \cite{gllmt04}).
There are many
interesting problems related to the ${\cal ALM}$ entailment, including that of
finding a sound and complete set of inference
rules for it. We hope to address these problems in our
future work.

\section{Language ${\cal ALM}$}\label{alm}
In this section we use examples to introduce the syntax of
theories and system descriptions of ${\cal ALM}$ and define their
semantics. (The full grammar for the language can be seen in \ref{grammar}.)

We begin with describing \emph{unimodule} system descriptions,
i.e. system descriptions whose theories consist of exactly one module.

\subsection{Unimodule System Descriptions}
We start with a comparatively simple problem of formalizing
the domain described by the following story:

\begin{example}[A Travel Domain]\label{travel_domain}
Consider a travel domain
in which there are
two $agents$, Bob and John, and three locations,
New York, Paris, and Rome. Bob and John can move from one location to another
if the locations are connected.
\end{example}
If we were to represent this knowledge in $\mathcal{AL}$ we would
start with identifying objects of the domain including actions such as,
say, $go(bob,paris,rome)$ and write $\cal{AL}$ axioms describing
the relationships between these objects. The use of  $\cal{ALM}$
suggests a very different methodology.

\medskip
\noindent\fbox{%
    \parbox{\textwidth}{%
{\bf Methodology of Describing Dynamic Domains in ${\cal ALM}$:}
\begin{enumerate}
\item Determine what \emph{sorts} of objects are relevant to the domain
    of discourse and how these sorts can be organized into an
    inheritance hierarchy.

\item Use ${\cal ALM}$ to describe the basic action theory for
this type of domains. This should be done in two steps:

\begin{itemize}
\item  Describe the action signature
         of our abstraction by declaring
    sorts (together with their attributes and the inheritance
    hierarchy), basic and defined statics and fluents.
		(Notice that this signature normally will not contain particular objects of our story. 
		 It would have no mention of Bob,
    Paris, etc. However, the signature may include some object constants
		pertinent to the {\em general} domain of the story --
                see for instance the Monkey and Banana Problem in Section \ref{mb}.)
		
\item  Use this action signature to formulate axioms
          of the theory.
\end{itemize}
\item Populate sorts of your hierarchy with objects relevant to your
  story and describe these objects and their sort membership in ${\cal ALM}$.

\end{enumerate}
}
}

\smallskip
As is the case with other problem solving methodologies, we begin by
choosing a proper level of abstraction for our example.
Since the example is used for illustrative purposes we opted
for using the following simple abstraction:

Our domains will contain {\em things} and
{\em discrete points} in space. Certain things, called {\em agents},
will be able to move from one point to another if the two points are connected.
We are interested in the {\em relations between points}
and the {\em locations of things}, including changes of these
locations caused by a sequence of given {\em moves}.
(Note that our abstraction does not allow a location to be a part of another
location, e.g., we will not be able to express that Paris is located in
France. It ignores the means of transportation, the possibility that
locations may have restrictions on the number of things they can
contain, etc.)

Accordingly, our basic action theory containing commonsense
knowledge about motion formulated in these terms will include sorts $things$,
$agents$, $points$, and $move$, together with special sorts
$universe$ and $actions$, which belong to every action signature.

\begin{figure}[!htbp]
\centering
\includegraphics[scale=0.5]{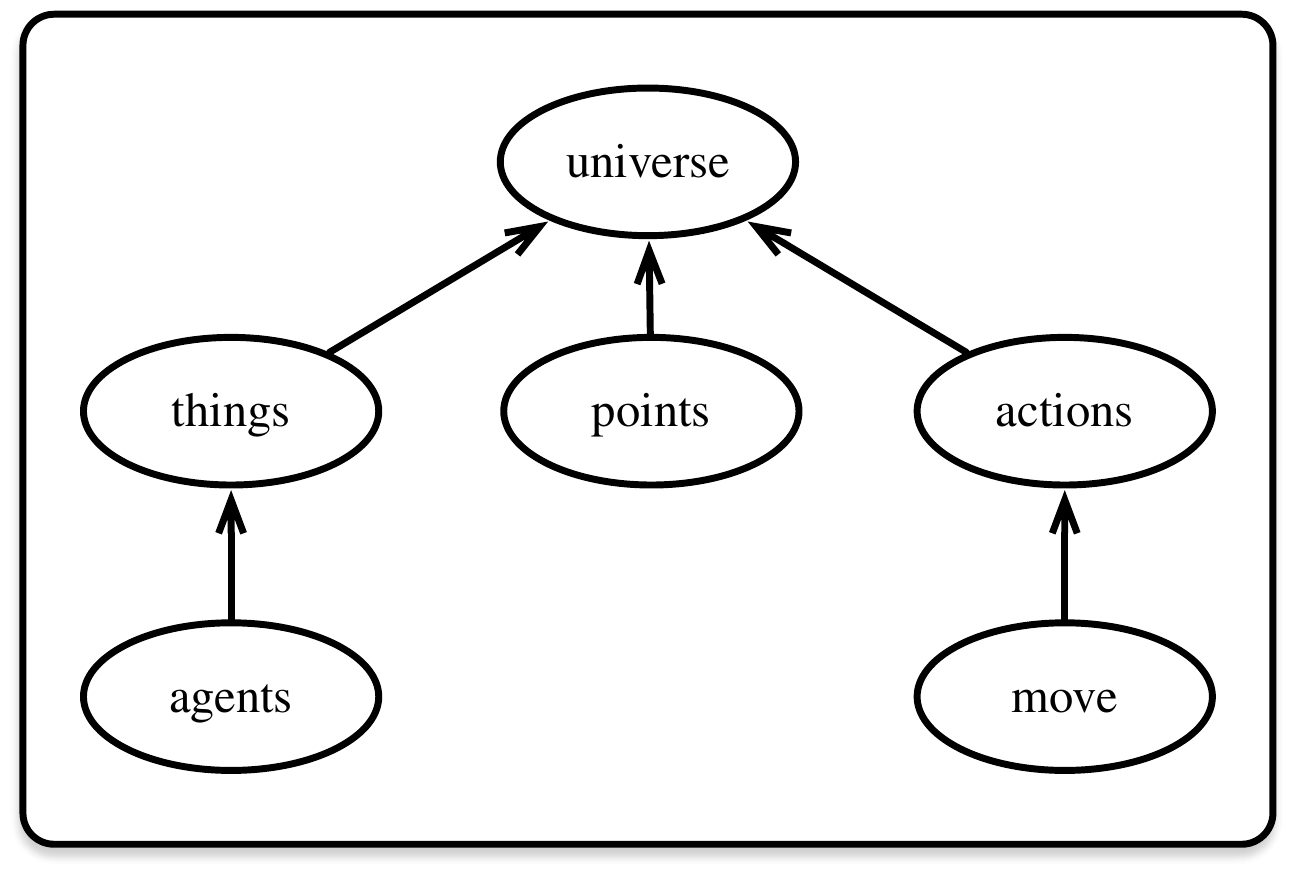}
\caption{Sort Hierarchy for $T_{bm}$}
\label{fig1}
\end{figure}

We call this basic action theory $T_{bm}$.
The sorts of $T_{bm}$ will be organized in a hierarchy
depicted in Figure \ref{fig1}.

\medskip
Our next step is to describe $T_{bm}$ in $\mathcal{ALM}$.

\begin{example}[Motion Theory in $\mathcal{ALM}$]\label{ex2}
The description of a theory in $\mathcal{ALM}$ starts with the keyword
{\bf theory} and is followed by a collection of modules.
Our theory, called $basic\_motion$,  consists of
only one module $moving$


\st
$
\begin{array}{l}
{\bf theory}\ basic\_motion\\
\ \ \ {\bf module} \ moving\\
\ \ \ \ \ \ \langle module\ body\rangle
\end{array}
$

\medskip
\noindent
where $\langle module\ body\rangle$ stands for the declarations of sorts,
functions, and axioms of the theory.
We assume that $things$, $points$, and $agents$ have
no attributes, while actions from the sort $move$ may come with
attribute $actor$ indicating the agent involved in the action, and
attributes $origin$ and $destination$ (abbreviated as $dest$) describing the locations of the
actor before and after the execution of the action.
Syntactically, all this information is specified as:

\bigskip

$
\begin{array}{l}
\ \ \ \ {\bf sort\ declarations}\\
\ \ \ \ \ \ \ \ points, things\ :: \ universe\\
\ \ \ \ \ \ \ \ agents \ :: \ things
\end{array}
$

\smallskip

$
\begin{array}{l}
\ \ \ \ \ \ \ \ move \ :: \ actions\\
\ \ \ \ \ \ \ \ \ \ \ \ {\bf attributes}\\
\ \ \ \ \ \ \ \ \ \ \ \ \ \ \ \ actor : agents\\
\ \ \ \ \ \ \ \ \ \ \ \ \ \ \ \ origin : points\\
\ \ \ \ \ \ \ \ \ \ \ \ \ \ \ \ dest : points
\end{array}
$

\bigskip
The construct $\ :: \ $ is called \emph{specialization}
and corresponds to the links of the sort hierarchy; for instance,
the link from $agents$ to $things$ in Figure \ref{fig1} is
recorded by the statement $agents \ :: \ things$.
Multiple links going into the same sort can be recorded by a single statement,
as in $points, things \ :: \ universe$.
Note that the special sorts $universe$ and $actions$ do not
have to be declared.
In case of a sort hierarchy with multiple links from $c$ to
$pc_1, \dots, pc_k$ we will use a specialization statement of the form
$c \ :: \ pc_1, \dots, pc_k$. In describing the attributes of actions
of the sort $move$ we use a shorthand. Attributes of $move$ are functions defined
on elements of the sort $move$, which means that the definition of, say,
attribute $actor$ should be written as $actor : move \rightarrow agents$.
After some deliberation however, we decided to allow to write it simply
as $actor : agents$. The same agreement holds for attributes with a
larger number of parameters; an attribute of a sort $c$ that has the form
$attr\_name : c \times c_0 \times \dots \times c_n \rightarrow
c_{n+1}$ can be written as $attr\_name : c_0 \times \dots \times c_n \rightarrow
c_{n+1}$.
This completes the description of the syntactic representation of
our sort hierarchy in ${\cal ALM}$.

\bigskip
The next step is to syntactically describe functions in the signature.
One of the functions mentioned in our informal description
specifies whether two points are connected or not.
Let us call it $connected$.
In general, the value of $connected$ can be changed by actions
(airports can be closed, roads blocked, etc.) and hence we define 
$connected$ to be a \emph{basic fluent}.
In some scenarios, the property $connected$ will be a symmetric 
relation but not in others;
similarly, it may be a transitive relation or not. 
To allow for elaboration tolerance, we introduce two basic static functions,
$symmetric\_connectivity$ and $transitive\_connectivity$
to characterize the property $connected$.
The other function relevant to our domain
maps things into points at which they are located. Let us
call it  $loc\_in$.
The value of the function can be changed by actions of our domain, hence it is a \emph{fluent}.
It is not defined in terms of other functions, thus it is a \emph{basic
fluent}.
It is also a \emph{total} function, as we assume that the location of every $thing$
is defined in every state.
In ${\cal ALM}$ these functions are syntactically declared
as:

\medskip\noindent
$
\begin{array}{l}
\ \ \ \ {\bf function\ declarations}\\
\ \ \ \ \ \ \ \ {\bf statics} \\
\ \ \ \ \ \ \ \ \ \ \ \ {\bf basic}\\
\ \ \ \ \ \ \ \ \ \ \ \ \ \ symmetric\_connectivity : booleans\\
\ \ \ \ \ \ \ \ \ \ \ \ \ \ transitive\_connectivity : booleans\\
\ \ \ \ \ \ \ \ {\bf fluents} \\
\ \ \ \ \ \ \ \ \ \ \ \ {\bf basic}\\
\ \ \ \ \ \ \ \ \ \ \ \ \ \ \ connected : points \times points \rightarrow booleans\\
\ \ \ \ \ \ \ \ \ \ \ \ \ \ \ {\bf total} \ loc\_in : things \rightarrow points\\
\end{array}
$

\st
In this example the keywords ${\bf function \ declarations}$ are followed by the lists
of statics and fluents. Elements from each list are divided into basic
and defined with each total function in the list preceded by the
keyword ${\bf total}$.  Naturally, the declaration of a sort, static,
or fluent in a module should be unique.

\medskip\noindent
This concludes our description of action signature of $T_{bm}$\footnote{
The description does not mention object constants, which can be 
declared in $\mathcal{ALM}$ by statements
$o : c \ $ and
$r(c_1,\dots,c_n) : c$.
The first statement defines object constant $o$ of sort $c$;
the second defines the collection of object constants of the form 
$r(x_1,\dots,x_n)$ where $x_1,\dots,x_n$ are object constants 
from sorts $c_1,\dots,c_n$. Example of the latter can be found in
module \emph{climbing} of Monkey and Banana representation from
section \ref{mb}.
}.

\medskip
Now we are ready to define the collection of axioms of $T_{bm}$.
In ${\cal ALM}$, we precede this collection by the keyword ${\bf axioms}$.
Each axiom will be ended by a period (.),
as in:

\medskip
$
\begin{array}{l}
\ \ \ \ {\bf axioms}
\end{array}
$

$
\begin{array}{lllll}
\ \ \ \ \ \ \ \ occurs(X) & \causes & loc\_in(A) = D & \lif & instance(X, move), \\
                          &         &                &      & actor(X) = A,\\
                          &         &                &      & dest(X) = D\mbox{.}
\end{array}
$

$
\begin{array}{lll}
\ \ \ \ \ \ \ \ connected(X, X)\mbox{.} & &  \\
\ \ \ \ \ \ \ \ connected(X, Y) & \lif & connected(Y, X),\\
                                &      & symmetric\_connectivity\mbox{.} \\
\ \ \ \ \ \ \ \ \neg connected(X, Y) & \lif & \neg connected(Y, X),\\
                                &       & symmetric\_connectivity \mbox{.} \\
\ \ \ \ \ \ \ \ connected(X, Z) & \lif & connected(X, Y), \\
                                &      & connected(Y, Z),\\
                                &      & transitive\_connectivity\mbox{.}
\end{array}
$

$
\begin{array}{llll}
\ \ \ \ \ \ \ \ \impossible & occurs(X) & \lif & instance(X, move), \\
                                    &   &      & actor(X) = A,\\
                                    &   &      & loc\_in(A) \neq origin(X)\mbox{.}
\end{array}
$

$
\begin{array}{llll}
\ \ \ \ \ \ \ \ \impossible & occurs(X) & \lif & instance(X, move), \\
                                    &   &      & actor(X) = A,\\
                                    &   &      & loc\_in(A) = dest(X)\mbox{.}
\end{array}
$

$
\begin{array}{llll}
\ \ \ \ \ \ \ \ \impossible & occurs(X) & \lif & instance(X, move), \\
                                    &   &      & actor(X) = A,\\
                                    &   &      & loc\_in(A) = O,\\
                                    &   &      & dest(X) = D,\\
                                    &   &      & \neg connected(O, D)\mbox{.}
\end{array}
$

\smallskip
\noindent
The keyword ${\bf total}$ in the declaration of the basic fluent $loc\_in$
stands for the axiom

\medskip
$
\begin{array}{lll}
\ \ \ \ \ \ \ \ false & \lif & \neg dom_{loc\_in}(X)\mbox{.}
\end{array}
$

\st
that would otherwise have to be included among the axioms above.
In general, the keyword ${\bf total}$ included in the declaration of a function
$f : c_0 \times \dots \times c_n \rightarrow c$
stands for the axiom

\medskip
$
\begin{array}{lll}
\ \ \ \ \ \ \ \ false & \lif & \neg dom_f(X_0, \dots, X_n)\mbox{.}
\end{array}
$

\bigskip\noindent
This completes our description of  the basic action theory $T_{bm}$ in
$\mathcal{ALM}$.

\end{example}

Note that the \emph{semantics of the unimodule
$\mathcal{ALM}$ theory $basic\_motion$ is given by
the basic action theory $T_{bm}$ defined by it}.
In the following sections
we will present other examples of basic action theories and their
interpretations represented
in $\mathcal{ALM}$. (Whenever possible we will make no distinction between
these theories and their $\mathcal{ALM}$ representations.)

\medskip

As discussed above, a basic action theory $T$
is used to define the collection of its models --- transition diagrams
representing dynamic domains
with shared ontology and properties.
Usually, a knowledge engineer is interested in  one such domain,
characterized by particular objects, sorts, and values of statics.
If the engineer's knowledge about this domain is complete,
the domain will be represented by a unique
model of $T$. Otherwise there can be several alternative models.

The syntactic construct of  ${\cal ALM}$ used to define such knowledge
is called a \emph{structure} and has the form

\medskip
$
\begin{array}{l}
{\bf structure}\ name  \\
\ \ \ \langle structure\ body\rangle
\end{array}
$

\medskip
\noindent
where $\langle structure\ body\rangle$ stands for the definition of
objects in the hierarchy of $T_{bm}$ and the values of its statics.
Let us illustrate the use of this construct by the following example:

\begin{example}[$\mathcal{ALM}$'s Representation of a Specific Basic
  Motion Domain.]\label{ex3}
{
Let us consider the $\mathcal{ALM}$ theory ${basic\_motion}$ from Example
\ref{ex2}, which encodes the basic action theory $T_{bm}$,
and use $\mathcal{ALM}$ to specify the particular basic motion domain
from Example \ref{travel_domain}.

\smallskip
The $\mathcal{ALM}$ definition of the structure used to describe this domain
starts with the header:

\medskip\noindent
$\ \ {\bf structure}\ Bob\_and\_ John$

\smallskip\noindent
followed by the definition of $agents$ and $points$:

\smallskip

$
\begin{array}{l}
\ \ \ \ {\bf instances}\\
\ \ \ \ \ \ \ \ bob, john \ {\bf in} \ agents\\
\ \ \ \ \ \ \ \ new\_york, paris, rome \ {\bf in} \ points
\end{array}
$

\st
To specify particular actions of our domain we expand our list of
instances by

\smallskip
$
\begin{array}{l}
\ \ \ \ \ \ \ \ go(X, P_1, P_2) \ {\bf in} \ move \ {\bf where} \ P_1 \neq P_2\\
\ \ \ \ \ \ \ \ \ \ \ \ actor = X\\
\ \ \ \ \ \ \ \ \ \ \ \ origin = P_1\\
\ \ \ \ \ \ \ \ \ \ \ \ dest = P_2
\end{array}
$

\smallskip\noindent
Note that the last definition describes several instances simultaneously via the use
of variables; we call this type of definition an {\em instance schema}.
The instance schema defining $go(X, P_1, P_2)$ stands for the collection
of instance definitions:

\smallskip
$
\begin{array}{l}
\ \ \ \ \ \ \ \ go(bob, new\_york, paris) \ {\bf in} \ move\\
\ \ \ \ \ \ \ \ \ \ \ \ actor = bob\\
\ \ \ \ \ \ \ \ \ \ \ \ origin = new\_york\\
\ \ \ \ \ \ \ \ \ \ \ \ dest = paris\\
\ \ \ \ \ \ \ \ \dots\\
\ \ \ \ \ \ \ \ go(john, paris, rome) \ {\bf in} \ move\\
\ \ \ \ \ \ \ \ \ \ \ \ actor = john\\
\ \ \ \ \ \ \ \ \ \ \ \ origin = paris\\
\ \ \ \ \ \ \ \ \ \ \ \ dest = rome
\end{array}
$

\smallskip
\noindent
The condition $\ {\bf where}\ P_1 \neq P_2\ $ ensures that
Bob and John do not move to a destination identical to the origin.

\smallskip\noindent
The following would also be a valid instance schema:

\smallskip
$
\begin{array}{l}
\ \ \ \ \ \ \ \ go(X, P) \ {\bf in} \ move\\
\ \ \ \ \ \ \ \ \ \ \ \ actor = X\\
\ \ \ \ \ \ \ \ \ \ \ \ dest = P
\end{array}
$

\smallskip\noindent
if we were interested only in the destinations of Bob and John's movements,
but not in their origins.



\smallskip\noindent
In our example connectivity between points is both 
symmetric and transitive:
This is captured syntactically by the following:
\footnote{If a theory 
  contains an object constant $o$ then its value, say $y$, can be declared as:

\smallskip\noindent 
$
\begin{array}{l}
\ \ \ \ {\bf object \ constants}\\ 
\ \ \ \ \ \ \ \ o = y 
\end{array}
$

\smallskip\noindent 
If the structure contains no assignment of value to constant $o$, we 
assume that $o$ belongs to the structure's universe and is mapped 
into itself. 
}

\smallskip
$
\begin{array}{l}
\ \ \ \ {\bf values\ of\ statics}\\
\ \ \ \ \ \ \ \ symmetric\_connectivity\mbox{.} \\
\ \ \ \ \ \ \ \ transitive\_connectivity\mbox{.} \\
\end{array}
$

\smallskip\noindent
This concludes our definition of $Bob\_and\_John$ structure.

\medskip
To syntactically relate a theory with its structure, we use the construct of
$\mathcal{ALM}$ called \emph{system description}. In our case it will
look as follows:

\medskip
$
\begin{array}{l}
{\bf system \ description\ } travel\\
\ \ \ \ {\bf theory}  \ basic\_motion\\
 \ \ \ \ \ \ \ \ {\bf module} \ moving\\
\ \ \ \ \ \ \ \ \ \ \ \ \langle module\ body\rangle\\
\ \ \ \ {\bf structure}\ Bob\_and\_John\\
\ \ \ \ \ \ \ \ \langle structure \ body\rangle
\end{array}
$

\medskip
\noindent
where $\langle module\ body\rangle$ and $\langle structure \
body\rangle$ are defined in Examples \ref{ex2} and \ref{ex3}.
The system description \emph{travel} contains all the information
we considered relevant to our particular travel domain.
It is not difficult to see that this knowledge is complete and
therefore describes exactly one model (i.e., one transition diagram) 
of \emph{basic\_motion}. This is exactly the model we
\emph{intended} for our domain.
A part of this model can be seen in Figure \ref{fig3}.
We only show fluent $loc\_in$ and assume that in every state of the part
of the diagram shown in the picture Paris and Rome are connected to
each other, but neither of them is connected to New York; 
we use shorthands $b$, $j$, $ny$, $p$, and $r$ for
$bob$, $john$, $new\_york$, $paris$, and $rome$ respectively; 
and we only show arcs that
are labeled by a single action.

\begin{figure}[!htbp]
\centering
\includegraphics[scale=0.7]{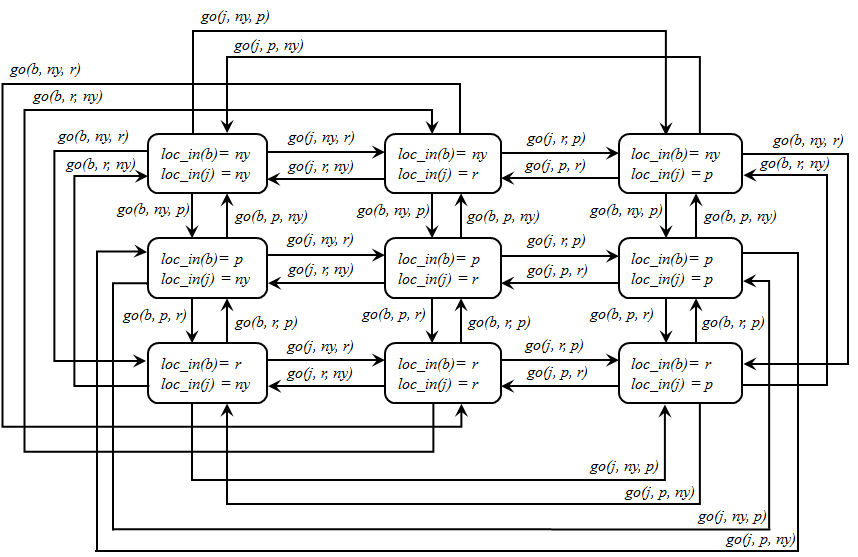}
\caption{(Partial) Transition Diagram for System Description $travel$}
\label{fig3}
\end{figure}

\noindent
The model is unique because we specified the membership of our objects in the source
nodes of the hierarchy. This information is sufficient to uniquely define the
universe and the interpretations of all the sorts.

}
\end{example}





The next example illustrates how incomplete information about a
domain can lead to multiple models of the system
description of this domain:

\begin{example}[System Description with Multiple Models]\label{ex4}
{

Consider a system description \emph{underspecified\_hierarchy}
consisting of a theory \emph{professors} and a structure \emph{alice}:

\medskip\noindent
$
\begin{array}{l}
{\bf system\_description}\ underspecified\_hierarchy\\
\ \ {\bf theory} \ professors\\
\ \ \ \ \ {\bf module} \ professors\\
\ \ \ \ \ \ \ \ {\bf sort\ declarations}\\
\ \ \ \ \ \ \ \ \ \ \ professor \ :: \ person\\
\ \ \ \ \ \ \ \ \ \ \ assistant, associate, full\ \ :: \ professor
\end{array}
$

\noindent
$
\begin{array}{l}
\ \ \ \ \ \ \ \ {\bf axioms}\\
\ \ \ \ \ \ \ \ \ \ \ false \lif instance(X,C_1),\ \ instance(X,C_2),\\
\ \ \ \ \ \ \ \ \ \ \ \ \ \ \ \ \ \ \ \ \  link(C_1,professor),\ \ link(C_2,professor),\\
\ \ \ \ \ \ \ \ \ \ \ \ \ \ \ \ \ \ \ \ \  C_1 \not= C_2.
\end{array}
$

\noindent
$
\begin{array}{l}
\ \ {\bf structure}\ alice\\
\ \ \ \ \ \ {\bf instances}\\
\ \ \ \ \ \ \ \ \ \ alice \ {\bf in} \ professor
\end{array}
$

\smallskip\noindent
The theory describes a simple hierarchy. The structure
populates the hierarchy with one member, $Alice$ (see Figure ~\ref{fig:Alice}).
Unfortunately all we know about $Alice$ is that she is a professor.
It is not difficult to check that this system description has three
models. In the first one $Alice$ is an assistant professor, in the
second she is an associate professor, and in the third one - a full professor.

\begin{figure}[!htbp]
\centering 
\includegraphics[scale=0.3]{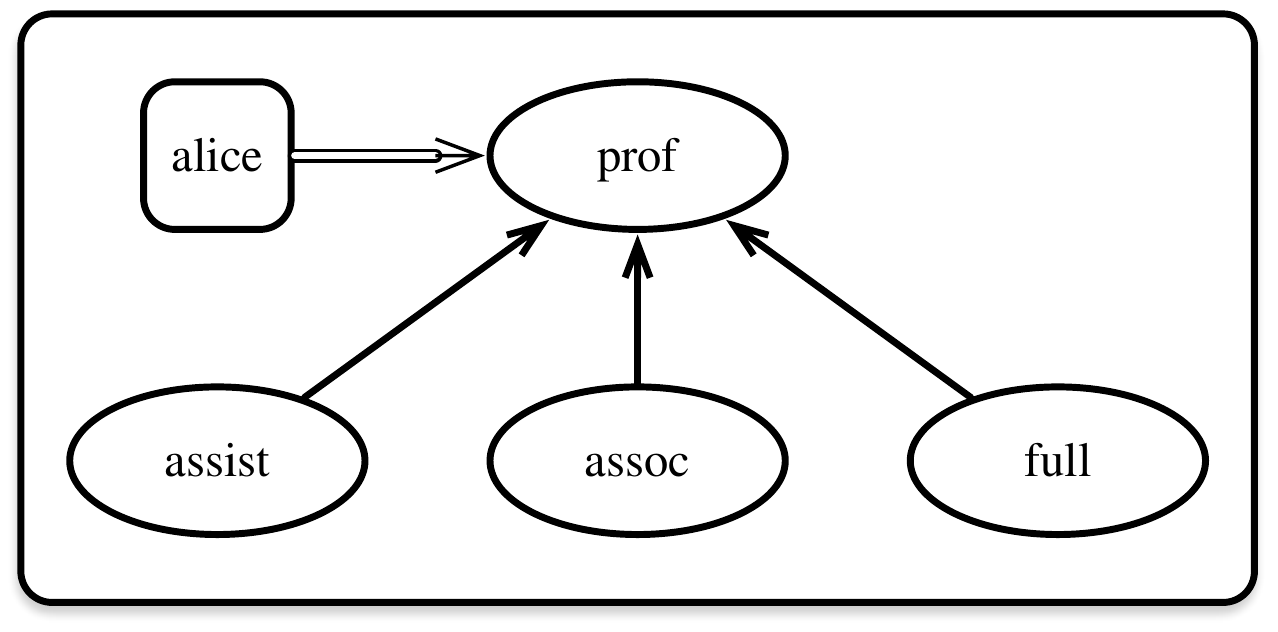}
\caption{Underspecified Hierarchy}
\label{fig:Alice}
\end{figure}

}
\end{example}

We hope that these examples gave the reader a sufficient insight in
the meaning of unimodule ${\cal ALM}$ theories and system
descriptions.
In
general, the semantics of a syntactically correct unimodule theory
${\cal T}$ of ${\cal ALM}$ is
given by the unique ${\cal BAT}$ defined by ${\cal T}$.
Similarly,
the semantics of a system description $D$ of ${\cal ALM}$ is given
by models of the ${\cal BAT}$ theory defined by ${\cal T}$ and by the set
of interpretation defined by the structure of $D$.

\subsection{Organizing Knowledge into Modules}
\label{alm_modules}
So far we only considered very simple ${\cal ALM}$ theories consisting of one
module. To create theories containing a larger body of knowledge
we need  multiple modules organized into a module hierarchy.
To illustrate this concept let us consider an extension of  basic action theory
$T_{bm}$ of motion by an additional sort of things called \emph{carriables},
which can be \emph{carried} between connected points by \emph{agents}
that are holding them. Recall from Example \ref{ex2}
that we represented the original $T_{bm}$ as an ${\cal ALM}$ theory called $basic\_motion$,
with a unique module $moving$.
We will use the name $motion$ for the $\cal ALM$ theory that will
specify the extension of $T_{bm}$. The new theory
will contain the $moving$ module developed above as well as
a new module called $carrying\_things$:

\smallskip
\noindent
$
\begin{array}{l}
{\bf theory}\ motion\\
\ \ \ {\bf module} \ moving\\
\ \ \ \ \ \ \langle module\ body\rangle \\
\ \ \ {\bf module} \ carrying\_things\\
\ \ \ \ \ \ \langle module\ body\rangle
\end{array}
$

\smallskip
\noindent
In addition to sorts, fluents, and
axioms from module $moving$, the signature of the new module
$carrying\_things$ will contain two new sorts, $carriables$ and $carry$;
a new inertial fluent, ${holding}$; and a defined fluent, $is\_held$.
Informally, $holding$ will be understood as
  \emph{having in one's hands} and
 $carry$ as \emph{moving while holding},
which will allow us to define $carry$ as a special case of $move$.

The dependency of $carrying\_things$ on $moving$ is
expressed in $\mathcal{ALM}$ by the syntactic construct ${\bf depends \ on}$ called
\emph{module dependency} as follows:

\smallskip
\noindent
$
\begin{array}{l}
{\bf module} \ carrying\_things\\
\ \ \ \ {\bf depends \ on} \ moving\\
\end{array}
$

\smallskip
\noindent
This says that the sorts and functions explicitly declared in $carrying\_things$
depend on sorts and functions declared in the module $moving$.
We say that the declarations of $moving$ are {\em implicit} in module $carrying\_things$.
We \emph{require all sorts and functions appearing in a module to be either explicitly
or implicitly declared in that module}. By means of the module dependency construct,
a theory of ${\cal ALM}$ can be structured into a \emph{hierarchy of
  modules}. \emph{The dependency relation of this hierarchy should form a DAG}.
Now we define the body of the new module:

\smallskip
\noindent
$
\begin{array}{l}
\ \ \ \ {\bf sort \ declarations}\\
\ \ \ \ \ \ \ \ carriables :: things
\end{array}
$

\noindent
$
\begin{array}{l}
\ \ \ \ \ \ \ \ carry :: move\\
\ \ \ \ \ \ \ \ \ \ \ \ {\bf attributes}\\
\ \ \ \ \ \ \ \ \ \ \ \ \ \ \ \ carried\_object : carriables
\end{array}
$

\smallskip
\noindent
Note that, since $carry$ is defined as a special case of $move$, it
automatically inherits the attributes of $move$;
hence those attributes do not have to be repeated in
the declaration of $carry$.
Next, the module contains the declarations of functions:

\smallskip
\noindent
$
\begin{array}{l}
\ \ \ \ {\bf function \ declarations}\\
\ \ \ \ \ \ \ {\bf fluents}\\
\ \ \ \ \ \ \ \ \ {\bf basic}\\
\ \ \ \ \ \ \ \ \ \ \ {\bf total}\ holding : agents \times things \rightarrow booleans\\
\ \ \ \ \ \ \ \ \ {\bf defined} \\
\ \ \ \ \ \ \ \ \ \ \ \  is\_held : things \rightarrow booleans
\end{array}
$

\smallskip
\noindent
and the new axioms:

\smallskip

$
\begin{array}{l}
\ \ \ \ {\bf axioms}
\end{array}
$

$
\begin{array}{lll}
\ \ \ \ \ \ \ \ loc\_in(C)=P & \lif & holding(T, C),\\
                                   &       & loc\_in(T)=P\mbox{.}\\
\ \ \ \ \ \ \ \ loc\_in(T)=P & \lif & holding(T, C),\\
                                   &       & loc\_in(C)=P\mbox{.}\\
\end{array}
$

$
\begin{array}{lll}
\ \ \ \ \ \ \ \ is\_held(X) & \lif & holding(T, X)\mbox{.}
\end{array}
$

$
\begin{array}{llll}
\ \ \ \ \ \ \ \ \impossible & occurs(X) & \lif & instance(X, move),\\
                            &   &      & actor(X) = A,\\
                            &   &      & is\_held(A)\mbox{.}
\end{array}
$

$
\begin{array}{llll}
\ \ \ \ \ \ \ \ \impossible & occurs(X) & \lif & instance(X, carry),\\
                            &   &      & actor(X) = A,\\
                            &   &      & carried\_object(X) = C,\\
                            &   &      & \neg holding(A, C)\mbox{.}
\end{array}
$

\st
The first two axioms say that an agent and an object he is holding have
the same location. The next defines fluent $is\_held(X)$ -- object $X$
is held by someone or something. The first executability condition states that to
move an actor should be free (i.e., not held). The second states that
it is impossible to carry a thing without holding it.

\medskip

Structuring a theory of ${\cal ALM}$ into a hierarchy of modules has several advantages.
\emph{First, this supports the stepwise development of a knowledge base by
allowing parts of its theory to be developed and tested
independently from other parts.
Second, it increases the readability of ${\cal ALM}$ theories, due to the
more manageable size of their modules.\footnote{For greater readability,
we recommend maintaining a balance between a manageable module size and a
relatively shallow module dependence hierarchy.}
And finally, this approach facilitates the creation of knowledge libraries.}
Theories
containing very general information can be stored in a library and \emph{imported}
from there when constructing system descriptions. For instance,
imagine that our $motion$ theory
is stored in a library called $commonsense\_library$.
The system description $travel$ could then be re-written by importing this
theory as follows:

\smallskip
$
\begin{array}{l}
{\bf system \ description\ } travel\\
\ \ \ \ {\bf import\ theory} \ motion \ {\bf from} \ commonsense\_library\\
\ \ \ \ {\bf structure}\ Bob\_and\_John\\
\ \ \ \ \ \ \ \ \langle structure \ body\rangle
\end{array}
$

\medskip

We hope that these examples gave the reader some insight into the
meaning of theories of ${\cal ALM}$
that have more than one
module. The accurate semantics for such a theory $T$ is given by its
\emph{flattening}, i.e., by translating $T$ into the unimodular theory with the same
intuitive meaning.

First, we will give the semantics of theories satisfying the semantic conditions
given in the following definition, theories that we call {\em semantically coherent}.

\begin{definition}[Semantically Coherent Theory]
\label{sem_cond}
A theory of ${\cal ALM}$ is {\em semantically coherent} if it satisfies
the following conditions:
\begin{itemize}
\item All sorts and functions appearing in a module of $T$ are (explicitly
or implicitly) declared in that module.
\item The module hierarchy of $T$ defined by relation
``\emph{depends on}'' forms a DAG, $G$.
(The nodes of $G$ correspond to modules of
$T$. An arc  $\langle M_2,M_1 \rangle$ is in $G$ if and only if
module $M_2$ contains the statement ``\emph{depends on $M_1$}''.)
\item No two modules of a theory contain different declarations of the same
sort or the same function name.
\end{itemize}
\end{definition}

The last condition in Definition \ref{sem_cond} can be weakened to
allow the use of the same name for a function and its restriction on a
smaller sort. This and other similar features however can somewhat
distract from the main ideas of ${\cal ALM}$ and will not be included
in the original version of ${\cal ALM}$.

\medskip
The flattening $f(T)$ of an ${\cal ALM}$ theory $T$ is constructed
by the following algorithm:

\begin{enumerate}
\item  Select modules $M_1$ and $M_2$ of $T$ such that $M_1$
contains the statement ``\emph{depends on $M_2$}''.
\item Replace $M_1$ and $M_2$ by the new module $M$
obtained by uniting \emph{depends on} statements, sort declarations, object
constant declarations, function
declarations, and axioms of $M_2$ with those of $M_1$.
\item Remove the statement ``\emph{depends on $M_2$}''  from $M$.
 \item Replace $M_1$ and $M_2$ in all the statements of $T$ of the form
``\emph{depends on $M_1$}'' and ``\emph{depends on $M_2$}'' by $M$.
\item Repeat until no dependent modules exist.
\item Construct a new module with declarations and axioms defined as
  unions of the corresponding declarations and axioms of the remaining
  modules.
\item Return the resulting unimodule theory $f(T)$.
\end{enumerate}

\medskip
The second condition in Definition \ref{sem_cond} guarantees that the algorithm will
terminate. The first and second conditions ensure that the result of the algorithm
does not contain the \emph{depends on} statement and that
all sorts and functions within module $M$ of step 2
have unique (explicit or implicit) declarations.
Thanks to condition three this property is preserved by step 6 of the
algorithm and hence $f(T)$ is indeed a unimodule
theory.

\smallskip
As expected, \emph{the semantics of an ${\cal ALM}$ theory $T$ with more than one
module is given by the semantics of  the unimodule theory $f(T)$.}

\medskip
For illustrative purposes we give
the result of applying the flattening algorithm to the $motion$ theory given
above:

\st
$
\begin{array}{l}
{\bf theory} \ flat\_motion \\
\ \ {\bf module} \ flat\_motion\\
\end{array}
$

\smallskip\noindent
$
\begin{array}{l}
\ \ \ \ {\bf sort\ declarations}\\
\ \ \ \ \ \ \ \ points, things\ :: \ universe\\
\ \ \ \ \ \ \ \ agents, carriables \ :: \ things
\end{array}
$

\smallskip\noindent
$
\begin{array}{l}
\ \ \ \ \ \ \ \ move \ :: \ actions\\
\ \ \ \ \ \ \ \ \ \ \ \ {\bf attributes}\\
\ \ \ \ \ \ \ \ \ \ \ \ \ \ \ \ actor : agents\\
\ \ \ \ \ \ \ \ \ \ \ \ \ \ \ \ origin : points\\
\ \ \ \ \ \ \ \ \ \ \ \ \ \ \ \ dest : points
\end{array}
$

\smallskip\noindent
$
\begin{array}{l}
\ \ \ \ \ \ \ \ carry \ :: \ move\\
\ \ \ \ \ \ \ \ \ \ \ \ {\bf attributes}\\
\ \ \ \ \ \ \ \ \ \ \ \ \ \ \ \ carried\_object : carriables
\end{array}
$

\smallskip\noindent
$
\begin{array}{l}
\ \ \ \ {\bf function\ declarations}\\
\ \ \ \ \ \ \ \ {\bf statics} \\
\ \ \ \ \ \ \ \ \ \ \ \ {\bf basic}\\
\ \ \ \ \ \ \ \ \ \ \ \ \ \ symmetric\_connectivity : booleans\\
\ \ \ \ \ \ \ \ \ \ \ \ \ \ transitive\_connectivity : booleans
\end{array}
$

\noindent
$
\begin{array}{l}
\ \ \ \ \ \ \ \ {\bf fluents} \\
\ \ \ \ \ \ \ \ \ \ \ \ {\bf basic}\\
\ \ \ \ \ \ \ \ \ \ \ \ \ \ \ {\bf total} \ loc\_in : things \rightarrow points\\
\ \ \ \ \ \ \ \ \ \ \ \ \ \ \ {\bf total} \ holding : agents \times
things \rightarrow booleans\\
\ \ \ \ \ \ \ \ \ \ \ \ \ \ \ connected : points \times points \rightarrow booleans
\end{array}
$

\noindent
$
\begin{array}{l}
\ \ \ \ \ \ \ \ \ \ \ \ {\bf defined} \\
\ \ \ \ \ \ \ \ \ \ \ \ \ \ \ is\_held : things \rightarrow booleans
\end{array}
$

\medskip\noindent
$
\begin{array}{l}
\ \ \ \ {\bf axioms}
\end{array}
$

\noindent
$
\begin{array}{lllll}
\ \ \ \ \ \ \ \ occurs(X) & \causes & loc\_in(A) = D & \lif & instance(X, move), \\
                          &         &                &      & actor(X) = A,\ dest(X) = D\mbox{.}
\end{array}
$

\noindent
$
\begin{array}{lll}
\ \ \ \ \ \ \ \ connected(X, X)\mbox{.} & &  \\
\ \ \ \ \ \ \ \ connected(X, Y) & \lif & connected(Y, X),\ \ 
                                symmetric\_connectivity\mbox{.} \\
\ \ \ \ \ \ \ \  \neg connected(X, Y) & \lif & \neg connected(Y, X),\ \ 
																		symmetric\_connectivity\mbox{.}\\
\ \ \ \ \ \ \ \ connected(X, Z) & \lif & connected(X, Y), \ \ 
																connected(Y, Z),\\
			        &     & transitive\_connectivity\mbox{.}
\end{array}
$

\noindent
$
\begin{array}{lll}
\ \ \ \ \ \ \ \ loc\_in(C)=P & \lif & holding(T, C), \ loc\_in(T)=P\mbox{.}\\
\ \ \ \ \ \ \ \ loc\_in(T)=P & \lif & holding(T, C), \ loc\_in(C)=P\mbox{.}\\
\end{array}
$

\noindent
$
\begin{array}{lll}
\ \ \ \ \ \ \ \ is\_held(C) & \lif & holding(T, C)\mbox{.}
\end{array}
$

\noindent
$
\begin{array}{llll}
\ \ \ \ \ \ \ \ \impossible & occurs(X) & \lif & instance(X, move), \ actor(X) = A,\\
                                    &   &      & origin(X) \neq loc\_in(A)\mbox{.}
\end{array}
$

\noindent
$
\begin{array}{llll}
\ \ \ \ \ \ \ \ \impossible & occurs(X) & \lif & instance(X, move), \ actor(X) = A,\\
                                    &   &      & dest(X) = loc\_in(A)\mbox{.}
\end{array}
$

\noindent
$
\begin{array}{llll}
\ \ \ \ \ \ \ \ \impossible & occurs(X) & \lif & instance(X, move), \ actor(X) = A,\\
                                    &   &      & loc\_in(A) = O,\ dest(X) = D,\\
																		&   &      & \neg connected(O, D)\mbox{.}
\end{array}
$

\noindent
$
\begin{array}{llll}
\ \ \ \ \ \ \ \ \impossible & occurs(X) & \lif & instance(X, move),\\
                            &   &      & actor(X) = A,\  is\_held(A)\mbox{.}
\end{array}
$

\noindent
$
\begin{array}{llll}
\ \ \ \ \ \ \ \ \impossible & occurs(X) & \lif & instance(X, carry),\ actor(X) = A,\\
                            &   &      & carried\_object(X) = C,\ \neg holding(A, C)\mbox{.}
\end{array}
$

\medskip
 For readability, we selected the same names
for the theory and its module. This theory will be used in
\ref{ALMandMAD}, for the purpose of comparing ${\cal ALM}$ and $MAD$.

\medskip
Finally, \emph{the semantics of a system description with a theory $T$
consisting of multiple modules is given by the collection of models
of the ${\cal BAT}$ defined by $f(T)$ and the collection of
interpretations defined by the system's structure.}

\medskip
This concludes our introduction to the syntax and semantics of $\mathcal{ALM}$.

\section{Methodology of Language Use}

In this section we further illustrate the methodology of using $\mathcal{ALM}$
for knowledge representation and for solving various computational
tasks.

\subsection{Representing Knowledge in $\mathcal{ALM}$}
\label{mb}

We exemplify the methodology of representing knowledge
in ${\cal ALM}$ by considering a benchmark commonsense example
from the field of reasoning about action and change --- the Monkey and
Banana Problem \cite{mc63,mc68}. (Another, more realistic, example of
the use of ${\cal ALM}$ can be found in \ref{aristotle}.)

\begin{problem}[Monkey and Banana]
\emph{A monkey is in a room.
Suspended from the ceiling is a bunch of bananas, beyond the monkey's reach.
In the room there is also a box.
The ceiling is just the right height so that a monkey standing on the box
under the bananas can reach the bananas.
The monkey can move around, carry other things around, climb on the box, and
grasp the bananas. What is the best sequence of actions for the monkey
to get the bananas?}
\end{problem}

In accordance with the basic methodology of declarative programming,
we will first
\emph{represent knowledge about the problem domain and then
reduce the problem's solution to reasoning with this knowledge}.
Based on our current experience, we recommend to divide the process
of representation into the following steps:

\medskip
\noindent\fbox{%
    \parbox{\textwidth}{%
{\bf Methodology of Creating Modular Representations in ${\cal ALM}$:}
\begin{itemize}
\item Build a hierarchy of \emph{actions} pertinent to the domain.
\item Starting from the top of the hierarchy  \emph{gradually} build
  and  \emph{test} modules
capturing properties of its actions. 
If necessary, add \emph{general} non-action modules (e.g. a module
defining a sequence of actions). 
Whenever feasible, use existing library modules.
\item Build a module \emph{main} containing \emph{specific} information needed for the
problem solution.
\item Populate the hierarchy with the domain's objects.
\end{itemize}
}
}

\medskip\noindent
Here are a few comments about the second step listed above: 
When deciding how many actions to describe
in one module, consider balancing the size of the module with the depth of the (part of the)
hierarchy that it captures; also consider the resulting depth of the module dependency hierarchy.
For instance, an action and its opposite are normally included in the
same module. So are actions that usually occur together and share
common fluents and sorts.
To facilitate the discovery of relevant library modules, 
we assume that a dictionary indexed by action classes
will be available to knowledge engineers. 
Action classes will be associated with the library modules in which they are described.
The signature and axioms of library modules will be viewable by the knowledge engineer.

\begin{figure}[!hbtp]
\centering
\includegraphics[scale=0.4]{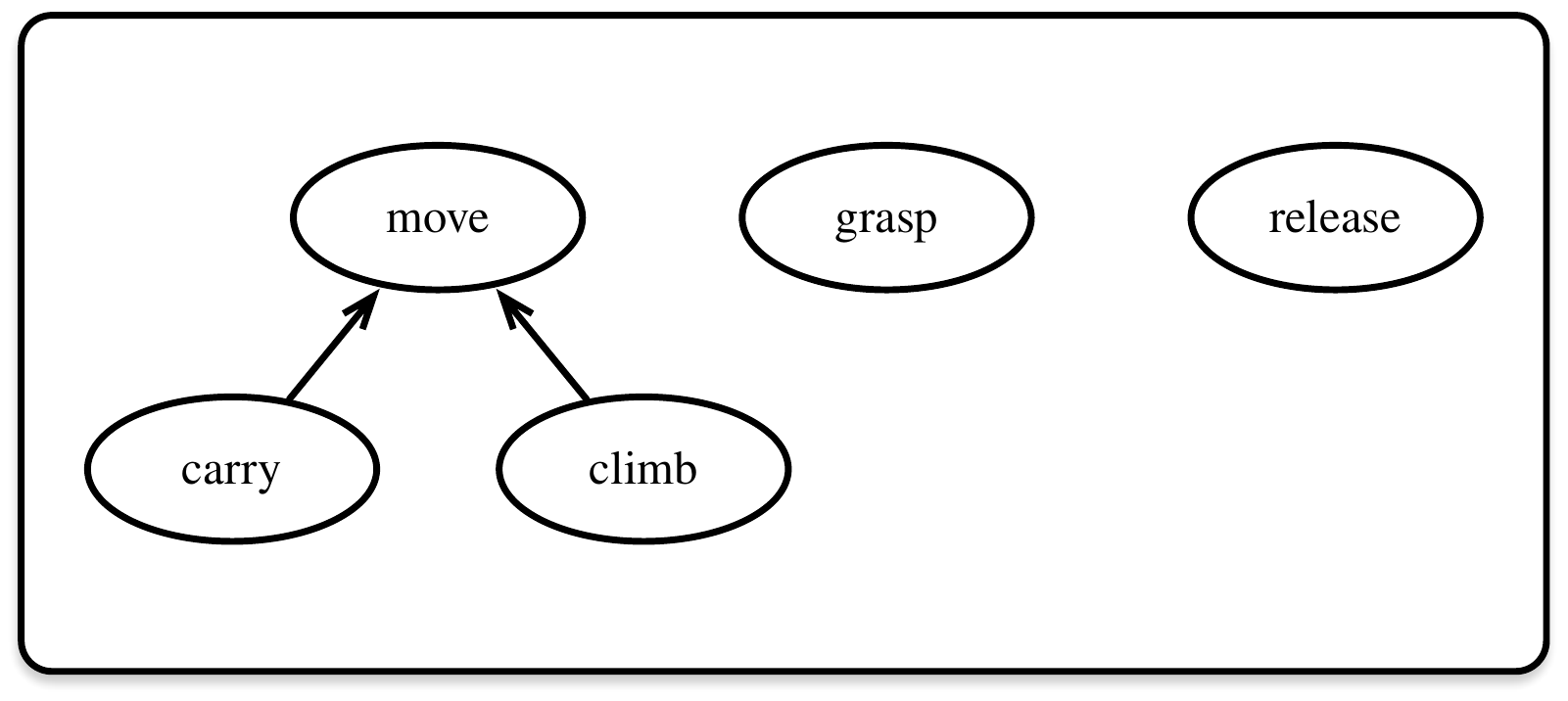}
\caption{Action Hierarchy for the Monkey and Banana Problem}
\label{fig:action-hierarchy}
\end{figure}

\smallskip
Let us illustrate the methodology by solving the Monkey and
Banana problem.
The story is clearly about an agent moving around, and grasping and  carrying things between
various points.  The hierarchy of actions pertinent to the story is illustrated
in Figure~\ref{fig:action-hierarchy}.

Note that, unlike other actions, action \emph{release} does not
explicitly appear in the story.  However, it is often advisable to consider
actions together with their opposites, so our hierarchy contains
$release$ together with $grasp$.

To gradually build a theory
$monkey\_and\_banana$ containing the knowledge needed to solve the Monkey
and Banana problem, we start with selecting a root of the action
hierarchy -- in our case action \emph{move}. The inheritance hierarchy
pertinent to \emph{move} appears in
Figure \ref{fig1}. We already discussed the module $moving$
describing the properties of \emph{move}. The theory consisting of this
module can be tested on a number of specific domains using ASP-based methods
discussed in the next section.
Next we select three actions \emph{carry}, \emph{grasp}, and
\emph{release} understood as \emph{move while holding},
\emph{take and hold}, and \emph{stop holding} respectively. 
Since these actions share a fluent \emph{holding}\footnote{For simplicity 
we assume that an agent can only hold one thing at a time.
A more general module may allow to grasp a collection of things up to a certain capacity.}
and sorts \emph{things} and \emph{agents}, 
and since a things-carrying agent usually also executes 
actions \emph{grasp} and \emph{release}, knowledge about these actions can be put in the same module.
To do that we extend the inheritance
hierarchy by a  subclass \emph{carriables} of
\emph{things} and expand module \emph{carrying\_things} from section
\ref{alm_modules} by information about another two actions.
Sort declarations of  \emph{carrying\_things} from \ref{alm_modules} 
will now also include

\smallskip\noindent 
$
\begin{array}{l}
\ \ \ \ \ \ \ \ grasp \ :: \ actions\\
\ \ \ \ \ \ \ \ \ \ \ \ {\bf attributes}\\
\ \ \ \ \ \ \ \ \ \ \ \ \ \ \ \ grasper : agents\\
\ \ \ \ \ \ \ \ \ \ \ \ \ \ \ \ grasped\_thing : things 
\end{array}
$

\noindent
and

\smallskip\noindent 
$
\begin{array}{l}
\ \ \ \ \ \ \ \ release \ :: \ actions\\
\ \ \ \ \ \ \ \ \ \ \ \ {\bf attributes}\\
\ \ \ \ \ \ \ \ \ \ \ \ \ \ \ \ releaser : agents\\
\ \ \ \ \ \ \ \ \ \ \ \ \ \ \ \ released\_thing : things 
\end{array}
$

\smallskip\noindent 
The section {\bf function declarations} of the new module will
contain the additional function \emph{can\_reach} needed as a precondition for the 
executability of \emph{grasp}. The function will be defined in terms
of locations of things. 

\smallskip\noindent 
$
\begin{array}{l}
\ \ \ \ \ \ \ \ \ \ \ {\bf defined}\\
\ \ \ \ \ \ \ \ \ \ \ \ \ can\_reach :agents \times things \rightarrow booleans 
\end{array}
$

\smallskip\noindent 
The set of {\bf axioms} will be expanded as follows.
The first two axioms below describe the direct effects of our new actions: 
action $grasp$ results in the grasper holding the thing he grasped;
this is no longer true after the thing is released. 

\smallskip\noindent 
$
\begin{array}{ll}
\ \ \ \ \ \ \ \ occurs(A) \ {\bf causes} \ holding(X, Y) \ \lif & instance(A, grasp),\\
                 & grasper(A) = X,\\
          & grasped\_thing(A) = Y\mbox{.}
\end{array}
$

\smallskip\noindent 
$
\begin{array}{ll}
\ \ \ \ \ \ \ \ occurs(A)\  {\bf causes}\  \neg holding(X, Y)\  \lif & instance(A, release),\\
                 & releaser(A) = X,\\
                & released\_thing(A) = Y\mbox{.}
\end{array}
$

\smallskip\noindent 
The constraint 

\smallskip\noindent 
$
\begin{array}{lll}
\ \ \ \ \ \ \ \ \neg holding(X, Y_2) & \lif & holding(X,Y_1), Y_1 \not= Y_2 
\end{array}
$

\smallskip\noindent 
ensures that only one thing can be held at a time (and hence to grasp 
a thing an agent must have his hands free).

\smallskip\noindent 
This is followed by the executability conditions: one cannot grasp a 
thing he is already holding or a thing that is out of his reach; one 
cannot release a thing unless he is holding it. 

\smallskip\noindent 
$
\begin{array}{ll}
\ \ \ \ \ \ \ \ {\bf impossible} \ occurs(A) \ \lif & instance(A, grasp),\\
                                     & grasper(A) = X,\\
                                 & grasped\_thing(A) = Y,\\
                                      & holding(X, Y)\mbox{.}
\end{array}
$

\noindent 
$
\begin{array}{ll}
\ \ \ \ \ \ \ \ {\bf impossible} \ occurs(A) \ \lif & instance(A, grasp),\\
                           & grasper(A) = X,\\
                                 & grasped\_thing(A) = Y,\\
                                   & \neg can\_reach(X, Y)\mbox{.}
\end{array}
$

\noindent 
$
\begin{array}{ll}
\ \ \ \ \ \ \ \ {\bf impossible} \ occurs(A) \ \lif & instance(A, release),\\
                          & releaser(A) = X,\\
                              & released\_thing(A) = Y,\\
                                & \neg holding(X, Y)\mbox{.}
\end{array}
$

\smallskip\noindent 
We also need a simple definition of \emph{can\_reach} --
an agent can always reach an object he shares a location with. 

\smallskip\noindent 
$
\begin{array}{ll}
\ \ \ \ \ \ \ \  can\_reach(M, O) \ \lif & loc\_in(M) = loc\_in(O)\mbox{.}
\end{array}
$

\smallskip\noindent 
This definition will later be expanded to describe the specific 
geometry of our domain.

\smallskip\noindent 
This completes our construction of the new module \emph{carrying\_things}.

\medskip
After testing the  theory consisting of
\emph{moving} and \emph{carrying\_things} we proceed to
constructing a new module, $climbing$, which
axiomatizes  action $climb$ understood as \emph{moving from
the bottom of a thing to its top}.  We assume that one
can climb only on tops of a special type of things called
\emph{elevations}, which will be added to our hierarchy as a subset of \emph{things}.
The corresponding declarations look as follows:


\smallskip
$
\begin{array}{l}
{\bf module} \ climbing\\
\ \ \ \ {\bf depends \ on} \  \ moving\\
\end{array}
$

\smallskip\noindent
$
\begin{array}{l}
\ \ \ \ {\bf sort \ declarations}\\
\ \ \ \ \ \ \ \ elevations \ :: \ things\\
\ \ \ \ \ \ \ \ climb \ :: \ move\\
\ \ \ \ \ \ \ \ \ \ \ \ {\bf attributes}\\
\ \ \ \ \ \ \ \ \ \ \ \ \ \ \ \ elevation : elevations\\
\end{array}
$

\smallskip\noindent
Now we introduce notation for \emph{points associated with
the tops of elevations}. The points are represented
by \emph{object constants} of the form $top(E)$ where $E$ is an
elevation. In ${\cal ALM}$ this is expressed by the following: 


\noindent
$
\begin{array}{l}
\ \ \ \ {\bf object \ constants}\\
\ \ \ \ \ \ \ \ top(elevations) :  points 
\end{array}
$
 
%




\noindent
(Notice that $top$ here is not a function symbol;
if $e$ is an elevation, then $top(e)$ is simply a point.)

\smallskip
The module contains axioms saying that $top(E)$ is the
destination of climbing an elevation $E$:

\smallskip\noindent
$
\begin{array}{l}
\ \ \ \ \ \ \ \ dest(A) = top(E) \ \lif \  elevation(A) = E \mbox{.}
\end{array}
$

\smallskip\noindent
and that a thing cannot be located on its own top:

\smallskip\noindent
$
\begin{array}{l}
\ \ \ \ \ \ \ \ false \lif loc\_in(E) = top(E) \mbox{.}
\end{array}
$

\smallskip\noindent
The last axiom prohibits an attempt by an agent to
climb an elevation from a distance:

\smallskip\noindent
$
\begin{array}{ll}
\ \ \ \ \ \ \ \ {\bf impossible} \ occurs(X) \  \lif & instance(X, climb),\\
  & actor(X) = A,\\
  & elevation(X) = O,\\
  & loc\_in(O) \not= loc\_in(A)\mbox{.}
\end{array}
$


\medskip
After testing the existing modules
we concentrate on the specific information needed for the
problem solution. It will be presented in a module called $main$.


\medskip\noindent
$
\begin{array}{l}
\ \ \ \ \ \ \ \ {\bf module}\ main\\
\ \ \ \ \ \ \ \ \ \ \ \ {\bf depends\ on}\ carrying\_things,
climbing\\ 
\end{array}
$

\medskip\noindent
The main goal of the module is to define when the monkey can reach the
banana.
We start by dividing our sort \emph{points} into three parts:
\emph{floor\_points}, \emph{ceiling\_points}, and \emph{movable\_points}:

\smallskip\noindent
$
\begin{array}{l}
\ \ \ \ {\bf sort \ declarations}\\
\ \ \ \ \ \ \ \  \ floor\_points, ceiling\_points, movable\_points \ :: \ points\\
\end{array}
$

\smallskip\noindent
where the latter correspond to tops of movable objects.
We will see the use of these sorts a little later.
Now we move to function declarations.
The story is about three particular entities: the monkey, the banana,
and the box.
They will be defined as constants of our module.

\smallskip\noindent
$
\begin{array}{l}
\ \ \ \ \ \ \ \ \ \ \ \ {\bf object\ constants}\\
 \ \ \ \ \ \ \ \ \ \ \ \ \ \ \ \ \ monkey : agents\\
 \ \ \ \ \ \ \ \ \ \ \ \ \ \ \ \ \ box : carriables, elevations\\
 \ \ \ \ \ \ \ \ \ \ \ \ \ \ \ \ \ banana : carriables
\end{array}
$

\smallskip\noindent
We will also need a function \emph{under}, such that $under(P,T)$ is true when
point $P$ is located under the thing $T$. Note that, if we consider
this function to be defined for arbitrary points, it will be dynamic --
$under(top(box),banana)$ can be true in one state and false in
another.
This will force us to declare this function as a fluent, causing an
unnecessary complication. Instead we define $under$ for floor points
only, which is sufficient for our purpose and is substantially simpler.

\smallskip\noindent
$
\begin{array}{l}
\ \ \ \ \ \ \ \ \ \ \ \ {\bf function\ declarations}\\
\ \ \ \ \ \ \ \ \ \ \ \ \ \ \ \ {\bf statics}\\
\ \ \ \ \ \ \ \ \ \ \ \ \ \ \ \ \ \ \ \ {\bf basic}\ under : floor\_points
\times things \rightarrow booleans
\end{array}
$

\smallskip\noindent
To define our function \emph{can\_reach} we need the
following axiom:

\smallskip\noindent
$
\begin{array}{lll}
\ \ \ \ \ \ \ \ \ \ \ \ {\bf axioms}& & \\
\ \ \ \ \ \ \ \ \ \ \ \ \ \ \ \ can\_reach(monkey,banana) & \lif & loc\_in(box) = P,\\
                       & & under(P,banana),\\
                      & & loc\_in(monkey) = top(box)\mbox{.}

\end{array}
$

\smallskip\noindent
Finally, we need the following axioms for the basic fluent $connected$:

\smallskip\noindent
$
\begin{array}{rll}
\ \ \ \ \ \ \ \ \ \ \ \ \ \ \ \ connected(top(box), P) & \lif & loc\_in(box) = P,\\ 
        & & instance(P, floor\_points)\mbox{.}\\
\ \ \ \ \ \ \ \ \ \ \ \ \ \ \ \ \neg connected(top(box), P) & \lif & loc\_in(box) \neq P,\\
        & & instance(P, floor\_points)\mbox{.}\\
\ \ \ \ \ \ \ \ \ \ \ \ \ \ \ \ connected(P_1,P_2) & \lif & instance(P_1,floor\_points),\\
           &   & instance(P_2,floor\_points) \mbox{.}\\
\ \ \ \ \ \ \ \ \ \ \ \ \ \ \ \ \neg connected(P_1,P_2) & \lif & instance(P_1,ceiling\_points),\\
           &   & instance(P_2,points),\\
					 &   & P_1 \neq P_2\mbox{.}
\end{array}
$

\smallskip\noindent
This completes the construction of module $main$ as well as theory
$monkey\_and\_banana$ that we will use to solve the Monkey and Banana problem.
It is easy to see that the theory is semantically coherent, as it satisfies
the conditions in Definition \ref{sem_cond}.

\begin{figure}[!htbp]
\centering
\includegraphics[scale=0.5]{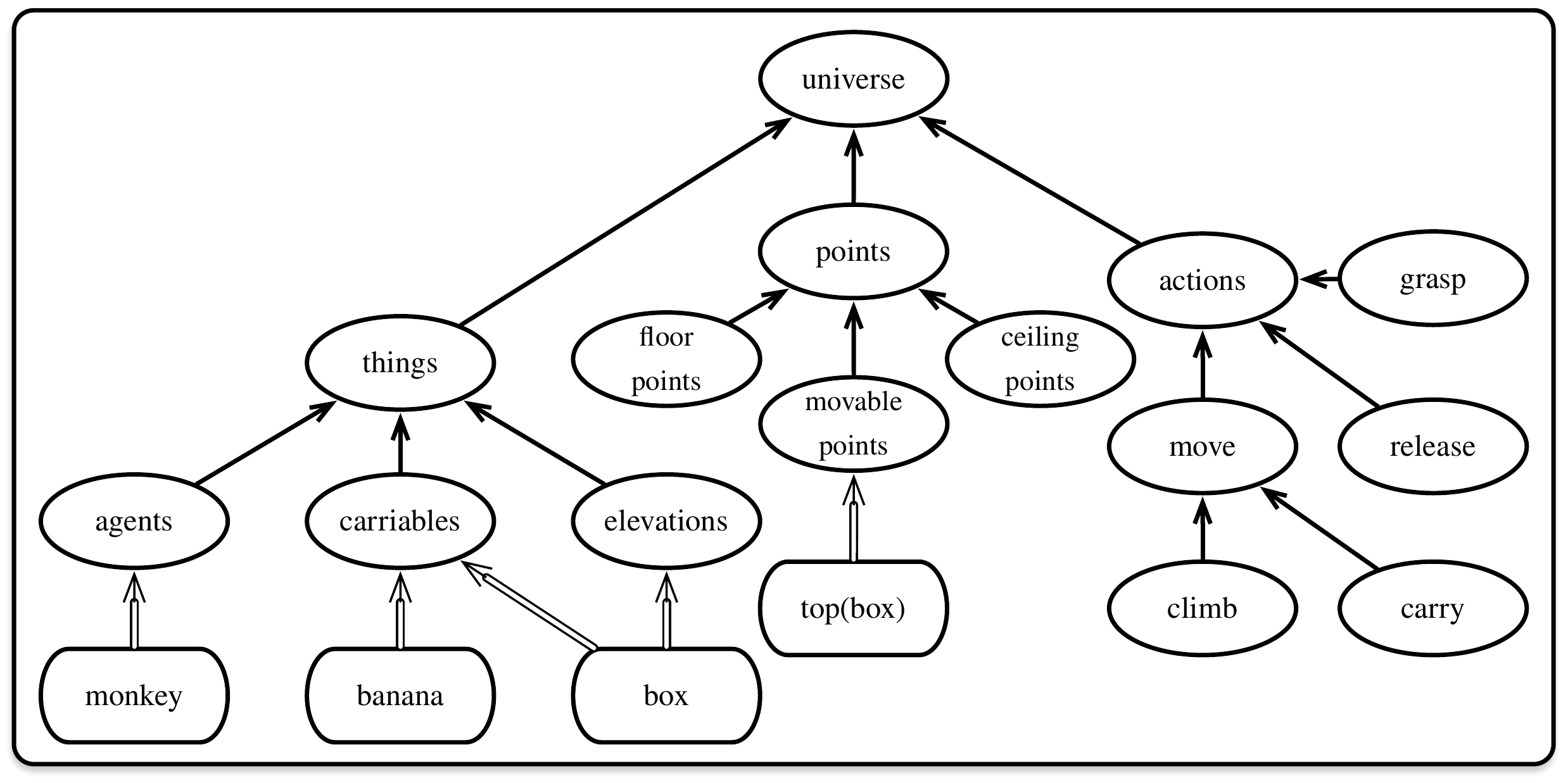}
\caption{Sort Hierarchy for the Monkey and Banana Problem}
\label{fig4:hierarchy}
\end{figure}

\begin{figure}[!htbp]
\centering
\includegraphics[scale=0.5]{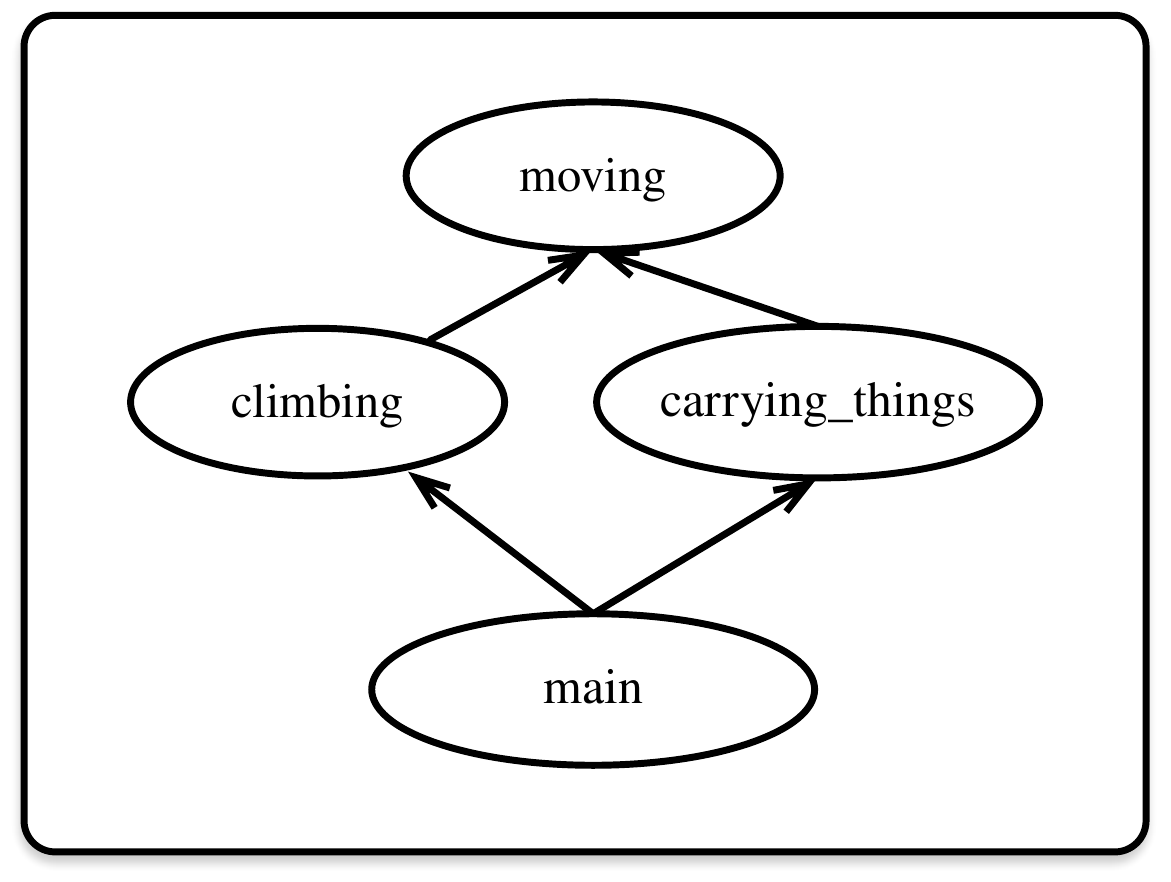}
\caption{Module Hierarchy for the Monkey and Banana Problem}
\label{fig5:hierarchy}
\end{figure}

\st
Figure \ref{fig4:hierarchy} and \ref{fig5:hierarchy}
represent the sort hierarchy and module hierarchy of
this theory, respectively.

\medskip
To complete the description of our domain we introduce
the structure containing three points located on the
floor of the room and one point located on the ceiling, 
as well as movable points and particular actions mentioned
in the story:

\smallskip

$
\begin{array}{l}
\ \ \ \ {\bf structure}\ monkey\_and\_banana\\
\ \ \ \ \ \ \ \ {\bf instances}\\
\ \ \ \ \ \ \ \ \ \ \ \ under\_banana, initial\_monkey, initial\_box\ {\bf in}\ floor\_points\\
\ \ \ \ \ \ \ \ \ \ \ \ initial\_banana\ {\bf in}\ ceiling\_points\\
\ \ \ \ \ \ \ \ \ \ \ \ top(box) \ {\bf in} \  movable\_points
\end{array}
$

\smallskip

\smallskip

$
\begin{array}{l}
\ \ \ \ \ \ \ \ \ \ \ \ move(P)\ {\bf in}\ move\ {\bf where}\ instance(P, points)\\
\ \ \ \ \ \ \ \ \ \ \ \ \ \ \ \ actor = monkey\\
\ \ \ \ \ \ \ \ \ \ \ \ \ \ \ \ dest = P
\end{array}
$

\smallskip

$
\begin{array}{l}
\ \ \ \ \ \ \ \ \ \ \ \ carry(box, P)\ {\bf in}\ carry\ {\bf where}\ instance(P,floor\_points)\\
\ \ \ \ \ \ \ \ \ \ \ \ \ \ \ \ actor = monkey\\
\ \ \ \ \ \ \ \ \ \ \ \ \ \ \ \ carried\_object = box\\
\ \ \ \ \ \ \ \ \ \ \ \ \ \ \ \ dest = P
\end{array}
$

\smallskip

$
\begin{array}{l}
\ \ \ \ \ \ \ \ \ \ \ \ grasp(C)\ {\bf in}\ grasp\ {\bf where}\ instance(C,carriables)\\
\ \ \ \ \ \ \ \ \ \ \ \ \ \ \ \ grasper = monkey\\
\ \ \ \ \ \ \ \ \ \ \ \ \ \ \ \ grasped\_thing = C
\end{array}
$

\smallskip

$
\begin{array}{l}
\ \ \ \ \ \ \ \ \ \ \ \ release(C)\ {\bf in}\ release\ {\bf where}\ instance(C,carriables)\\
\ \ \ \ \ \ \ \ \ \ \ \ \ \ \ \ releaser = monkey\\
\ \ \ \ \ \ \ \ \ \ \ \ \ \ \ \ released\_thing = C
\end{array}
$

\smallskip
$
\begin{array}{l}
\ \ \ \ \ \ \ \ \ \ \ \ climb(box)\ {\bf in}\ climb\ \\
\ \ \ \ \ \ \ \ \ \ \ \ \ \ \ \ actor = monkey\\
\ \ \ \ \ \ \ \ \ \ \ \ \ \ \ \ elevation = box\\
\end{array}
$

\smallskip
$
\begin{array}{l}
\ \ \ \ {\bf values\ of\ statics}\\
\ \ \ \ \ \ \ \ under(under\_banana,banana)\mbox{.}\\
\ \ \ \ \ \ \ \ symmetric\_connectivity\mbox{.}\\
\ \ \ \ \ \ \ \ \neg transitive\_connectivity\mbox{.}
\end{array}
$

\st
The structure specifies that the relation $connected$ is symmetric,
but not transitive. The latter prevents the monkey from moving 
from its initial location directly on top of the box. 

The theory and structure described above can be combined into a system
description \emph{monkey\_and\_banana} as follows:

\medskip
$
\begin{array}{l}
{\bf system \ description\ } monkey\_and\_banana\_problem\\
\ \ \ \ {\bf theory}  \ monkey\_and\_banana\\
\ \ \ \ \ \ \ \ {\bf import}\ motion\ {\bf from}\ commonsense\_library\\
\ \ \ \ \ \ \ \ {\bf module} \ main\\
\ \ \ \ \ \ \ \ \ \ \ \ \langle module\ body\rangle\\
\end{array}
$

\noindent
$
\begin{array}{l}
\ \ \ \ {\bf structure}\ monkey\_and\_banana\\
\ \ \ \ \ \ \ \ \langle structure \ body\rangle
\end{array}
$

\medskip\noindent
Note that the import statement above is a directive to import all of the modules of the
library theory $motion$ into the theory $monkey\_and\_banana$.

The system describes a unique hierarchy and a unique transition
diagram, $\tau$. Note that the hierarchy contains properly typed constants
$monkey$, $box$, and $banana$ declared in our
module \emph{main}; and that some of our functions, e.g. $under$,
are partial. 

\medskip
It is not difficult to check that there is a 
path in $\tau$ that starts with the initial state of our problem and
is generated by actions
$move(initial\_box)$, $grasp(box)$, $carry(box, under\_banana)$, 
$release(box)$, $climb(box)$, $grasp(banana)$. The final state of this
path will contain a fluent $holding(monkey,banana)$.
In the next section we discuss how ASP based reasoning can be
used to automatically find such sequences.

\subsection{${\cal ALM}$'s Use in Solving Computational Tasks}
\label{solving_comp_tasks}

A system description of ${\cal ALM}$ describes a collection of transition diagrams
that specifies some dynamic system.
System descriptions can be used to solve computational tasks such as
temporal projection or planning, using a methodology similar to that developed
for non-modular action languages like ${\cal AL}$ (see, for instance,
\cite{gk14}).


\subsubsection{Temporal Projection}
\label{tp}

Normally, system descriptions of ${\cal ALM}$ are used
in conjunction with the description of the \emph{system's recorded history} ---
a collection of facts about the values of fluents and the occurrences of actions
at different time steps in a trajectory. (Since we are only dealing
with discrete systems such steps are represented by non-negative integers).
Together, the system description and the history
define the collection of possible trajectories of the system
up to the current step.
In our methodology of solving temporal projection tasks,
possible trajectories are obtained by computing the answer sets of a logic program.
To formally describe this methodology, we need the following definitions.

\begin{definition}[History -- adapted from \cite{bg03}]
\label{def-history}
By the \emph{recorded history} $\Gamma_n$ of a system description
${\cal D}$ up to time step $n$
we mean a collection of observations, i.e., facts of the form:
\begin{itemize}
\item[1.] $observed(f(\overline{t}), v, i)$ -- fluent $f(\overline{t})$ was observed to have
value $v$ at time step $i$, where $0 \leq i \leq n$.
\item[2.] $happened(a, i)$ -- action $a$ was observed to happen at time step $i$,
where $0 \leq i < n$.
\end{itemize}
(There are two small differences between this and the definition
of a history by Balduccini and Gelfond \citeyear{bg03}: the latter only allows boolean
fluents and observations that have the form $observed(l,i)$ where $l$
is a fluent or its negation. Similarly for the next definitions in this subsection.)

\medskip
We say that the {\em initial situation} of $\Gamma_n$ is {\em complete} if,
for every user-defined basic fluent $f$ and any sequence of ground
terms $\overline{t}$ such that
$observed(dom_f(\overline{t}), true, 0) \in \Gamma_n$,
$\Gamma_n$ also contains a fact of the form $observed(f(\overline{t}), v, 0)$.
\end{definition}
\begin{example}[History]\label{ex-history}
{
A possible recorded history for the system description $monkey\_and\_banana\_problem$
in Section \ref{mb} may look as follows:
$$
\begin{array}{lll}
\Gamma_1 & =_{def} & \{ observed(loc\_in(monkey), initial\_monkey, 0),\\
         &         & \ \ observed(loc\_in(box), initial\_box, 0),\\
         &         & \ \ happened(move(initial\_box), 0)\}
\end{array}
$$
which says that, initially, the monkey was at point $initial\_monkey$ and the box was at $initial\_box$;
the monkey went to the initial location of the box.
}
\end{example}

The semantics of a history $\Gamma_n$ is given by the following
definition:

\begin{definition}[Model of a History -- adapted from \cite{bg03}]
\label{model_of_history}
{Let $\Gamma_n$ be a history of a system description ${\cal D}$ up to time step $n$.
\begin{itemize}
\item[(a)] A trajectory $\langle \sigma_0, a_0, \sigma_1, \dots, a_{n-1}, \sigma_n \rangle$
is a \emph{model of} $\Gamma_n$ if:

1. $a_i = \{ a \ :\ happened(a, i) \in \Gamma_n \}$, for every $0 \leq i < n$.

2. if $observed(f(\overline{t}), v, i) \in \Gamma_n$ then $f(\overline{t}) = v \in \sigma_i$,
for every $0 \leq i \leq n$.

\item[(b)] $\Gamma_n$ is \emph{consistent} if it has a model.

\item[(c)] An atom $f(\overline{t}) = v$ \emph{holds} in a model $M$ of
$\Gamma_n$ at time $0 \leq i \leq n$ if $f(\overline{t}) = v \in \sigma_i$;\\
A literal $f(\overline{t}) \neq v$ \emph{holds} in a model $M$ of
$\Gamma_n$ at time $0 \leq i \leq n$ if $dom_f(\overline{t}) = true \in \sigma_i$ and
$f(\overline{t}) = v \notin \sigma_i$;\\
$\Gamma_n$ \emph{entails} a literal $l$ at time step $0 \leq i \leq n$ if,
for every model $M$ of $\Gamma_n$, $l$ holds in $M$.
\end{itemize}
}
\end{definition}
\begin{example}[Model of a History]\label{ex-model-hist}
{
History $\Gamma_1$ from Example \ref{ex-history} is consistent. 
Its model is the trajectory:
$$
\begin{array}{lrl}
M & = \langle \ \{ & loc\_in(monkey) = initial\_monkey,\ loc\_in(box) = initial\_box, \ \dots\ \}, \\
  &              & move(initial\_box),\\
  &           \{ & loc\_in(monkey) = initial\_box,\ loc\_in(box) = initial\_box,\ \dots\ \} \rangle.
\end{array}
$$
(We do not show the values of $connected$ since they are unchanged by
our actions).
$\Gamma_1$ entails, for example, $loc\_in(monkey) = initial\_box$ at time step 1.
}
\end{example}

Note that a consistent history may have more than one model
if non-deterministic actions are involved or the initial situation is
not complete.

\bigskip
Next, we define some useful vocabulary.

\begin{definition}[Set of Literals Defining a Sequence -- adapted from \cite{bg03}]
{
Let $\Gamma_n$ be a history of ${\cal D}$ and $A$ be a set of literals over
signature $\Sigma$.
We say that $A$ \emph{defines} the sequence
$$\langle \sigma_0, a_0, \sigma_1, \dots, a_{n-1}, \sigma_n \rangle $$
if:
\begin{itemize}
\item[(a)]
\noindent
$
\begin{array}{ll}
\sigma_i = & \{ f(t_0, \dots, t_n) = t :
f(t_0, \dots, t_n) = t \in A \mbox{ and } f \mbox{ is a static or attribute}\}\  \cup \\
           & \{ f(t_0, \dots, t_n) = t :
f(t_0, \dots, t_n, i) = t \in A \mbox{ and } f \mbox{ is a fluent}\}
\end{array}
$

for any  $0 \leq i \leq n$, and
\item[(b)] $a_k = \{ a : occurs(a, k) \in A\}$ for any $0 \leq k < n$.
\end{itemize}
}
\end{definition}

\begin{definition}[Program $\Omega_{tp}$ -- adapted from \cite{bg03}]
{
If $\Gamma_n$ is a history of system description ${\cal D}$ up to time step $n$,
then by $\Omega_{tp}$ we denote the ASP\{f\} program constructed as follows:
\begin{itemize}
\item[1.] For every action $a$ such that $happened(a, i) \in \Gamma_n$,
$\Omega_{tp}$ contains:

\medskip
$occurs(a, i) \leftarrow happened(a, i).$
\medskip

\item[2.] For every expression $observed(f(\overline{t}), v, 0) \in \Gamma_n$,
$\Omega_{tp}$ contains:

\medskip
$f(\overline{t}, 0) = v \leftarrow observed(f(\overline{t}), v, 0).$

\medskip
\item[3.] For every expression $observed(f(\overline{t}), v, i) \in \Gamma_n$,
$i > 0$,
$\Omega_{tp}$ contains the \emph{reality check} axiom:

\medskip
$
\begin{array}{ll}
\leftarrow & observed(f(\overline{t}), v, i),\\
           & dom_f(\overline{t}, i),\\
           & f(\overline{t}, i) \neq v.
\end{array}
$
\end{itemize}
}
\end{definition}

\bigskip
Our methodology of finding trajectories by computing answer sets of a logic program
is designed for system descriptions that match the intuition that
defined functions are only shorthands, and their values are fully determined
by those of basic statics and fluents. We call such system descriptions
well--founded and define them formally as follows.

\begin{definition}[Well--founded System Description -- adapted from \cite{gi13}]
Let ${\cal D}$ be a system description whose theory encodes the
${\cal BAT}$ theory $T$, and whose structure defines a collection ${\cal S}$ of
models of $T$.
${\cal D}$ is {\em well--founded} if, for every model ${\cal M}$ in ${\cal S}$,
and every interpretation ${\cal I}$ with static part ${\cal M}$,
the program $S_{\cal I}$ (defined as in Section \ref{sec_semantics})
has at most one answer set.
\end{definition}

The system description $monkey\_and\_banana\_problem$ from Section \ref{mb} is well--founded.
An example of a system description that is {\em not well--founded} is $n\_w\_f$
shown below and adapted from \cite{gi13}.
The two defined fluents of $n\_w\_f$ are not defined in terms
of basic statics or fluents but rather in terms of one another by
mutually recursive axioms.

\smallskip
$
\begin{array}{l}
{\bf system\ description}\ n\_w\_f\\
\ \ \ \ {\bf theory}\ n\_w\_f\\
\ \ \ \ \ \ \ \ {\bf module}\ main\\
\ \ \ \ \ \ \ \ \ \ \ \ {\bf sort\ declarations}\\
\ \ \ \ \ \ \ \ \ \ \ \ \ \ \ \ c :: universe\\
\ \ \ \ \ \ \ \ \ \ \ \ {\bf fluent\ declarations}\\
\ \ \ \ \ \ \ \ \ \ \ \ \ \ \ \ {\bf defined}\\
\ \ \ \ \ \ \ \ \ \ \ \ \ \ \ \ \ \ \ \ f : c \rightarrow booleans\\
\ \ \ \ \ \ \ \ \ \ \ \ \ \ \ \ \ \ \ \ g : c \rightarrow booleans\\
\ \ \ \ \ \ \ \ \ \ \ \ {\bf axioms}
\end{array}
$

$
\begin{array}{lll}
\ \ \ \ \ \ \ \ \ \ \ \ \ \ \ \ f(X) & \lif & \neg g(X)\mbox{.}\\
\ \ \ \ \ \ \ \ \ \ \ \ \ \ \ \ g(X) & \lif & \neg f(X)\mbox{.}
\end{array}
$

$
\begin{array}{l}
\ \ \ \ {\bf structure}\ n\_w\_f\\
\ \ \ \ \ \ \ \ {\bf instances}\\
\ \ \ \ \ \ \ \ \ \ \ \ x \ {\bf in} \ c
\end{array}
$

\smallskip
\noindent
In the case of the non-modular action language ${\cal AL}$, there is a known syntactic
condition that guarantees that a system description is well--founded \cite{gi13}. This condition
can be easily expanded to ${\cal ALM}$ due to close connections between ${\cal ALM}$ and ${\cal AL}$.

\bigskip
Trajectories of a dynamic system specified by a well--founded system description
are computed using a logic program $\Pi$ that consists of the ASP\{f\} encoding of the system
description, the system's recorded history, and the program $\Omega_{tp}$ connecting the
recorded history with the system description.

To simplify the presentation, in what follows we limit ourselves to
well--founded system descriptions that describe domains in which there is {\em complete}
information about the sort memberships of objects of the domain.\footnote{This is not a serious restriction;
it can be easily lifted by adding to the ASP encoding of the ${\cal ALM}$ system description rules of the type
$$is\_a(x, c)\ or\ \neg is\_a(x,c)$$ for every object $x$ and every source node $c$ in the hierarchy of sorts.}
Let us consider system description ${\cal D}$ that meets this requirement,
and let ${\cal M}$ be a model of ${\cal D}$'s theory. Then, the program $P_{\cal M}$ obtained from
the theory of ${\cal D}$ and ${\cal M}$ as described in section \ref{sec_semantics}
will be used as the ASP\{f\} encoding of ${\cal D}$.

\begin{definition}[Program $\Pi_{tp}({\cal D})$]
If $\Gamma_n$ is a history of ${\cal D}$ up to step $n$, then
$\Pi_{tp}({\cal D})$ is the logic program defined as
$$\Pi_{tp}({\cal D}) =_{def} P_{\cal M} \cup \Gamma_n \cup \Omega_{tp}$$
such that the sort $step$ in the signature of $\Pi_{tp}({\cal D})$
ranges over the set $\{0, \dots, n\}$.
\end{definition}

\begin{proposition}
If $\Gamma_n$ is a \emph{consistent} history of ${\cal D}$ 
such that the initial situation of $\Gamma_n$ is complete, then
$M$ is a model of $\Gamma_n$ iff
$M$ is defined by some answer set of program $\Pi_{tp}({\cal D})$.
\end{proposition}

This proposition can be proven using techniques similar to the ones employed in
Lemma 5 in \cite{bg03}.\footnote{The proof and text of Lemma 5 appear on
page 29 of the version of \cite{bg03}
available at http://arxiv.org/pdf/cs/0312040v1.pdf. Retrieved on August 3, 2014.}

We used the above methodology of solving temporal projection tasks
to create a question answering system in the context of the Digital Aristotle project \cite{ig11}.
Our system was capable of answering complex end-of-the-chapter questions on cell division,
extracted from a well-known biology textbook.

\subsubsection{Planning}

In planning problems, in addition to the history of
the dynamic system up to the current time point, information about the
goal to be achieved is also provided.
Given a system description of ${\cal ALM}$ whose theory describes a basic action theory $T$,
a \emph{goal} is a collection $G$ of ground user-defined fluent literals
over the signature of $T$. For instance, for the Monkey and Banana problem in Section \ref{mb}, the goal is
$G_{mb} = \{holding(monkey,banana)\}$. Goals can be encoded as logic programming rules,
as described in the following definition:

\smallskip
\begin{definition}[Goal Encoding]
Given a goal $G$, we call {\em encoding of $G$}, denoted by $lp(G)$ the
rule
$$goal(I) \ \leftarrow \ body$$
where $body$ is defined as follows:
$$body =_{def} \{f(\overline{t}, I) = v\ :\ \ f(\overline{t}) = v\ \in\ G\}\ \cup\
\{f(\overline{t}, I) \neq v\ :\ \ f(\overline{t}) \neq v\ \in\ G\}.$$
\end{definition}

\medskip
In order to solve planning problems, a slightly different logic programming module will be needed
than for solving temporal projection tasks. This module is defined in CR-Prolog \cite{bg03a},
an extension of ASP designed to handle, among other things, rare events.
In addition to regular ASP rules, programs in CR-Prolog may contain {\em consistency restoring}
rules that have the following syntax:
$$h_1 \ or\ \dots\ or\ h_k \ \ \stackrel{+}{\leftarrow}\ \ l_1, \dots, l_m, \no l_{m+1}, \dots, \no l_n.$$
Informally, this statement says that
an intelligent agent who believes $l_1,\dots, l_m$ and has no reason to believe
$l_{m+1}, \dots, l_n$ may believe one of $h_i$'s, $1 \leq i \leq k$,
but only if no consistent set of beliefs can be formed otherwise.
For the formal semantics of CR-Prolog, we refer the reader to \cite{bg03a}.
An extension of ASP\{f\} by consistency restoring rules is defined in \cite{bg12}.
Solvers for CR-Prolog are described in \cite{bal07b} and \cite{bgz12}.

\begin{definition}[Planning Module \cite{bal04,gk14}]
Given a goal $G$, the planning module $\Omega_{pl}$ extends module $\Omega_{tp}$ from Section \ref{tp}
by the following rules:
$$
\begin{array}{llrl}
& success & \leftarrow & goal(I),\ I \leq n\\
&         & \leftarrow & \no success\\
r_1(A, I) :& occurs(A, I) & \stackrel{+}{\leftarrow} & instance(A, actions)\\ 
& smtg\_happened(I) & \leftarrow & occurs(A, I)\\
& & \leftarrow & \no smtg\_happened(I),\\
& &            & smtg\_happened(I+1) .
\end{array}
$$
$\Omega_{pl}$ computes minimal plans of maximum length $n$ by the use of the consistency restoring rule $r_1$
and the two regular rules that follow it.
\end{definition}

\medskip
The actual program for computing plans is constructed similarly as before.

\begin{definition}[Program $\Pi_{pl}({\cal D})$]
If $\Gamma_n$ is a history of ${\cal D}$ up to step $n$ and $G$ is a goal over ${\cal D}$, then
$\Pi_{pl}({\cal D}, G)$ is the logic program defined as
$$\Pi_{pl}({\cal D}, G) =_{def} P_{\cal M} \cup \Gamma_n \cup \Omega_{pl} \cup lp(G)$$
such that the sort $step$ in the signature of $\Pi_{pl}({\cal D}, G)$
ranges over the set $\{0, \dots, n\}$.
\end{definition}

\medskip
The following proposition specifies how answer sets of the logic program defined above
can be mapped into plans for achieving given goals.

\begin{proposition}
If $\Gamma_n$ is a \emph{consistent} history of ${\cal D}$ 
such that the initial situation of $\Gamma_n$ is complete and $G$ is a
goal over ${\cal D}$, then the collection of atoms of the form $occurs(a, i)$ from an answer set
of $\Pi_{pl}({\cal D}, G)$ defines a minimal plan for achieving goal
$G$, and every such plan is represented by the $occurs$ atoms of some
answer set of $\Pi_{pl}({\cal D}, G)$.

\end{proposition}

\begin{example}[Planning in the Monkey and Banana Problem]
If we consider the Monkey and Banana problem with the initial situation
$$
\begin{array}{ll}
\Gamma_{mb} = \{& observed(loc\_in(monkey), initial\_monkey, 0),\\
                & observed(loc\_in(box), initial\_box, 0)
\end{array}
$$
and the goal $$G_{mb} = \{holding(monkey,banana)\}$$ defined earlier,
then an answer set of program $\Pi_{pl}(monkey\_and\_banana\_problem, G_{mb})$
will contain the following $occurs$ atoms:

$
\begin{array}{llll}
\{ & occurs(move(initial\_box), 0), & occurs(grasp(box), 1), &\\
   & occurs(carry(box, under\_banana), 2), & occurs(release(box), 3), &\\
	 & occurs(climb(box), 4), & occurs(grasp(banana), 5) & \}
\end{array}
$

\noindent
defining a minimal plan
$\langle\ move(initial\_box)$, $grasp(box)$, $carry(box, under\_banana)$,
$release(box)$, $climb(box)$, $grasp(banana)\ \rangle$
resulting in the monkey holding the banana at time step 6.
The program will also find the second minimal plan in which
$carry(box, under\_banana)$ at step $2$ is replaced by
$move(under\_banana)$. Since the first action is more specific than
the second one the first plan seems to be preferable. This can easily
be expressed by a slightly modified planning module allowing only most
specific actions.

\end{example}

\section{Related Work}

Many ideas of ${\cal ALM}$, such as the notions of action language, module,
sort hierarchy, attribute defined  as a partial function, etc., are well-known from the
literature on programming languages and knowledge representation.
Some of the basic references to these notions were given in the text.
In this section we briefly comment on the relationship between ${\cal ALM}$ and the 
previously existing modular action languages $MAD$ \cite{lr06,el06,ds07}, 
TAL-C \cite{gk04}, and the
earlier version of ${\cal ALM}$ \cite{gi09}. 

We start with summarizing the differences between the two versions of ${\cal ALM}$.
There are a number of changes in the syntax of the
language. For instance, theories of the new version of $\mathcal{ALM}$
may contain
non-boolean fluents\footnote{In the field of logic programming,
an early discussion on the introduction of functions appears in \cite{mh94}.
} and constants that substantially simplify $\mathcal{ALM}$'s
use for knowledge representation. Axioms of a theory, which in the old
version were included in the theory's  declarations, are now put in a
separate section of the theory. This removed the problem of deciding
which fluent or action declaration should contain an axiom, and
improved the readability of the language. There are also substantial
improvements in the syntax of axioms, etc. Another collection of
changes is related to the semantics of the language. First, the new
semantics, based on the notions of basic action theory and its models,
clarified and generalized the
old definition and allowed the introduction of the entailment relation.
Second, the semantics is now defined for
structures with possibly underspecified membership relations of its objects in
the sort hierarchy, which simplifies reasoning with incomplete
information.
Third, the semantics  was initially
given in terms of action language ${\cal AL}$ \cite{tur97,bg00},
where the ${\cal AL}$ semantics is defined by a translation into ASP;
now, we give the semantics of our language directly in ASP -- in fact, in an extension of ASP
with non-Herbrand functions, ASP\{f\} \cite{bal13}.
We believe that decoupling ${\cal ALM}$ from ${\cal AL}$ will
allow us to combine ${\cal ALM}$ with action languages
that correspond to other intuitions.

Another modular language is TAL-C \cite{gk04}, which allows definitions of
classes of objects that are somewhat similar to those in ${\cal ALM}$. TAL-C,
however, seems to have more ambitious goals: the language is used to
describe and reason about various dynamic scenarios, whereas in ${\cal ALM}$
the description of a scenario and that of reasoning tasks are not
viewed as part of the language.
The more rigid structure of ${\cal ALM}$ supports the \emph{separation of concerns}
design principle and makes it easier to give a formal semantics of the
language.
These differences led to vastly distinct knowledge representation
styles reflected in these languages.

There are smaller, but still very substantial, differences between
${\cal ALM}$ and $MAD$. The two languages are based on non-modular action languages with
substantially different semantics and underlying assumptions,
use very different constructs for creating modules and for defining
actions as special cases, etc. A more detailed comparison between the
two approaches can be found in \ref{ALMandMAD}.

\section{Conclusions and Future Work}

In this paper, we have presented a methodology of representing and
reasoning about dynamic systems. A knowledge engineer following this
methodology starts with finding a proper generalization of a particular
dynamic system $D$, finds the sorts of
objects pertinent to this generalization, organizes these sorts into an
inheritance hierarchy and uses causal laws, definitions, and
executability conditions to specify relevant properties of the sorts elements.
The resulting basic action theory, say $T$, gives the first mathematical
model of the system. In the next step of the development,
a knowledge engineer refines this model by
providing its description in the high level action language ${\cal ALM}$.
The language has means for precisely representing the signature of $T$
including its sort hierarchy. It is characterized by a modular
structure, which improves readability and supports the step-wise
development of a knowledge base, reuse of knowledge, and creation of
knowledge libraries. ${\cal ALM}$'s description of $T$ can be used to
specify multiple dynamic systems with different collections of objects
and statics. A particular system $D$ can be specified
by populating sorts of $T$ by objects of $D$ and defining values of
$D$'s statics. This step is also supported
by ${\cal ALM}$, which clearly separates the definition
of \emph{sorts of objects} of the domain (given in $T$)
from the definition of \emph{instances} of
these sorts (given by an ${\cal ALM}$ structure).
This, together with the means for defining objects
of the domain as special cases of previously defined ones,
facilitates the stepwise development and testing of the knowledge base
and improves its elaboration tolerance.

A close relationship between ${\cal ALM}$ and Answer Set Programming
allows the use of ${\cal ALM}$ system descriptions for non-trivial
reasoning problems including temporal projection, planning, and
diagnosis. This is done by an automatic translation of an  ${\cal
  ALM}$ system description into logic programs whose answer sets
correspond to solutions of the corresponding problems.
The existence of efficient answer set solvers that allow to compute
these answer sets substantially increases the practical value of this approach.

The above methodology has been illustrated by two examples:
the well-known benchmark Monkey and Banana
problem and  a more practical problem of formalization
of knowledge and answering questions
about biological processes such as the cell division (see \ref{aristotle}).
It is possible (and even likely) that further experience with ${\cal ALM}$
will suggest some useful extensions of the language but
the authors believe that the version presented in this paper
will remain relatively stable and provide a good basis for such
extensions.

We conclude by briefly outlining a number of questions about
${\cal ALM}$ that we believe deserve further investigation:

\begin{itemize}
\item
Investigating mathematical properties of ${\cal ALM}$ and its
entailment relation. This includes but is not limited to studying
compositional properties of ${\cal ALM}$ modules,
axiomatizing its entailment relation, and establishing a closer
relationship between ${\cal ALM}$ and modular logic programming.

\item
Developing more efficient reasoning algorithms
exploiting the modular structure of ${\cal ALM}$'s theories
and the available information about the sorts of objects
in ${\cal ALM}$'s system descriptions. Among other things it
is worth investigating the possible use of modular logic programming as
well as the methods from \cite{GebserSS11},
\cite{GebserGKS11}, and \cite{bgz12}.
It may also be interesting to see if the implementation could benefit
from hybrid approaches combining description logics with ASP (e.g. \cite{eilst08})
or from typed logic programming (e.g. \cite{Pfenning92}).

\item
Designing and implementing a development environment to
facilitate the use of ${\cal ALM}$ in applications,
the creation and storage of libraries, and the testing and debugging of theories and modules.

\item
Extending ${\cal ALM}$ with the capability of representing
knowledge about {\em hybrid} domains, i.e., domains that allow both discrete
and continuous change.
In particular, it may be a good idea to combine ${\cal ALM}$ with action language ${\cal H}$
\cite{cgw05,sc12}.

\item
Developing the core of an ${\cal ALM}$
library of commonsense knowledge. (In particular we would like to
create an ${\cal ALM}$ library module containing a theory of
intentions in the style of \cite{bgb14}.)
This work would allow us to extend our study on the
capabilities of our language, while simultaneously providing a tool for
members of our community to use when building their reasoning systems.

\end{itemize}

\section*{Acknowledgments}

We are grateful to Evgenii Balai, Justin Blount, 
Vinay Chaudhri, Vladimir Lifschitz, Yana Todorova, 
and the anonymous reviewers for useful comments and discussions.
This work was partially supported by NSF grant IIS-1018031.

\label{firstpage}

\newpage

\begin{appendix}
\section{Grammar of ${\cal ALM}$}\label{grammar}
\label{grammar}

$
\begin{array}{l}
\langle boolean\rangle \is \mbox{true} \br \mbox{false}\\
\langle non\_zero\_digit\rangle \is 1\br...\br9\\
\langle digit\rangle \is 0 \br \langle non\_zero\_digit\rangle \\
\langle lowercase\_letter\rangle \is \mbox{a}\br...\br \mbox{z}\\
\langle uppercase\_letter\rangle \is \mbox{A}\br...\br \mbox{Z}\\
\langle letter \rangle \is \langle lowercase\_letter \rangle \br 
      \langle uppercase\_letter\rangle\\
\langle identifier \rangle \is \langle lowercase\_letter\rangle  \br
      \langle identifier\rangle \langle letter\rangle  \br
			\langle identifier\rangle \langle digit\rangle \\
\langle variable\rangle \is \langle uppercase\_letter\rangle  \br
      \langle variable\rangle \langle letter\rangle  \br \langle variable\rangle \langle digit\rangle \\
\langle positive\_integer\rangle \is \langle non\_zero\_digit\rangle  \br
      \langle positive\_integer\rangle \langle digit\rangle \\
\langle integer\rangle \is 0 \br \langle positive\_integer\rangle  \br
      - \langle positive\_integer\rangle \\
\langle arithmetic\_op\rangle \is + \br - \br * \br / \br \mbox{mod}\\
\langle comparison\_rel\rangle \is >  \br >= \br <  \br <= \\
\langle arithmetic\_rel\rangle \is \langle eq \rangle  \br \langle neq \rangle \br \langle comparison\_rel\rangle \\
\langle basic\_arithmetic\_term\rangle \is \langle variable\rangle  \br
     \langle identifier\rangle  \br \langle integer\rangle \\
\langle basic\_term\rangle \is \langle basic\_arithmetic\_term\rangle  \br \langle boolean\rangle \\
\langle function\_term\rangle \is \langle identifier\rangle \langle function\_args\rangle \\
\langle function\_args\rangle \is (\langle term\rangle \langle remainder\_function\_args\rangle )\\
\langle remainder\_function\_args\rangle \is \epsilon \br
     , \langle term\rangle \langle remainder\_function\_args\rangle \\
\langle arithmetic\_term\rangle \is \langle basic\_arithmetic\_term\rangle  \langle arithmetic\_op\rangle
    \langle basic\_arithmetic\_term\rangle\\
\langle term\rangle \is \langle basic\_term\rangle  \br
		\langle arithmetic\_term\rangle
\end{array}
$


\noindent
$
\begin{array}{l}
\langle positive\_function\_literal\rangle \is \langle function\_term\rangle  \br
      \langle function\_term\rangle  \langle eq\rangle  \langle term\rangle \\
\langle function\_literal\rangle \is \langle positive\_function\_literal\rangle  \br
      \neg \langle function\_term\rangle  \br \\
			\ \ \ \ \ \ \langle function\_term\rangle  \langle neq\rangle  \langle term\rangle \\
\langle literal\rangle \is \langle function\_literal\rangle  \br 
\langle arithmetic\_term\rangle  \langle arithmetic\_rel\rangle  \langle arithmetic\_term\rangle \\
\end{array}
$

\noindent
$
\begin{array}{l}
\langle var\_id\rangle \is \langle variable\rangle  \br \langle identifier\rangle \\
\langle body\rangle \is \epsilon \br , \langle literal \rangle\langle body\rangle \\
\langle dynamic\_causal\_law\rangle \is \mbox{occurs} (\langle var\_id\rangle ) \mbox{ causes }
     \langle positive\_function\_literal\rangle  \mbox{ if }\\
		 \ \ \ \ \ \ \mbox{instance}(\langle var\_id\rangle , \langle var\_id\rangle )\langle body\rangle  .\\

\langle state\_constraint\rangle \is \langle sc\_head\rangle \mbox{ if } \langle body\rangle  .\\
\langle sc\_head\rangle \is \mbox{false} \br \langle positive\_function\_literal\rangle \\

\langle definition\rangle \is \langle function\_term\rangle \mbox{ if } \langle body\rangle  .\\

\langle executability\_condition\rangle \is \mbox{imposible occurs}(\langle var\_id\rangle ) \mbox{ if}\\
     \ \ \ \ \ \ \mbox{instance}(\langle var\_id\rangle , \langle var\_id\rangle )\langle extended\_body\rangle  .\\
\langle extended\_body\rangle \is \epsilon \br , \langle literal \rangle \langle body\rangle  \br
     , \mbox{ occurs}(\langle var\_id\rangle )\langle extended\_body\rangle  \br\\
		 \ \ \ \ \ \ ,\ \neg \mbox{occurs}(\langle var\_id\rangle )\langle extended\_body\rangle \\
\end{array}
$

\noindent
$
\begin{array}{l}
\langle system\_description\rangle \is \mbox{system description }\langle identifier\rangle\
      \langle theory\rangle  \langle structure\rangle \\
\end{array}
$

\noindent
$
\begin{array}{l}
\langle theory\rangle \is \mbox{theory } \langle identifier\rangle  \langle set\_of\_modules\rangle \br
     \mbox{import }\langle identifier\rangle \mbox{ from } \langle identifier\rangle \\
\langle set\_of\_modules\rangle \is \langle module\rangle  \langle remainder\_modules\rangle \\
\langle remainder\_modules\rangle \is \epsilon \br \langle module\rangle  \langle remainder\_modules\rangle \\
\end{array}
$

\noindent
$
\begin{array}{l}
\langle module\rangle \is \mbox{module } \langle identifier\rangle  \langle module\_body\rangle  \br \\
     \ \ \ \ \ \ \mbox{import }\langle identifier\rangle \mbox{.}\langle identifier\rangle \mbox{ from } \langle identifier\rangle \\
\langle module\_body\rangle \is \langle sort\_declarations\rangle  
      \langle constant\_declarations\rangle 
			\langle function\_declarations\rangle  
			\langle axioms\rangle \\
\end{array}
$

\noindent
$
\begin{array}{l}
\langle sort\_declarations\rangle \is \epsilon \br \mbox{sort declarations } \langle one\_sort\_decl\rangle
     \langle remainder\_sort\_declarations\rangle \\
\langle remainder\_sort\_declarations\rangle \is \epsilon \br \langle one\_sort\_decl\rangle
     \langle remainder\_sort\_declarations\rangle \\

\langle one\_sort\_decl\rangle \is \langle identifier\rangle \langle remainder\_sorts\rangle  ::
     \langle sort\_name \rangle \langle remainder\_sort\_names\rangle  \langle attributes\rangle \\
\langle remainder\_sorts\rangle \is \epsilon \br , \langle identifier\rangle \langle remainder\_sorts\rangle \\
\langle remainder\_sort\_names\rangle \is \epsilon \br , \langle sort\_name\rangle \langle remainder\_sorts\rangle \\
\langle sort\_name \rangle \is \langle identifier \rangle \br [ \ \langle integer \rangle \mbox{..} 
     \langle integer \rangle \ ]\\

\langle attributes\rangle \is \epsilon \br \mbox{attributes } \langle one\_attribute\_decl\rangle
     \langle remainder\_attribute\_declarations\rangle \\
\langle one\_attribute\_decl\rangle \is \langle identifier\rangle  : \langle arguments\rangle
     \langle identifier\rangle \\
\langle arguments\rangle \is \epsilon \br \langle identifier\rangle  \langle remainder\_args\rangle  \rightarrow \\
\langle remainder\_args\rangle \is  \epsilon \br \times \langle identifier\rangle  \langle remainder\_args\rangle \\

\langle remainder\_attribute\_declarations\rangle \is \epsilon \br\\
     \ \ \ \ \ \ \langle one\_attribute\_decl\rangle \langle remainder\_attribute\_declarations\rangle \\
\end{array}
$

\noindent
$
\begin{array}{l}
\langle constant\_declarations\rangle \is \epsilon \br \mbox{object constants }\langle one\_constant\_decl\rangle
      \langle remainder\_constant\_declarations\rangle \\
\langle one\_constant\_decl\rangle \is \langle identifier\rangle 
      \langle constant\_params \rangle  : \langle identifier\rangle \\
\langle remainder\_constant\_declarations\rangle \is \epsilon \br \langle one\_constant\_decl\rangle
      \langle remainder\_constant\_declarations\rangle \\
\langle constant\_params \rangle \is ( \ \langle identifier\rangle \langle remainder\_constant\_params \rangle\ )\\
\langle remainder\_constant\_params \rangle \is \epsilon \br , \ \langle identifier\rangle 
      \langle remainder\_constant\_params \rangle
\end{array}
$

\noindent
$
\begin{array}{l}
\langle function\_declarations\rangle \is \epsilon \br \mbox{function declarations }
     \langle static\_declarations\rangle  \langle fluent\_declarations\rangle \\
		 
\langle static\_declarations\rangle \is \epsilon \br \mbox{statics }\langle basic\_function\_declarations\rangle
     \langle defined\_function\_declarations\rangle \\
\langle fluent\_declarations\rangle \is \epsilon \br \mbox{fluents }\langle basic\_function\_declarations\rangle
     \langle defined\_function\_declarations\rangle \\

\langle basic\_function\_declarations\rangle \is \epsilon \br \mbox{basic }\langle one\_function\_decl\rangle
     \langle remainder\_function\_declarations\rangle \\
\langle defined\_function\_declarations\rangle \is \epsilon \br \mbox{defined }\langle one\_function\_decl\rangle
     \langle remainder\_function\_declarations\rangle \\

\langle one\_function\_decl\rangle \is \langle total\_partial\rangle  \langle one\_f\_decl\rangle \\
\langle total\_partial\rangle \is \epsilon \br \mbox{total}\\
\langle one\_f\_decl\rangle \is \langle identifier\rangle  : \langle identifier\rangle  \langle remainder\_args\rangle \rightarrow
     \langle identifier\rangle \\

\langle remainder\_function\_declarations\rangle \is \epsilon \br \langle one\_function\_decl\rangle
      \langle remainder\_function\_declarations\rangle
\end{array}
$

\noindent
$
\begin{array}{l}
\langle axioms\rangle \is \epsilon \br \mbox{axioms }\langle one\_axiom\rangle  \langle remainder\_axioms\rangle \\
\langle one\_axiom\rangle \is \langle dynamic\_causal\_law\rangle  \br \langle state\_constraint\rangle  \br
       \langle definition\rangle  \br \langle executability\_condition\rangle \\
\langle remainder\_axioms\rangle \is \epsilon \br \langle axiom\rangle  \langle remainder\_axioms\rangle \\
\end{array}
$

\noindent
$
\begin{array}{l}
\langle structure\rangle \is \mbox{structure }\langle identifier\rangle  \langle constant\_defs\rangle
       \langle instance\_defs\rangle  \langle statics\_defs\rangle \\
\end{array}
$

\noindent
$
\begin{array}{l}
\langle constant\_defs\rangle \is \epsilon \br \mbox{constants }\langle one\_constant\_def\rangle
      \langle remainder\_constant\_defs\rangle \\
\langle one\_constant\_def\rangle \is \langle identifier\rangle  = \langle value\rangle \\
\langle value \rangle \is \langle identifier\rangle  \br \langle boolean \rangle \br \langle integer \rangle\\

\langle remainder\_constant\_defs\rangle \is \epsilon \br \langle one\_constant\_def\rangle  \langle remainder\_constant\_defs\rangle \\
\end{array}
$

\noindent
$
\begin{array}{l}
\langle instance\_defs\rangle \is \epsilon \br \mbox{instances }\langle one\_instance\_def\rangle
       \langle remainder\_instance\_defs\rangle \\
\langle one\_instance\_def\rangle \is \langle object\_name\rangle \langle remainder\_object\_names\rangle  \mbox{ in }\\
       \ \ \ \ \ \ \langle identifier\rangle  \langle cond\rangle  \langle attribute\_defs\rangle \\

\langle object\_name\rangle \is \langle identifier\rangle \langle object\_args\rangle \\
\langle object\_args\rangle \is \epsilon \br (\langle basic\_term\rangle \langle remainder\_object\_args\rangle )\\
\langle remainder\_object\_args\rangle \is \epsilon \br , \langle basic\_term\rangle \langle remainder\_object\_args\rangle \\

\langle remainder\_object\_names\rangle \is \epsilon \br , \langle object\_name\rangle \langle remainder\_object\_names\rangle \\
\langle cond\rangle \is \epsilon \br \mbox{ where }\langle literal\rangle \langle remainder\_cond\rangle \\
\langle remainder\_cond\rangle \is \epsilon \br , \langle literal\rangle \langle remainder\_cond\rangle

\end{array}
$

\noindent
$
\begin{array}{l}
\langle attribute\_defs\rangle \is \epsilon \br \langle one\_attribute\_def\rangle  \langle remainder\_attribute\_defs\rangle \\
\langle one\_attribute\_def\rangle \is \langle identifier\rangle \langle object\_args\rangle  = \langle basic\_term\rangle \\
\end{array}
$

\noindent
$
\begin{array}{l}
\langle statics\_defs\rangle \is \epsilon \br \mbox{ values of statics }\langle one\_static\_def\rangle  \langle remainder\_statics\_defs\rangle \\
\langle one\_static\_def\rangle \is \langle function\_literal\rangle  \mbox{ if } \langle body\rangle  .\\
\langle remainder\_statics\_defs\rangle \is \epsilon \br \langle one\_static\_def\rangle  \langle remainder\_statics\_defs\rangle
\end{array}
$


\mbox{}\vfill\eject
\section{${\cal ALM}$ and the Digital Aristotle}
\label{aristotle}

The reader may have noticed that the ${\cal ALM}$ examples included in the body of the paper
are relatively small, which is understandable given that their purpose was to illustrate the syntax
and semantics of our language and the methodology of representing knowledge in ${\cal ALM}$.
In this section, we show how the reuse of knowledge in ${\cal ALM}$
can potentially lead to the creation of larger practical systems.
We present an application of our language to the task of question answering,
in which ${\cal ALM}$'s conceptual separation between an abstract theory and its structure
played an important role in the reuse of knowledge.
The signature of the theory and its structure provided the vocabulary 
for the logic form translation
of facts expressed in natural language
while the theory axioms contained the background knowledge needed
for producing answers. The theory representing the biological
domain remained unchanged and was coupled
with various structures corresponding to particular questions
and representing the domain at different levels of granularity.
In addition to demonstrating the reuse of knowledge in ${\cal ALM}$,
this application also shows the elaboration tolerance of our language,
as only minor changes to the structure had to be made when the domain
was viewed in more detail, while the theory stayed the same.
In what follows, we present the application in more detail.

\medskip

After designing our language, we tested and confirmed its
adequacy for knowledge representation
in the context of a practical question answering application: Project Halo (2002-2013)
sponsored by Vulcan Inc.\footnote{http://www.allenai.org/TemplateGeneric.aspx?contentId=9}
The goal of Project Halo was the creation of a Digital Aristotle --- {\em
``an application containing large volumes of
scientific knowledge and capable of applying sophisticated
problem-solving methods to answer novel questions''} \cite{projecthalo10}.
Initially, the Digital Aristotle was only able to reason and answer questions about \emph{static}
domains. It lacked a methodology for answering questions
about \emph{dynamic} domains, as it was not clear how to represent and reason
about such domains in the language of the Digital Aristotle.
Our task within Project Halo was to create a methodology for answering questions
about temporal projection in dynamic domains. We had two objectives. First, we wanted to
see if the use of ${\cal ALM}$ for knowledge representation
facilitated the task of encoding extensive amounts of
scientific knowledge through its means for the reuse of knowledge.
Second, we investigated whether
provable correct and efficient logic programming algorithms
could be developed to use the resulting ${\cal ALM}$ knowledge base
in answering non-trivial questions.

Our target scientific domain was biology, specifically the biological process of
{\em cell division} (also called {\em cell cycle}).
Cell cycle refers to the phases
a cell goes through from its ``birth"
to its division into two daughter cells.
Cells consist of a number of parts, which in turn consist of other parts
(e.g., eukaryotic cells contain organelles, cytoplasm, and a nucleus;
the nucleus contains chromosomes, and the description
can continue with more detailed parts).
The eukaryotic cell cycle consists of a growth phase (interphase)
and a duplication/division phase (mitotic phase),
both of which are conventionally described as sequences of sub-phases.
Depending on the level of detail
of the description, these sub-phases may be simple events
or sequences of other sub-phases (e.g., the mitotic phase is described
in more detail as a sequence of two sub-phases: mitosis and cytokinesis;
mitosis, in turn, can be seen as a sequence of five sub-phases, etc.).
Certain chemicals, if introduced in the cell, can
interfere with the ordered succession of events that is
the cell cycle.

In order to be useful in answering complex questions,
the ${\cal ALM}$ representation of cell cycle had to capture
(1) non-trivial specialized biological knowledge about
the structure of the cell at different stages of the cell cycle
and (2) the dynamics of \emph{naturally evolving process} (such as cell cycle),
which consist of a series of phases and sub-phases that follow one another in a specific order,
unless interrupted. We represented such processes
as \emph{sequences of actions intended by nature} and used
a commonsense \emph{theory of intentions} \cite{bg05i} to reason about them.

Our ${\cal ALM}$ cell cycle knowledge base consisted of two library modules.
One of them was a general commonsense module describing sequences,
in particular sequences of actions. The other module was a specialized one
formalizing the biological phenomenon of cell division.
We begin with the presentation of our commonsense module
describing sequences, useful in modeling naturally evolving
processes such as cell division.
The equality $component(S, N) = E$ appearing in the axioms of module $sequence$
is supposed to be read as ``the $N^{th}$ component of sequence $S$ is $E$''.
The library module $sequence$ is stored in a general library called $commonsense\_lib$.

\st

\noindent
$
\begin{array}{l}
{\bf module} \ sequence \\
\ \ \ \ {\bf sort \ declarations} \\
\ \ \ \ \ \ \ \ sequences \ :: \ universe\\
\ \ \ \ \ \ \ \ \ \ \ \ {\bf attributes}\\
\ \ \ \ \ \ \ \ \ \ \ \ \ \ \ \ length : positive\_natural\_numbers \\
\ \ \ \ \ \ \ \ \ \ \ \ \ \ \ \ component : [0 \mbox{..} length] \rightarrow universe 
\end{array}
$

\smallskip

\noindent
$
\begin{array}{l}
\ \ \ \ \ \ \ \ action\_sequences \ :: \ sequences
\end{array}
$

\smallskip

\noindent
$
\begin{array}{l}
\ \ \ \ {\bf axioms} \\
\end{array}
$


\noindent
$
\begin{array}{lll}
\ \ \ \ \ \ \ \ false & \lif & component(S, N) = E,\\
     & & instance(S, action\_sequences),\\
     & & \neg instance(E, actions),\\
     & & \neg instance(E, action\_sequences)\mbox{.}
\end{array}
$

\medskip\noindent
The axiom ensures proper typing for the domain of an attribute $component$.

\medskip

Next, we present our formalization of cell cycle,
given in a library module called $basic\_cell\_cycle$
stored in a general $cell\_cycle\_lib$ library.
We started by modeling the eukaryotic cell, consisting of various parts
that in turn consist of other parts. Together, they form a ``\emph{part of}"
hierarchy, say $H_{cell}$, which can be viewed as a tree.
Nodes of this hierarchy were captured by a new sort, $types\_of\_parts$, while
links in the hierarchy were represented by an attribute, $is\_part\_of$,
defined on elements of the new sort (e.g., $is\_part\_of(X) = Y$ indicates that
$Y$ is the father of $X$ in $H_{cell}$).
We modeled the transitive closure of $is\_part\_of$ by introducing a boolean function,
$part\_of$, where $part\_of(X, Y)$ is true if $X$ is a descendant of $Y$ in $H_{cell}$.

In the type of questions we addressed,
at any given stage of the cell cycle process,
all cells in the experimental sample had the same number of nuclei;
similarly for the other inner components.
As a result, we could assume that,
at every stage and for each link from a child $X$ to its parent $Y$ in $H_{cell}$,
this link was assigned a particular number indicating the number of elements of type $X$ in one
element of type $Y$. The states of our domain were described by
a basic fluent,
$num : types\_of\_parts \times types\_of\_parts \rightarrow natural\_numbers$,
where $num(P_1, P_2) = N$ holds
if the number of elements of type $P_1$ in one element of type $P_2$ is $N$.
For instance, $num(nucleus, cell) = 2$ indicates that, at the
current stage of the cell cycle, each cell in the environment has two nuclei.

To describe the cell cycle we needed two action classes: $duplicate$ and $split$.
$Duplicate$, which acts upon an $object$ that is an element from sort $types\_of\_parts$,
doubles the number of every part of this kind present in the environment.
$Split$ also acts upon an $object$ ranging over $types\_of\_parts$.
An action $a$ of this type with $object(a) = c_1$, where $c_2$ is a child of $c_1$
in $H_{cell}$, duplicates the number of elements of type $c_1$ in the environment
and cuts in half the number of elements of type $c_2$ in one element of type $c_1$.
For example, if the experimental environment consists of one cell with two nuclei,
the occurrence of an instance $a$ of action $split$ with $object(a) = cell$
increases the number of cells to two and decreases the number of nuclei per
cells to one, thus resulting in an environment consisting of two cells with only one nucleus each.
In addition to these two actions we had an exogenous action, $prevent\_duplication$,
with an attribute $object$ with the range $types\_of\_parts$.
The occurrence of an instance action $a$ of $prevent\_duplication$
with $object(a) = c$ nullifies the effects of duplication and splitting for the
type $c$ of parts.
We made use of this exogenous action in representing
external events that interfere with the
normal succession of sub-phases of cell cycle.
All this knowledge is represented by the following module:

\st

$
\begin{array}{l}
{\bf module} \ basic\_cell\_cycle \\
\ \ \ \ {\bf sort \ declarations} \\
\ \ \ \ \ \ \ \ types\_of\_parts \ :: \ universe\\
\ \ \ \ \ \ \ \ \ \ \ \ {\bf attributes}\\
\ \ \ \ \ \ \ \ \ \ \ \ \ \ \ \ is\_part\_of : types\_of\_parts
\end{array}
$

\smallskip

$
\begin{array}{l}
\ \ \ \ \ \ \ \ duplicate \ :: \ actions \\
\ \ \ \ \ \ \ \ \ \ \ \ {\bf attributes}\\
\ \ \ \ \ \ \ \ \ \ \ \ \ \ \ \ object : types\_of\_parts
\end{array}
$

\smallskip

$
\begin{array}{l}
\ \ \ \ \ \ \ \ split \ :: \ duplicate
\end{array}
$

\smallskip

$
\begin{array}{l}
\ \ \ \ \ \ \ \ prevent\_duplication \ :: \ actions\\
\ \ \ \ \ \ \ \ \ \ \ \ {\bf attributes}\\
\ \ \ \ \ \ \ \ \ \ \ \ \ \ \ \ object : types\_of\_parts
\end{array}
$

\smallskip

$
\begin{array}{l}
\ \ \ \ {\bf function \ declarations} \\
\ \ \ \ \ \ \ \ {\bf statics} \\
\ \ \ \ \ \ \ \ \ \ \ \ {\bf defined}\\
\ \ \ \ \ \ \ \ \ \ \ \ \ \ \ \ part\_of : types\_of\_parts \times types\_of\_parts \rightarrow booleans
\end{array}
$

$
\begin{array}{l}
\ \ \ \ \ \ \ \ {\bf fluents} \\
\ \ \ \ \ \ \ \ \ \ \ \ {\bf basic}\\
\ \ \ \ \ \ \ \ \ \ \ \ \ \ \ \ {\bf total} \ num : types\_of\_parts \times types\_of\_parts \rightarrow natural\_numbers\\
\ \ \ \ \ \ \ \ \ \ \ \ \ \ \ \ prevented\_dupl : types\_of\_parts \rightarrow booleans
\end{array}
$

$
\begin{array}{l}
\ \ \ \ {\bf axioms}
\end{array}
$

$
\begin{array}{lllll}
\ \ \ \ \ \ \ \ occurs(X) & {\bf causes} & num(P_{2}, P_{1}) = N_{2} & \lif & instance(X, duplicate),\\
  & & & & object(X) = P_{2},\\
  & & & & is\_part\_of(P_{2}) = P_{1},\\
  & & & & num(P_{2}, P_{1}) = N_{1},\\
  & & & & N_{1} * 2 = N_{2}.
\end{array}
$

$
\begin{array}{lllll}
\ \ \ \ \ \ \ \ occurs(X) & {\bf causes} & num(P_{2}, P_{1}) = N_{2} & \lif & instance(X, split),\\
  & & & & object(X) = P_{1},\\
  & & & & is\_part\_of(P_{2}) = P_{1},\\
  & & & & num(P_{2}, P_{1}) = N_{1},\\
  & & & & N_{2} * 2 = N_{1}.
\end{array}
$

$
\begin{array}{lllll}
\ \ \ \ \ \ \ \ occurs(X) & {\bf causes} & prevented\_dupl(P) & \lif & instance(X, prevent\_duplication),\\
  & & & & object(X) = P.
\end{array}
$

$
\begin{array}{lll}
\ \ \ \ \ \ \ \ part\_of(P_1, P_2) & \lif & is\_part\_of(P_1) = P2.\\
\ \ \ \ \ \ \ \ part\_of(P_1, P_2) & \lif & is\_part\_of(P_1) = P_3,\\
                                   &      & part\_of(P_3, P_2).
\end{array}
$

$
\begin{array}{l}
\ \ \ \ \ \ \ \ num(P, P) = 0.
\end{array}
$

$
\begin{array}{lll}
\ \ \ \ \ \ \ \ num(P_3, P_1) = N & \lif & is\_part\_of(P_3) = P_2,\\
                                  &      & part\_of(P_2, P_1),\\
                                  &      & num(P_2, P_1) = N_1,\\
                                  &      & num(P_3, P_2) = N_2,\\
                                  &      & N_1 * N_2 = N.
\end{array}
$

$
\begin{array}{llll}
\ \ \ \ \ \ \ \ {\bf impossible} & occurs(X) & \lif & instance(X, duplicate),\\
 & & & object(X) = P, \\
 & & & prevented\_dupl(P).
\end{array}
$

\st

Any model of cell cycle consists of a theory
importing the two library modules presented above and
a structure corresponding to the level of detail of that model.
Let us consider a first model, in which we view cell cycle
as a sequence consisting of interphase and the mitotic phase.
This is represented in the structure by adding the attribute assignments
$component(1) = interphase$ and $component(2) = mitotic\_phase$
to the definition of instance $cell\_cycle$.
We remind the reader that such attribute assignments are read as
``the $1^{st}$ component of $cell\_cycle$ is $interphase$'' and
``the $2^{nd}$ component of $cell\_cycle$ is $mitotic\_phase$''.
Interphase is considered an elementary action, while the mitotic phase
splits the cell into two.
We limit our domain to cells contained in an
experimental environment, called $sample$.

\st

$
\begin{array}{l}
{\bf system \ description} \ cell\_cycle(1)\\
\ \ \ \ {\bf theory} \\
\ \ \ \ \ \ \ \ {\bf import\ module} \ sequence \ {\bf from} \ commonsense\_lib\\
\ \ \ \ \ \ \ \ {\bf import\ module} \ basic\_cell\_cycle \ {\bf from} \ cell\_cycle\_lib
\end{array}
$

$
\begin{array}{l}
\ \ \ \ {\bf structure} \\
\ \ \ \ \ \ \ \ {\bf instances}\\
\ \ \ \ \ \ \ \ \ \ \ \ sample \ {\bf in} \ types\_of\_parts
\end{array}
$

$
\begin{array}{l}
\ \ \ \ \ \ \ \ \ \ \ \ cell \ {\bf in} \ types\_of\_parts\\
\ \ \ \ \ \ \ \ \ \ \ \ \ \ \ \ is\_part\_of = sample
\end{array}
$

\smallskip

$
\begin{array}{l}
\ \ \ \ \ \ \ \ \ \ \ \ cell\_cycle \ {\bf in} \ action\_sequences\\
\ \ \ \ \ \ \ \ \ \ \ \ \ \ \ \ length = 2\\
\ \ \ \ \ \ \ \ \ \ \ \ \ \ \ \ component(1) = interphase\\
\ \ \ \ \ \ \ \ \ \ \ \ \ \ \ \ component(2) = mitotic\_phase
\end{array}
$

\smallskip

$
\begin{array}{l}
\ \ \ \ \ \ \ \ \ \ \ \ interphase \ {\bf in} \ actions
\end{array}
$

\smallskip

$
\begin{array}{l}
\ \ \ \ \ \ \ \ \ \ \ \ mitotic\_phase \ {\bf in} \ split \\
\ \ \ \ \ \ \ \ \ \ \ \ \ \ \ \ object = cell
\end{array}
$

\st

This initial model of cell division is quite general. It was
sufficient to answer a number of the questions targeted by the Digital
Aristotle. There were, however, some questions which required a different
model.

Consider, for instance, the following question from \cite{cr01}:

\st

\noindent
$
\begin{array}{ll}
\mbox{\emph{12.9.}} & Text: \mbox{ In some organisms mitosis occurs without
cytokinesis occurring.}\\
      & Question: \mbox{ How many cells will there be in the
sample at the end of the}\\
& \mbox{cell cycle, and how many nuclei will each cell contain?}
\end{array}
$

\st

To answer it, the system needed to know more about the structure
of the cell and that of the mitotic phase. ${\cal ALM}$ facilitated
the creation of a refinement of our original model of cell division:
a new system description, $cell\_cycle(2)$, was easily created
by adding to the previous structure a few new instances:

\smallskip
$
\begin{array}{l}
\ \ \ \ nucleus \ {\bf in} \ types\_of\_parts\\
\ \ \ \ \ \ \ \ is\_part\_of = cell
\end{array}
$

$
\begin{array}{l}
\ \ \ \ mitosis \ {\bf in} \ duplicate\\
\ \ \ \ \ \ \ \ is\_part\_of = nucleus
\end{array}
$

$
\begin{array}{l}
\ \ \ \ cytokinesis \ {\bf in} \ split\\
\ \ \ \ \ \ \ \ is\_part\_of = cell
\end{array}
$

\smallskip
\noindent
and replacing the old definition of the instance $mitotic\_phase$ by a new one:

\smallskip
$
\begin{array}{l}
\ \ \ \ mitotic\_phase \ {\bf in} \ action\_sequences\\
\ \ \ \ \ \ \ \ length = 2\\
\ \ \ \ \ \ \ \ component(1) = mitosis\\
\ \ \ \ \ \ \ \ component(2) = cytokinesis
\end{array}
$

\smallskip
\noindent
Similarly, various other refinements of our original model of cell division
contained the same theory as the original formalization; only
the structure of our original model needed to be modified,
in an elaboration tolerant way. Matching questions with models of cell division
containing just the right amount of detail is computationally advantageous and,
in most cases, the matching can be done automatically.

{\em Our formalization of cell division illustrates ${\cal ALM}$'s
capabilities of creating large knowledge bases for practical systems
through its mechanisms for reusing knowledge.} In our example, the two
modules that formed the theory were directly imported from the library
into the system description.
This shows that our main goal for ${\cal ALM}$ -- the
reuse of knowledge -- was successfully achieved.

Additionally, the example demonstrates ${\cal ALM}$'s suitability
for modeling not only {\em commonsense} dynamic systems, but also
{\em highly specialized, non-trivial domains}.
It shows the importance of creating and using libraries
of knowledge in real-life applications, and
it demonstrates the ease of elaborating initial formalizations of
dynamic domains into more detailed ones.

\medskip

Our second task in Project Halo was to develop a proof-of-concept
question answering system that used ${\cal ALM}$ formalizations of
cell cycle in solving complex temporal projection questions like \emph{12.9} above.
To do that, we used the methodology described in Section 4.2,
expanded by capabilities for reasoning about naturally evolving processes.
This latter part was done by incorporating a theory of intentions \cite{bg05i}
and assuming that naturally evolving processes have the {\em tendency} (or the {\em intention})
to go through their sequence of phases in order, unless interrupted
(e.g., we can say that a cell {\em tends}/ {\em intends} to go through
its cell cycle, which it does unless unexpected events happen).

{\em In our question answering methodology, the structure of
our ${\cal ALM}$ system description for the cell cycle domain
provided the vocabulary for translating the questions expressed in natural language into a history.
The theory of the system description contained the axioms encoding the
background knowledge needed to answer questions about the domain.}

As an example, the information given in the text of \emph{12.9} above
would be encoded by a history that contains the facts

\st

$
\begin{array}{l}
observed(num(cell, sample), 1, 0)\\
observed(num(nucleus, cell), 1, 0)\\
intend(cell\_cycle, 0)\\
\neg happened(cytokinesis, I)
\end{array}
$

\st

\noindent
for every step $I$. Note that, unless otherwise specified,
it would be assumed that the experimental sample consists of
one cell with one nucleus.

The query in \emph{12.9} would be encoded by the ASP\{f\} rules:

\smallskip
\noindent
$
\begin{array}{lll}
answer(X, \mbox{``cells per sample''}) & \leftarrow & last\_step(I),\\
                                       & & num(cell, sample, I) = X.\\
answer(X, \mbox{``nuclei per cell''}) & \leftarrow & last\_step(I), \\
                                      & & num(nucleus, cell, I) = X.
\end{array}
$

\smallskip
\noindent
Our system, ${\cal ALMAS}$, would solve the question answering problem by first generating
a logic program consisting of the above facts and rules encoding the history and query, respectively;
the ASP\{f\} translation of the ${\cal ALM}$ system description $cell\_cycle(2)$;
and the temporal projection module described in Section 4.2.
Then, the system would compute answer sets of this program, which
correspond to answers to the question. For \emph{12.9} there would be a unique answer set,
containing:

\st

$
\begin{array}{ll}
intend(cytokinesis, 2) \ \ \ \ \ \ \ & \neg occurs(cytokinesis, 2) \\
intend(cytokinesis, 3)               & \neg occurs(cytokinesis, 3) \\
intend(cytokinesis, 4)               & \neg occurs(cytokinesis, 4) \\
\dots & \\
\end{array}
$

\st

\noindent
These facts indicate that the unfulfillable intention of executing action
$cytokinesis$ persists forever. Additionally,
the answer set would include atoms:

\st

$
\begin{array}{ll}
answer(1, \mbox{``cells per sample''}) \ \ \ \ \
                                       & holds(val(num(cell, sample), 1),2)\\
answer(2, \mbox{``nuclei per cell''})  & holds(val(num(nucleus, sample), 2),2)\\
last\_step(2)                          & holds(val(num(nucleus, cell), 2),2)
\end{array}
$

\st

\noindent
which indicate that at the end of the cell cycle there will be
one cell in the sample, with two nuclei. This is in fact the correct
answer to question \emph{12.9}.

{\em This question answering methodology
and the methodology of reasoning about naturally evolving processes using intentions
was successfully applied to other questions about cell division.}



\mbox{}\vfill\eject
\section{Comparison between Languages ${\cal ALM}$ and $MAD$}
\label{ALMandMAD}

In this section we give an informal discussion of the relationship
between ${\cal ALM}$ and the modular action language $MAD$
\cite{lr06,el06}. Both languages have similar goals but differ
significantly in the proposed ways to achieve these goals.
We believe that each language supports its own distinctive style of
representing knowledge about actions and change. The difference starts
with the non-modular languages that serve as the basis for ${\cal ALM}$
and $MAD$. The former is a modular expansion of action language ${\cal
  AL}$. The latter expands action language ${\cal C}$
\cite{giul98}. Even though these languages have a lot in common (see
\cite{gl12}) they differ significantly in the underlying assumptions
incorporated in their semantics.
 For example, the semantics of ${\cal AL}$ incorporates  the Inertia Axiom,
which says that ``\emph{Things normally stay the same}.''
Language ${\cal C}$ is based on a different assumption --
the Causality Principle -- which says that ``\emph{Everything true in the world must be caused}.''
Its underlying logical basis is causal logic \cite{mt97,gllmt04}.
In ${\cal C}$ the inertia axiom for a literal $l$ is expressed by a statement
$${\bf caused}\ l \lif l \ {\bf after}\ l,$$
read as ``there is a cause for $l$ to hold after a
transition if $l$ holds both before and after the transition''.
While  ${\cal AL}$ allows two types of fluents -- inertial and
defined --,  ${\cal C}$ can be used to define other types of fluents
(e.g., default fluents that, unless otherwise stated, take on the fixed
default values). The authors of this paper did not find these types of
fluents to be particularly useful and, in accordance with their
minimalist methodology, did not allow them in either ${\cal AL}$ or
${\cal ALM}$. Of course, the question is not settled and our opinion
can change with additional experience. On another hand, ${\cal AL}$
allows recursive state constraints and definitions, which are severely
limited in ${\cal C}$.
There is a close relationship between ASP and
${\cal C}$ but, in our judgment, the distance between ASP and ${\cal AL}$
is smaller than that between ASP and ${\cal C}$.
There is also a substantial
difference between modules of ${\cal ALM}$ and $MAD$.

To better understand the relationship let us consider the  ${\cal ALM}$ theory
$motion$ and the system description $travel$ from Section 3.2
and represent them in $MAD$.\footnote{
Although the ``Monkey and Banana'' problem presented in Section 4.1
has been encoded in $MAD$ as well \cite{thesiser08}, we are not considering it here because of
the length of its representation and, most importantly, because there
are substantial differences in how the problem was addressed in ${\cal ALM}$ versus $MAD$
from the knowledge representation point of view.
}
\begin{example}[A $MAD$ Version of the System Description \emph{travel}]
The ${\cal ALM}$ system description $travel$ is formed by the
 theory $motion$ and the structure $Bob\_and\_John$. The theory
consists of two modules,
$moving$ and
$carrying\_things$, organized into a module hierarchy in which the latter module
depends on the former.
Let us start with the $MAD$ representation of ${\cal ALM}$'s module
$moving$.

In general, the representation of an ${\cal ALM}$ module
$M$ in $MAD$ consists of two parts:
the \emph{declaration of sorts} of $M$ and their \emph{inclusion relation},
and the collection of $MAD$
\emph{modules} corresponding to $M$.
(In our first example a module of ${\cal
  ALM}$ will be mapped into a single module of $MAD$.) Note that
sorts can also be declared within the module but in this case they
will be local (i.e., invisible to other modules). Declarations 
given outside of a module can be viewed as global.

In our case, the ${\bf sorts}$ and ${\bf inclusions}$ sections of the
translation\\ 
$M_1=MAD(moving)$ consist of the following
statements (We remind the reader that in $MAD$ variables are identifiers starting with a lower-case letter
and constants are identifiers starting with an upper-case letter, the opposite of ${\cal ALM}$ ):

\smallskip\noindent
$
\begin{array}{l}
{\bf sorts}\\
\ \ Universe; Points; Things; Agents;\\ 
{\bf inclusions}\\
\ \ Points << Universe; \\
\ \ Things << Universe;\\
\ \ Agents << Things;
\end{array}
$

\st
The ${\bf sorts}$ part declares the sort $universe$ (which is pre-defined
and does not require declaration in ${\cal ALM}$)
together with the sorts of $moving$ that are not special cases of $actions$.
The ${\bf inclusions}$ part describes the specialization relations between these sorts.
The definition of a $MAD$ module starts with a title:

\medskip\noindent
$
\begin{array}{l}
{\bf module} \ M_1
\end{array}
$

\medskip\noindent
The body of a $MAD$ module consists of separate (optional) sections for the declarations of sorts
specific to the current module, objects, fluents, actions, and variables, in this order,
together with a section dedicated to axioms \cite{thesiser08}.
Our module $M_1$ starts with the declarations of fluents:

\medskip\noindent
$
\begin{array}{l}
\ \ {\bf fluents} \\
\ \ \ \ Symmetric\_connectivity : rigid;\\
\ \ \ \ Transitive\_connectivity : rigid;\\
\ \ \ \ Connected(Points, Points) : simple;\\
\ \ \ \ Loc\_in(Things) : simple(Points);
\end{array}
$

\medskip\noindent
{\em Rigid} fluents of $MAD$ are {\em basic statics} of ${\cal ALM}$.

\medskip
To declare the action class $move$ of $moving$ we need to model
its attributes. To do that we introduce variables with the same names
as the associated attributes in $moving$.
This will facilitate referring to those attributes later in axioms.
We also order attributes alphabetically as arguments of the action term
to ease the translation of special case action classes of $move$:

\medskip\noindent
$
\begin{array}{ll}
\ \ {\bf actions} \\
\ \ \ \ Move(Agents, Points, Points);
\end{array}
$

\medskip
The variable declaration and axiom part come next.
We will need to add extra axioms (and associated variables)
to say that $Loc\_in$ 
is an inertial fluent
(i.e., {\em basic} fluent in ${\cal ALM}$ terminology)
and that $Move(Agents, Points, Points)$ is an exogenous action
(i.e., it does not need a cause in order to occur; it may or may not occur
at any point in time).

\medskip\noindent
$
\begin{array}{l}
\ \ {\bf variables} \\
\ \ \ \ t, t_1, t_2 : Things;\\
\ \ \ \ actor : Agents;\\
\ \ \ \ origin, dest : Points;
\end{array}
$

\noindent
$
\begin{array}{l}
\ \ {\bf axioms} \\
\ \ \ \ {\bf inertial}\ Loc\_in(t);\\
\ \ \ \ {\bf exogenous}\ Move(actor, dest, origin);
\end{array}
$

\medskip\noindent
The causal law for $move$ can now be expressed in a natural way:

\medskip\noindent
$
\begin{array}{l}
\ \ \ \ Move(actor, dest, origin)\ {\bf causes}\ Loc\_in(actor) = dest;
\end{array}
$

\medskip\noindent
Similarly for the executability conditions:

\medskip\noindent
$
\begin{array}{lll}
\ \ \ \ {\bf nonexecutable} \ Move(actor, dest, origin) & {\bf if} &
Loc\_in(actor) \neq origin;
\end{array}
$

\smallskip\noindent
$
\begin{array}{lll}
\ \ \ \ {\bf nonexecutable} \ Move(actor, dest, origin) & {\bf if} & Loc\_in(actor) = dest;
\end{array}
$

\smallskip\noindent
$
\begin{array}{lll}
\ \ \ \ {\bf nonexecutable} \ Move(actor, dest, origin) & {\bf if} & Loc\_in(actor) = origin,\\
                                                        &          & \neg Connected(origin, dest);
\end{array}
$

\medskip\noindent
The situation becomes substantially more difficult for the definition
of $Connected$. The definition used in $moving$ is recursive
and therefore cannot be easily emulated by $MAD$'s causal laws.
The relation can, of course, be \emph{explicitly} specified later together with the
description of particular places, but this causes considerable
inconvenience.

\medskip
To represent module $carrying\_things$ from the theory $motion$ we
need a new (global) sort:

\medskip\noindent
$
\begin{array}{l}
{\bf sorts}\\
\ \ Carriables; \\
{\bf inclusions}\\
\ \ Carriables << Things;
\end{array}
$

\medskip\noindent
The module $M_2$ that corresponds to $carrying\_things$ contains
declarations of the new action $Carry$ and the corresponding
variables.

\medskip\noindent
$
\begin{array}{l}
{\bf module} \ M_2;\\
\ \ {\bf actions} \\
\ \ \ \ Carry(Agents, Carriables, Points, Points);\\
\ \ {\bf variables} \\
\ \ \ \ t : Things;\\
\ \ \ \ actor : Agents;\\
\ \ \ \ dest, origin, p : Points;\\
\ \ \ \ carried\_object, c : Carriables;
\end{array}
$

\medskip\noindent
Next we need to define axioms of the module.
Clearly we need to say that the action\\ $Carry(actor, carried\_object, dest, origin)$
is a special case of the action\\ $Move(actor, dest, origin)$. Since
${\cal ALM}$ allows action sorts, no new mechanism is required
to do that in $carrying\_things$. In $MAD$, while there is a built-in
sort \emph{action}, special case actions are not sorts and the
special constructs ${\bf import}$ and ${\bf is}$ are introduced to achieve this goal.
Special case actions are declared in $MAD$ by importing the module
containing the original action and renaming the original action as the
special case action as follows:

\medskip\noindent
$
\begin{array}{l}
\ \ {\bf import}\ M_1;\\
\ \ \ \ Move(actor, dest, origin)\ {\bf is}\ Carry(actor, carried\_object, dest, origin);
\end{array}
$

\medskip\noindent
Intuitively, this import statement says that the action
$Carry(actor, carried\_object,$ $dest, origin)$ has all properties that
are postulated for the action
 $Move(actor, dest,$ $origin)$ in the
module $M_1$. We also need an additional axiom declaring the action to
be exogenous, and state constraints, and executability conditions similar to those in $carrying\_things$:

\medskip\noindent
$
\begin{array}{l}
\ \ {\bf axioms}\\
\ \ \ \ {\bf exogenous}\ Carry(actor, carried\_object, dest, origin);
\end{array}
$

\medskip\noindent
$
\begin{array}{ll}
\ \ \ \ \mbox{\% State constraints:} &\\
\ \ \ \ Is\_held(c)\ \  \  {\bf if} & Holding(t, c);
\end{array}
$

\smallskip\noindent
$
\begin{array}{l}
\ \ \ \ \mbox{\% Executability conditions:} \\

\ \ \ \ {\bf nonexecutable} \ Carry(actor,carried\_object,dest,origin) \ {\bf
  if}\\
 \ \ \ \ \ \ \ \ \  \ \ \ \ \ \ \ \ \  \ \  \ \ \ \ \ \ \ \ \   \neg Holding(actor,carried\_object);
\end{array}
$

\noindent
$
\begin{array}{l}
\ \ \ \ {\bf nonexecutable} \ Move(actor, dest, origin)  \ {\bf if} \ Is\_held(actor);
\end{array}
$

\smallskip\noindent
Note, however, that the ${\cal ALM}$ module $carrying\_things$
also contained the {\em recursive} state constraints below,
saying that agents and the objects they are holding 
have the same location:

\noindent
$
\begin{array}{lll}
\ \ \ \ \ \ \ \ \ \ \ \ loc\_in(C) = P & \lif & holding(T, C), loc\_in(T) = P.\\
\ \ \ \ \ \ \ \ \ \ \ \ loc\_in(T) = P & \lif & holding(T, C), loc\_in(C) = P.
\end{array}
$

\noindent
Since this is not allowed in $MAD$, we have to use a less elaboration tolerant 
representation by adding an explicit causal law saying

\noindent
$
\begin{array}{l}
\ \ \ \ Move(actor, dest, origin)  \ {\bf causes} \ Loc\_in(c) = dest
                                   \ {\bf if} \ Holding(actor, c);
\end{array}
$

\noindent
In $MAD$ additional axioms
will be needed to rule out certain initial situations
(e.g., {\em ``John is holding his suitcase. He is in Paris. His suitcase is in Rome.''}) 
or to represent and reason correctly about more complex scenarios 
(e.g., {\em ``Alice is in the kitchen,
holding her baby who is holding a toy.
Alice goes to the living room.''}).

\medskip\noindent
This completes the construction of $M_2$.

\medskip
In general, special case actions are declared in $MAD$ by importing the module
containing the original action and renaming the original action as the special case action.
That is why we needed to place the $MAD$ representation of $carry$ in a new module
that we call $M_2$, in which we import module $M_1$ while renaming $Move(actor, dest, origin)$ as
$Carry(actor, carried\_object, dest, origin)$. In ${\cal ALM}$
the declarations of $move$ and its specialization $carry$ could be placed in the same module -- the
decision is up to the user -- whereas in $MAD$ they {\em must} be placed in separate modules.
This potentially leads to a larger number of smaller modules in $MAD$ than in ${\cal ALM}$
representations.

\medskip
Finally, we consider the structure of our ${\cal ALM}$ system
description. It contains two types of actions $go(Actor,Dest)$ and
$go(Actor,Dest,Origin)$. Let us expand the structure by a new object,
$suitcase$, and a new action $carry(Actor,suitcase,Dest)$.
For illustrative purposes, let us assume that we would like the $MAD$ representation to preserve
these names.

To represent this in $MAD$, we introduce a new module $S$.
It has the local
definitions of objects:

\medskip\noindent
$
\begin{array}{l}
{\bf module} \ S;\\
\ \ {\bf objects} \\
\ \ \ \ John, Bob : Agents;\\
\ \ \ \ New\_York, Paris, Rome : Points;\\
\ \ \ \ Suitcase : Carriables;
\end{array}
$

\medskip\noindent
and those of actions. The latter can be defined via the renaming
mechanism of $MAD$. This requires importing
the modules in which the action classes were declared.
Thus, module $S$ imports modules $M_1$ and $M_2$.

\medskip\noindent
$
\begin{array}{l}
\ \ {\bf actions} \\
\ \ \ \ Go(Agents, Points);\\
\ \ \ \ Go(Agents, Points, Points);\\
\ \ \ \ Carry(Agents, Carriables, Points);\\
\end{array}
$

\smallskip\noindent
$
\begin{array}{l}
\ \ {\bf variables} \\
\ \ \ \ actor : Agents;\\
\ \ \ \ origin, dest : Points;\\
\end{array}
$

\smallskip\noindent
$
\begin{array}{l}
\ \ {\bf import} \ M_1;\\
\ \ \ \ Move(actor, dest, origin) \ {\bf is} \ Go(actor, dest, origin);
\end{array}
$

\smallskip\noindent
$
\begin{array}{l}
\ \ {\bf import} \ M_1;\\
\ \ \ \ Move(actor, dest, origin) \ {\bf is}\ Go(actor, dest);\\
\end{array}
$

\smallskip\noindent
$
\begin{array}{ll}
\ \ {\bf import} \ M_2;\\
\ \ \ \ Carry(actor,Suitcase, dest, origin) \ {\bf is}\ Carry(actor, Suitcase, dest)
\end{array}
$

\st
This completes the construction of the $MAD$ representation of the system
description $travel$.
\end{example}

\medskip
Even this simple example allows to illustrate some important
differences between ${\cal ALM}$ and $MAD$. Here is a short summary:

\begin{itemize}
\item {\em Recursive definitions}

The representation of state constraints of an ${\cal ALM}$ system description
is not straightforward if the set of state constraints defines a {\em cyclic}
fluent dependency graph \cite{gl12}. For instance, the ${\cal ALM}$ state constraint:
$$p \lif p.$$
is not equivalent to the same axiom in $MAD$. The ${\cal ALM}$ axiom can be eliminated
without modifying the meaning of the system description; it says that
``in every state in which $p$ holds, $p$ must hold.'' Eliminating the same axiom
from a $MAD$ action description would not produce an equivalent action description; in $MAD$,
the axiom says that ``$p$ holds by default.''  This difference between
${\cal ALM}$ and $MAD$ is inherited from the similar difference
between ${\cal AL}$ and ${\cal C}$.

\item \emph{Separation of Sorts and Instances}

One of the most important features of ${\cal ALM}$ is its
support for a clear separation of the definition
of \emph{sorts of objects} of the domain (given in the system's theory)
from the definition of \emph{instances} of
these sorts (given by the system's structure).
Even though it may be tempting to view the first two modules, $M_1$
and $M_2$ above as a $MAD$ counterpart of the ${\cal ALM}$ theory
$motion$, the analogy does not hold. Unlike ${\cal ALM}$ where the
corresponding theory has a clear semantics independent of that of the
structure, no such semantics exists in $MAD$. Modules $M_1$ and $M_2$
only acquire their meaning after the addition of module $S$ that
corresponds to the ${\cal ALM}$'s structure.
We believe that the existence of the independent semantics of ${\cal ALM}$ theories
facilitates the stepwise development and testing of the knowledge base
and improves their elaboration tolerance.

\item {\em Action Sorts}

In ${\cal ALM}$, the pre-defined sort $actions$ is part of the sort hierarchy,
whereas in $MAD$ actions are not considered sorts.
Instead, $MAD$ has special constructs ${\bf import}$ and ${\bf is}$
(also known as \emph{bridge rules}), which are used to define actions
as special cases of other actions. No such special constructs are
needed in ${\cal ALM}$.

Moreover, in ${\cal ALM}$, an action class and its specialization can be part of the same module.
This is not the case in MAD where a special case of an action class
must be declared in a separate module by importing the module containing the original action class
and using renaming clauses. As a consequence, the $MAD$ representation of ${\cal ALM}$
system descriptions will generally contain more modules
that are smaller in size than the ${\cal ALM}$ counterpart.
On the other hand, note that ${\cal ALM}$ modules are not required to be large;
they can be as small as a user desires.

${\cal ALM}$ allows the definition of fluents on (or ranging over)
\emph{specific} action classes only, and not necessarily the whole pre-defined $actions$ sort,
for instance:
$$intended : agent\_actions \rightarrow booleans$$
where $agent\_actions$ is a special case of $actions$.
There is no equivalent concept in $MAD$, where
fluents must be defined on, and range over, either
primitive sorts or the built-in sort $action$, but not specific actions.

\item {\em Variable Declarations}

In ${\cal ALM}$, we do not define the sorts of variables used in the
axioms. This information is
evident from the atoms in which they appear. In $MAD$, variables need to be defined, which
may lead to larger modules and cause errors related to use of
variables of wrong types.

\item {\em Renaming Feature of $MAD$}

In $MAD$, sorts can be renamed by importing the module containing
the original declaration of a sort and using a renaming clause.
The meaning of such a renaming clause is that the two sorts are synonyms.
There is no straightforward way to define this synonymy in ${\cal ALM}$.
The closest thing is to use the specialization construct of our language
and declare the new sort as a special case of the original one. The
reverse (i.e., the original sort being a special case of the renamed sort)
cannot be added, as sort hierarchies of ${\cal ALM}$ are required to be
DAGs. This leads to further problems when the renamed sorts appear as
attributes in renamed actions of $MAD$.

\item {\em Axioms of $MAD$ that have no equivalent in ${\cal ALM}$}

Some axioms, allowed in $MAD$, are not directly expressible in ${\cal ALM}$.
For instance,
$MAD$ axioms of the type:
$$formula\ {\bf may\ cause }\ formula\ [\lif formula\ ]$$
or
$${\bf default}\ formula\ [\lif formula\ ] \ [\ {\bf after}\ formula\ ]$$
belong to this group. The first axiom allows to specify
non-deterministic effects of actions, while the second assignes default
values to fluents (and more complex formulas).
As discussed above, we are not yet convinced
that the latter type of axioms needs to be allowed in ${\cal ALM}$.
Non-determinism, however, is an important feature that one should be
able to express in an action formalism. It may be added to ${\cal
  ALM}$ (and to ${\cal AL}$) in a very natural manner, but it is not allowed in
${\cal AL}$ and the mathematical properties of ``non-deterministic''
${\cal AL}$  were not yet investigated.
Because of this we decided to add this feature in the next version of
${\cal ALM}$.

\end{itemize}

\medskip
We hope that this section gives the reader some useful insight in
differences between ${\cal ALM}$ and $MAD$.
We plan to extend the comparison between ${\cal ALM}$ and MAD in the future.
Formally investigating the relationship between the two languages can facilitate the
translation of knowledge modules from one language to another, and can identify
situations when one language is preferable to the other.
Readers interested in a formal translation of system descriptions of
${\cal ALM}$ to action descriptions of $MAD$ can consult \cite{thesisdi12}.

\end{appendix}

\newpage
\bibliography{alm}
\end{document}